\documentclass[%
reprint,
groupedaddress,
amsmath,amssymb,amsfont,aps,
prb,
floatfix,
]{revtex4-2}

\usepackage{graphicx}
\usepackage{dcolumn}
\usepackage{bm,dsfont}
\allowdisplaybreaks
\usepackage[breaklinks]{hyperref}
\hypersetup{colorlinks=true, linkcolor=blue, citecolor=blue, filecolor=blue, urlcolor=blue}
\usepackage[capitalise]{cleveref}
\graphicspath{{figures}}
\usepackage{tikz}
\usepackage{mathtools}
\usepackage{soul}
\usepackage{physics,comment}

\def\CH{\textcolor{black}}

\newcommand{\leftrarrows}{\mathrel{\raise.75ex\hbox{\oalign{%
			$\scriptstyle\leftarrow$\cr
			\vrule width0pt height.5ex$\hfil\scriptstyle\relbar$\cr}}}}
\newcommand{\lrightarrows}{\mathrel{\raise.75ex\hbox{\oalign{%
			$\scriptstyle\relbar$\hfil\cr
			$\scriptstyle\vrule width0pt height.5ex\smash\rightarrow$\cr}}}}
\newcommand{\Rrelbar}{\mathrel{\raise.75ex\hbox{\oalign{%
			$\scriptstyle\relbar$\cr
			\vrule width0pt height.5ex$\scriptstyle\relbar$}}}}

\makeatletter
\def\leftrightarrowsfill@{\arrowfill@\leftrarrows\Rrelbar\lrightarrows}
\newcommand{\xleftrightarrows}[2][]{\ext@arrow 3399\leftrightarrowsfill@{#1}{#2}}
\newcommand{\wj}{\textcolor{black}}

\begin{document}


\author{Ruizhe Shen}
\email{e0554228@nus.edu.sg}
\affiliation{Department of Physics, National University of Singapore, Singapore 117542}
\author{Wei Jie Chan}
\email{wj\_chan@nus.edu.sg}
\affiliation{Department of Physics, National University of Singapore, Singapore 117542}
\author{Ching Hua Lee}
\email{phylch@nus.edu.sg}
\affiliation{Department of Physics, National University of Singapore, Singapore 117542}


\title{Non-Hermitian Skin Effect Along Hyperbolic Geodesics}
\date{\today}
\begin{abstract}
Recently, it has been revealed that a variety of novel phenomena emerge in hyperbolic spaces, while non-Hermitian physics has significantly enriched the landscape of condensed matter physics. 
Building on these developments, we construct a geodesic-based method to study the non-Hermitian skin effect (NHSE) in non-reciprocal hyperbolic lattices.
Additionally, we develop a geodesic-periodic boundary condition (geodesic-PBC), akin to the Euclidean periodic boundary condition (PBC), that complements its open boundary condition. 
Importantly, we find that the non-reciprocal directionality within a hyperbolic regular polygon and the geodesic-based boundary determine the spectral sensitivity, and hence, the NHSE. 
Unlike in Euclidean models, however, we must utilize boundary localization to distinguish non-trivial skin modes from their trivial boundary counterpart due to the extensive boundary volume of hyperbolic lattices.
We also relate the spatial density profile with the finite-size scaling of hyperbolic lattices.
These aspects highlight the profound impact of hyperbolic geometry on non-Hermitian systems and offer new perspectives on the intricate relationship between the geometric characteristics of hyperbolic lattices and non-Hermitian physics.
\end{abstract}

\maketitle
\section{Introduction}
The non-Hermitian skin effect (NHSE) reveals intriguing non-local behavior that leads to non-trivial broken bulk-boundary correspondence, physically arising from non-Hermitian gain/loss or asymmetry \cite{kunst2018biorthogonal,xiong2018does,song2019non,longhi2019probing,xiao2020non,helbig2020generalized,okuma2020topological,li2020critical,arouca2020unconventional,pan2020non,yang2024non,shen2022non,yang2022designing,li2022non,zou2021observation,zhang2021observation,qin2022non,Rafi-Ul-Islam2022,Jiang2023,zhang2022real,qin2024kinked,qin2023universal,shen2023observation,shen2024enhanced,liu2024non,Yang2024eud,li2024observation,gliozzi2024many,liu2024localization,yoshida2024non,liu2024non,Gliozzi2024a,Li2022c,Sun2024a}.
Recent studies show that the manifestation of the NHSE is significantly influenced by spatial properties such as dimensionality, lattice geometry, and boundary configurations \cite{shen2021non,schindler2021dislocation,zhang2022review,shang2022experimental,Jiang2023,manna2022inner,gu2016holographic,zhong2024higher,Sun2024a,qin2024geometry,qin2024dynamical,xiong2024non,shang2024observation,Hamanaka2024,Hamanaka2024a}.
For instance, higher spatial dimensions give rise to the emergence of higher-order skin modes  \cite{zhang2022review,okugawa2020second,kawabata2020higher,zhang2022universal}, while unique spatial topology, such as lattice dislocations \cite{schindler2021dislocation} and fractal geometries \cite{manna2022inner,Hamanaka2024,Zhang2024} can induce unique localizations beyond conventional skin modes. 
Particularly, the geometry-dependent skin effect sheds light on the intricate relationship between the NHSE and spatial inhomogeneities \cite{fang2022geometry,zhou2023observation,wang2023experimental}, thus emphasizing the critical role of spatial geometry in determining the nature of the NHSE.

Recently, hyperbolic geometry has garnered much interest as the interplay of negatively curved spaces and quantum matter has intrigued the community \cite{song2021mobius,maciejko2021hyperbolic,stegmaier2022universality,liu2022chern,boettcher2022crystallography,maciejko2022automorphic,cheng2022band,bienias2022circuit,zhang2022observation,lenggenhager2022simulating,chen2022hyperbolic,bienias2022circuit,Sun2023,yu2020topological,Lux2023,Urwyler2022,Lenggenhager2024,Tummuru2023}.
Pioneering works on circuit boards \cite{lenggenhager2022simulating,Kollar2019,Zhang2022,Yuan2024,Pei2023,Zhang2023} and photonics \cite{Huang2024} have demonstrated the practical potential of such interesting constructs. 
Moreover, significant research efforts have delved into the intricacies of topological phases emerging on hyperbolic surfaces and lattices \cite{maciejko2021hyperbolic,stegmaier2022universality,liu2022chern,boettcher2022crystallography,maciejko2022automorphic,cheng2022band,bienias2022circuit,zhang2022observation,lenggenhager2022simulating,chen2022hyperbolic,bienias2022circuit,Sun2023,Lux2023,Lenggenhager2024}. 
Although there are several developments in the study of NHSE on hyperbolic lattices \cite{Lv2022,Sun2023,Chadha2024}, their enormous boundary volume makes it fundamentally difficult to establish the bulk-boundary correspondence, necessary for understanding the NHSE.
Saliently, the translation operators in hyperbolic (negatively curved) spaces involve rotations \cite{boettcher2022crystallography} which can obstruct the tracking of non-reciprocity, and consequently, obscure the directions of the NHSE. 
Thus, two pivotal inquiries emerge: (i) How can one construct the analogs of open and periodic boundary conditions (OBCs/PBCs) on hyperbolic lattices \cite{cheng2022band,stegmaier2022universality,lux2023converging}, notions crucial for characterizing the NHSE? (ii) How does the interference between non-reciprocal hoppings in different directions carry over in curved space, and what are the implications for the NHSE \cite{sun2023hybrid,lv2024hidden}?

To address these questions, our work first proposes a geodesic-based method for generating finite-size hyperbolic lattices to capture the non-reciprocal hopping directionalities effectively. 
Next, we establish an adaptive \emph{pseudo-periodic boundary condition} (geodesic-PBC), in place of the usual PBC, which complements the constructed OBC. 
Similar to geodesics in Euclidean lattices, which are straight lines, our intricately constructed hyperbolic lattices are built from a network of geodesics with the geodesic-PBCs achieved by connecting the ends of these geodesics.
This elaborate construction of hyperbolic lattices enables the precise identification of non-reciprocal directions in both the bulk and the boundary.
To substantiate this construction, we investigate the non-Hermitian skin modes in two distinct hyperbolic lattices.
This examination is based on the non-reciprocity observed in their respective hyperbolic regular polygons and along their boundaries, which dictates the characteristics of the NHSE.
Unlike in non-Hermitian Euclidean models, the extensive boundary volume in hyperbolic lattices blurs the distinction between the non-Hermitian skin modes from the trivial boundary modes.
Here, we separate the non-Hermitian skin modes from the trivial boundary modes by comparing the boundary localization of energy states with and without non-Hermiticity.
Ultimately, we explicitly relate the finite-size scaling with the existence of these non-Hermitian skin modes. 

The remaining work is organized as follows, with \cref{sec1} beginning with a foundational overview of hyperbolic lattices. 
We detail our elaborate methodology for generating finite-size lattice models under both OBCs and fully closed geodesic loops; i.e. geodesic-PBCs, analogous to the PBCs in Euclidean lattices.
Building upon this framework, we discuss how to construct a non-Hermitian tight-binding Hamiltonian, with non-reciprocal hopping translating along geodesics.
In \cref{sec2}, we show how non-reciprocity within a hyperbolic regular polygon and on the boundary affects the NHSE in our constructed hyperbolic lattices through their energy spectra and we also compare with two relevant Euclidean square lattices. 
Next, the process of distinguishing the non-trivial non-Hermitian skin modes from the trivial boundary modes is shown through the boundary localization of eigenstates under zero and finite non-Hermiticity.
Furthermore, we find intrinsic boundary modes in our proposed hyperbolic models, which arise from the curved spatial nature and the unique scaling laws.
The key findings are then summarized in \cref{sec:con}.
Finally, the finite-size scaling laws, Hamiltonians and shortest path distributions of relevant Euclidean models are also in \cref{scale,ap1,appx:manhattan}.

\section{Geodesic loops on Hyperbolic lattices \label{sec1}}
\begin{figure}
	\centering
	\includegraphics[width=0.99\linewidth]{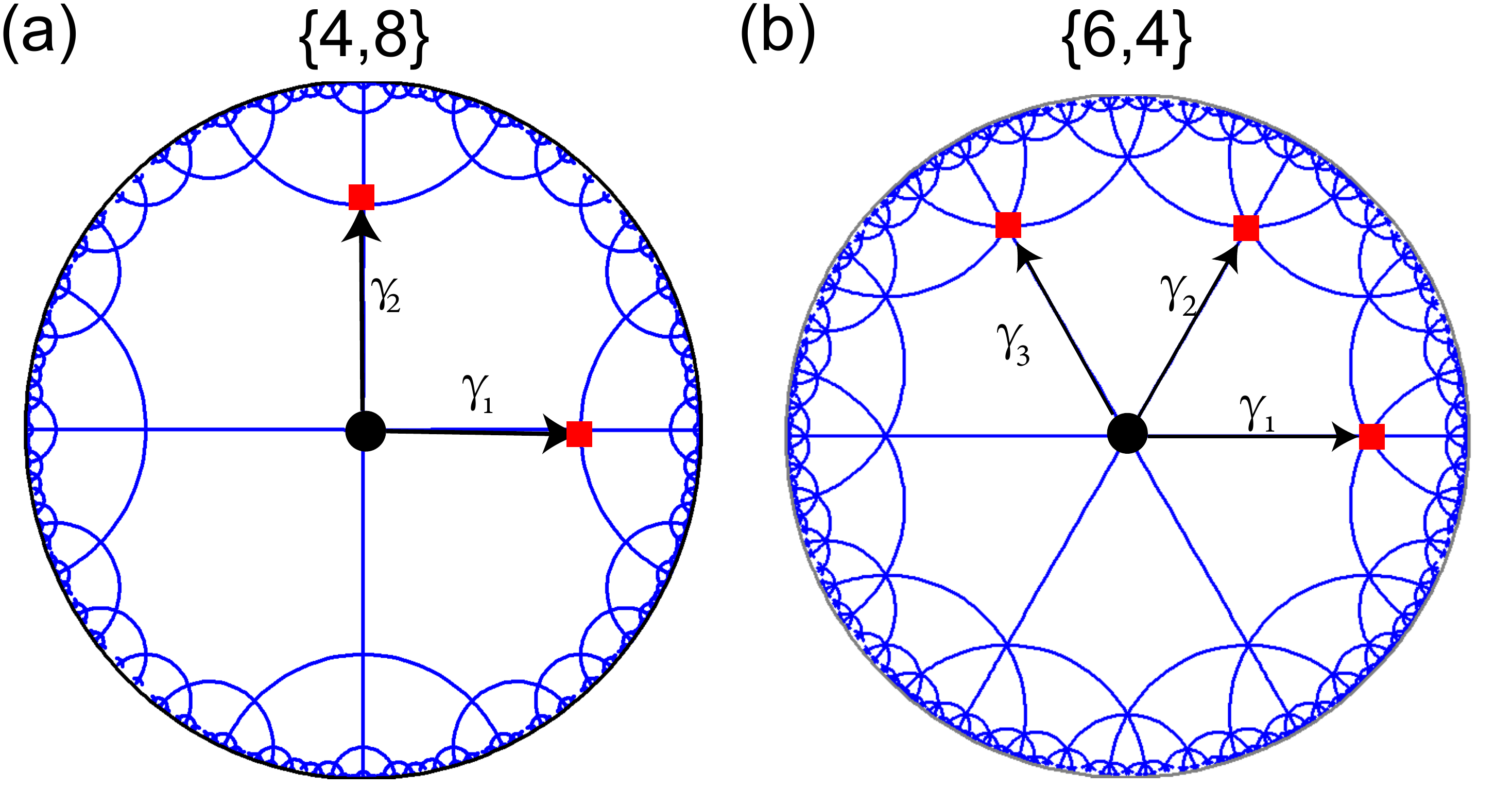}
	\caption{The structure of hyperbolic lattices featured in this work, characterized by $\{p,q\}=\{4,8\}$ (a) and $\{6,4\}$ (b). The translation operators, $\gamma_{\mu}$ (black arrows) generate the entire lattice from an initial site $z=0$ (black dot) on the Poincar\'e disk, with first-generation sites shown in red.
 }
	\label{fig:lattices}
\end{figure}

In this section, we first delve into the fundamentals of utilizing tiling patterns, which are formed by hyperbolic geodesics, to construct lattice models \cite{cannon1997hyperbolic,milnor1982hyperbolic,benedetti1992lectures,ramsay1995introduction}. 
These hyperbolic patterns are projected onto a Poincaré disk model, which is defined as $\{z\in \mathbb{C}:\abs{z}<1\}$ as shown in \cref{fig:lattices} (a) \cite{yu2020topological,brower2022hyperbolic,liu2022chern,boettcher2022crystallography}. 
Each pattern is uniquely defined by a pair of positive integers $p,q$ where $p$ is the coordination number of each vertex and $q$ is the number of sides of each polygon.
We label the pattern as $\{p,q\}$ in this work, which is also the Schl\"afli symbol \cite{zhu2021quantum,yu2020topological,boettcher2022crystallography} of its \emph{dual} hyperbolic lattice; $(p-2)(q-2)>4$ for hyperbolic lattices. 
We shall present the \{$4$,$8$\} and \{$6$,$4$\} hyperbolic lattices depicted in \cref{fig:lattices} as representative examples, and also compare their properties with the usual \{$4$,$4$\} square lattice in Euclidean (flat) space.


\subsection{Translations in Hyperbolic Plane}
We first describe the hyperbolic translation operators that are necessary for defining hyperbolic lattices \cite{cassels2001topics,coxeter2013generators,boettcher2022crystallography}. 
On the Poincar\'e disk, elementary translations on a hyperbolic $\{p,q\}$ tiling are realized by members of the Fuchsian translation group \cite{coxeter2013generators}, which can be expressed either as matrix operations or as fractional linear conformal mappings $z \rightarrow \gamma z$:
\begin{align}
    \gamma z := \begin{pmatrix}
        \gamma_{11}&\gamma_{12}\\\gamma_{12}^*&\gamma_{11}^*
    \end{pmatrix}z = \frac{\gamma_{11} z+ \gamma_{12}}{\gamma_{12}^*z+\gamma_{11}^*}, \quad \gamma_{11},\gamma_{12} \in \mathbb{C}.
\end{align}
As written, the matrix elements $\gamma_{11},\gamma_{12}$ and their conjugates encodes transformations such as Euclidean translations or rotations. 
In this form, the 
translation generators, \wj{$\gamma_1,\gamma_2,\dots$} of a generic $\{p,q\}$ hyperbolic lattice are explicitly given by
\begin{align}\label{ge1}
    \gamma_{1}=&\frac{1}{\sqrt{1-\sigma^{2}}} 
    \begin{pmatrix}
        1&\sigma\\\sigma&1
    \end{pmatrix},\quad \sigma=\sqrt{\frac{\left(\cos\frac{2\pi}{p}+\cos\frac{2\pi}{q}\right)}{\left(1+\cos\frac{2\pi}{q}\right)}},\nonumber \\ 
    \gamma_{\mu}&=R\left(\frac{(\mu-1)\pi}{p}\right) \gamma_{1} R\left(-\frac{(\mu-1)\pi}{p}\right),
\end{align} 
where $\gamma_\mu$, with \wj{non-negative} $\mu$, is related to $\gamma_1$ via a discrete rotation implemented by
\cite{magnus1974noneuclidean,balazs1986chaos,cassels2001topics,coxeter2013generators,boettcher2022crystallography}
\begin{equation}\label{ge2}	
	R\left(\frac{(\mu-1)\pi}{p}\right) =\left(\begin{array}{cc}
		e^{\mathrm{i}(\mu-1) \pi / p} & 0 \\
		0 & e^{-\mathrm{i}(\mu-1) \pi / p}
	\end{array}\right).
\end{equation}
While the $\gamma_\mu$ translation operators allow us to define the underlying Bravais lattice with respect to the primitive unit cell \cite{maciejko2021hyperbolic,Lenggenhager2023,Mosseri2023,Lux2023,Lux2023a,boettcher2022crystallography}, akin to Euclidean cases, vertices and connecting edges are simultaneously rotated as they are translated [\cref{ge1,ge2}]. 
This leads to the non-commutativity of translations since different sequences of translation operations can ultimately lead to the same point. 
In the following, we shall show how to consistently define directed loops for non-Hermitian pumping via these translation generators.  


\begin{figure}[ht!]
	\centering
	\includegraphics[width=.95\linewidth]{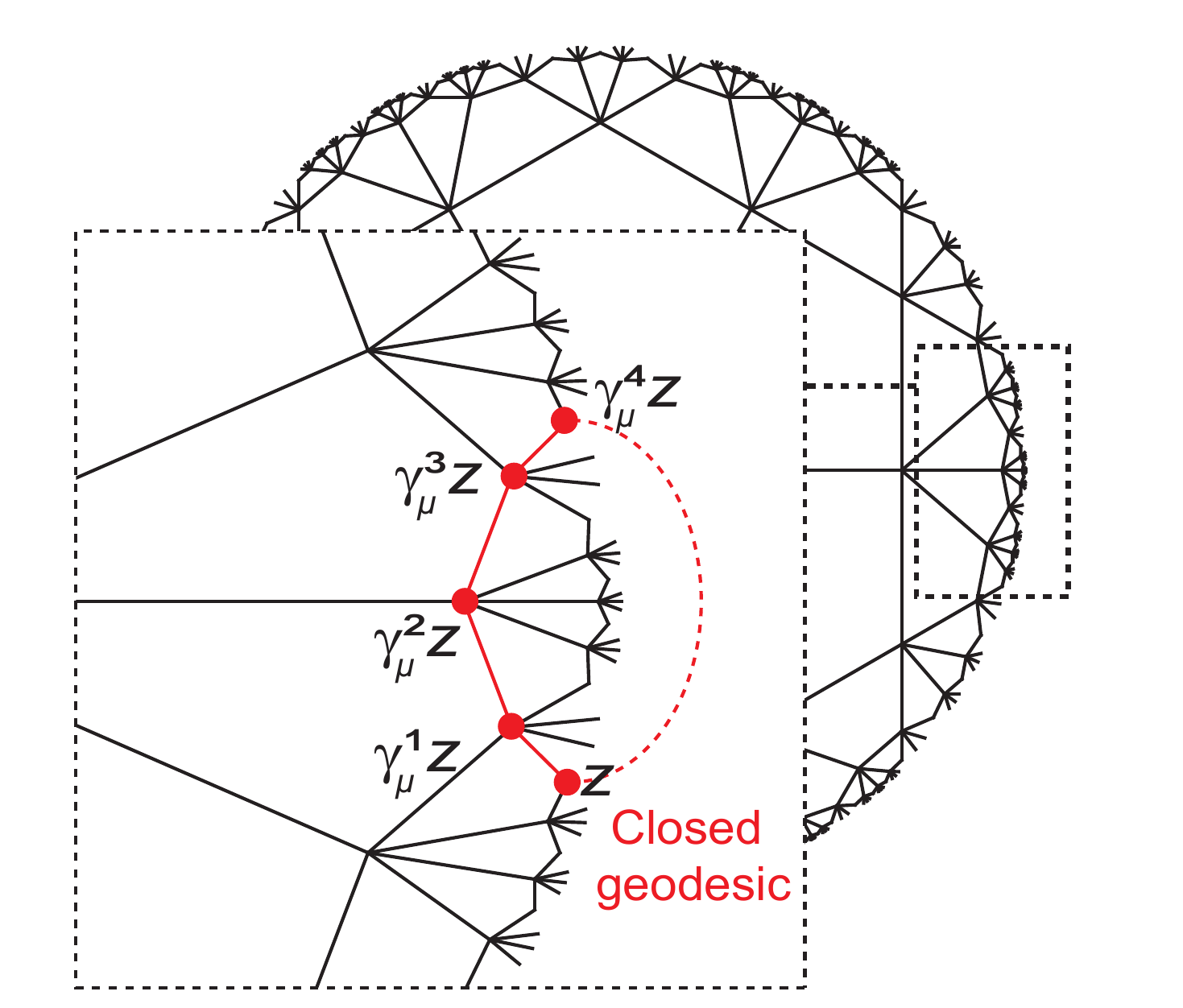}
	\caption{Illustration of constructing geodesic-PBCs by closing an open geodesic into a geodesic loop [see \cref{eqn:psuedoPBC}]. We show a representative geodesic in the \{$6$,$4$\} hyperbolic lattice. The red sequence of edges connects nodes from $z$ to $\gamma^{4}_\mu z$, with a length-$5$ geodesic loop formed by linking the two ends (red dashed line). The lattice experiences full geodesic-PBCs when all the geodesics are closed across all boundary nodes.
 }
	\label{fig:pbc}
\end{figure}

\subsection{OBCs and geodesic-PBCs through hyperbolic geodesics}
To set the stage for studying non-Hermitian skin pumping in hyperbolic geometry, we first define the bulk and boundary of a finite hyperbolic lattice. 
Starting from the center $z_0=(0,0)$,  we can apply translation operators $\gamma_\mu$ [\cref{ge1}] successively for up to $L$ times to generate the following set of lattice sites
\begin{equation}\label{gen}
	\{z \mid z= \overbrace{\gamma_{\mu} \cdots \gamma_{\mu'}}^{x} z_{0},\quad 0\leq x\leq L\},\
\end{equation}
where $\mu$, $\mu' \in [1,\mu_{\text{max}}]$ and $L$ is the number of total generations. Here, we only keep the composition of translation operators, $\gamma_{\mu} \cdots \gamma_{\mu'}$ that first generates the shortest path from the origin, $z_0$ to some $z$.
This guarantees the absence of duplicated sites in the generated hyperbolic lattices.
A site is defined to be an element of the bulk if it requires less than \CH{$L$} translations from $z_0$. 
Conversely, a boundary site requires exactly \CH{$L$} translations from $z_0$. This definition can be written as
\begin{equation}\label{bk}
	\begin{aligned}
		&\{\mathrm{bulk}\}=\{z \mid z=\overbrace{\gamma_{\mu} \cdots \gamma_{\mu'}}^{x} z_{0},\quad 0\leq x<L\},\\
		&\{\text{boundary}\}=\{z \mid z= \overbrace{\gamma_{\mu}\cdots\gamma_{\mu'}}^{x} z_{0}, \quad x=L\}.
	\end{aligned}
\end{equation}
The boundary set is well-defined for any hyperbolic lattice and provides \CH{an} intuitive definition for OBCs when we consider the NHSE. 
Note that this boundary so-defined differs from the smooth boundary of the Poincar\'e disk at $\abs{z}=1$. 


A concrete hyperbolic lattice implementation of PBCs is also required to investigate the NHSE. 
Recently, various notions of PBCs have been formulated via approaches such as hyperbolic band theory (HBT) \cite{maciejko2021hyperbolic,maciejko2022automorphic}, higher dimensional irreps \cite{cheng2022band}, taking the thermodynamic limit via the supercell method \cite{Lenggenhager2023}, the continued fractions method \cite{Mosseri2023} or by constructing a coherent sequence of normal subgroups \wj{in a Fushcian group,} each with trivial intersections \cite{Lux2023} 

For an explicit demonstration of directed NHSE amplification on a lattice defined with \cref{bk}, we diverge from these existing approaches and define so-called geodesic-PBCs by connecting two otherwise disconnected boundary sites that lie along the same geodesic [\cref{fig:pbc}]. 
Mathematically, such a pair of boundary sites can be written as $z$ and $\gamma^{l-1}_\mu z$, where $l-1$ is the length of the geodesic, which is generated by translating $l-1$ times by a generator $\gamma_{\mu}$. 
To implement geodesic-PBCs on this geodesic, we define an equivalence relation between a pair of sites in \cref{gen} such that $z_i \sim \gamma^{l'}_\mu z_i$ if and only if $z_i \equiv \gamma^{l'}_\mu z_i \pmod{l}$, where $l' \geq 0$ is the number of translations. 
Visually, this introduces an additional ``connection" [red dashed in \cref{fig:pbc}] which closes the loop and gives rise to geodesic-PBCs. 
Explicitly, this is represented by 
 \begin{align}\label{eqn:psuedoPBC}
     \text{geodesic-PBC: } \gamma^{l'}_\mu z \equiv z \pmod{l},
 \end{align} 
where a closed geodesic loop is formed after $l$ translations.
This closes the path to form a closed geodesic loop of length $l$ in the lattice (which does not need to be embedded in the Poincar\'e disk). 
In this way, all pairs of boundary points along existing geodesics can be selected at will and closed to form geodesic loops until all geodesics have been closed. 
This allows one to freely interpolate between full OBCs and full geodesic-PBCs. 
We note that the value of $l$ is not fixed, depending on the \CH{length} of the geodesic. 
While the lack of translation symmetry under geodesic-PBCs precludes a reciprocal-space description, 
they provide continuous loops for uninterrupted directed NHSE amplification in the hyperbolic lattice, akin to the transboundary loops in actual PBC lattices.


\subsection{Non-Reciprocal Tight-binding Models in Hyperbolic Space}
Our construction allows generic tight-binding models to be embedded into hyperbolic lattices:
\begin{equation}\label{tb}
	\hat{H}_{TB}=-\sum_\mu\sum_{j} A_\mu \hat{a}_{\gamma_{\mu} z_{j}}^{\dagger} \hat{a}_{z_{j}}.
\end{equation}
where $\hat{a}$ ($\hat{a}^{\dagger}$) denotes a particle annihilation (creation) operator, and $A_{\mu}$ denotes the hopping amplitude in the direction of $\gamma_\mu$. 
Consequently, a tight-binding chain along a geodesic path can be defined under both OBC\CH{s} and geodesic-PBCs in the following way, 
\begin{equation}\label{ge}
	\begin{aligned}
		\text{\bf OBC:\,}&\underbrace{z_{j} \xleftrightarrows[\text{$A_{\gamma_{\mu}}$}]{\text{$A_{\gamma_{\mu}^{-1}}$}}\gamma_{\mu}z_{j}\cdots\xleftrightarrows[\text{$A_{\gamma_{\mu}}$}]{\text{$A_{\gamma_{\mu}^{-1}}$}}\gamma_{\mu}^{l}z_{j}}_{l},\\
		\textbf{\parbox{4em}{Closed\\ geodesics}\,:\,}&\underbrace{z_{j} \xleftrightarrows[\text{$A_{\gamma_{\mu}}$}]{\text{$A_{\gamma_{\mu}^{-1}}$}}\gamma_{\mu}z_{j}\cdots\xleftrightarrows[\text{$A_{\gamma_{\mu}}$}]{\text{$A_{\gamma_{\mu}^{-1}}$}}\gamma_{\mu}^{l}z_{j}\xleftrightarrows[\text{$A_{\gamma_{\mu}}$}]{\text{$A_{\gamma_{\mu}^{-1}}$}}z_{j}}_{l+1}.  
	\end{aligned}
\end{equation}

In this work, we are concerned with non-Hermitian pumping from hopping asymmetry, so we define a non-Hermitian imaginary flux parameter $\alpha$ and a directional orientation $s_\mu=\pm 1$, where $A_{\gamma}=e^{s_\mu\alpha}$, $A_{\gamma^{-1}}=e^{-s_\mu\alpha}$ such that $A_{\gamma}A_{\gamma^{-1}}=e^{s_\mu\alpha}e^{-s_\mu\alpha}=1$. 
This orientation function $s_\mu$ indicates whether the hopping asymmetry is parallel or anti-parallel to the translation direction given by the generator $\gamma_\mu$ [\cref{ge1}], and are all fixed once the directions for every edge of a chosen polygon ($q$-gon) have been assigned.
\cref{tb} is thus specialized to
\begin{equation}\label{NH2}
	\hat{H}=-\sum_{\mu}\sum_j e^{s_\mu\alpha } \hat{a}_{\gamma_{\mu} z_{j}}^{\dagger} \hat{a}_{z_{j}}+e^{-s_\mu\alpha} \hat{a}_{z_{j}}^{\dagger} \hat{a}_{\gamma_{\mu} z_{j}}.
\end{equation}
Thus, the tight-binding chains along an open and closed geodesic path (geodesic-PBCs) are 
\begin{align}\label{arrow}
	\textbf{ OBCs:\,}&\underbrace{z_{j} \xleftrightarrows[e^{-s_\mu\alpha}]{e^{s_\mu\alpha}} \gamma_{\mu} z_{j} \dots \xleftrightarrows[e^{-s_\mu\alpha}]{e^{s_\mu\alpha}} \gamma_{\mu}^{l} z_{j}}_{l},\\
    \textbf{\parbox{4em}{Closed\\ geodesics}\,:\,}&\underbrace{z_{j} \xleftrightarrows[e^{-s_\mu\alpha}]{e^{s_\mu\alpha}} \gamma_{\mu} z_{j} \dots \xleftrightarrows[e^{-s_\mu\alpha}]{e^{s_\mu\alpha}} \gamma_{\mu}^{l} z_{j} \xleftrightarrows[e^{-s_\mu\alpha}]{e^{s_\mu\alpha}} z_j}_{l+1}.
\end{align}

\begin{figure}[htbp!]
	\centering
	\includegraphics[width=1\linewidth]{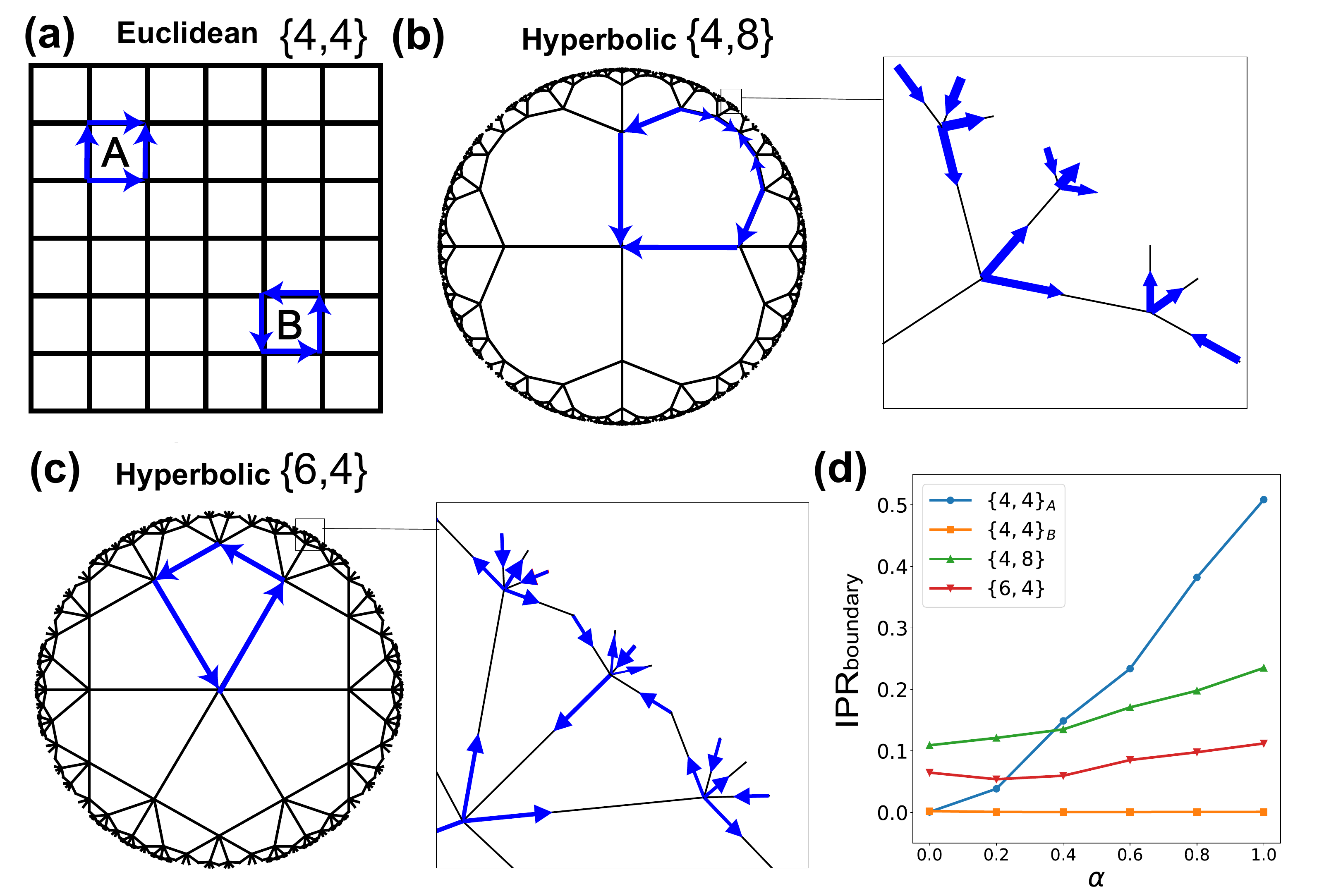}
	\caption{Assignment of non-reciprocal hopping directionalities and their resultant skin accumulation. 	
	(a-c) Illustration of the hopping directionalities within a single polygon ($q$-gon). The blue arrows indicate the direction of amplification represented by $e^{s_\mu\alpha}$ for the Euclidean $\{4,4\}_{\text{A,B}}$ [\cref{flat,flat2}] (a) and hyperbolic \{$4,8$\} (b) and \{$6,4$\} (c) lattices. 
    \wj{The chosen hopping orientations in a hyperbolic octagon for the \{$4,8$\} lattice are $s_{1}=s_{2}=-1$, while the chosen hopping orientations in a hyperbolic square for the \{$6,4$\} lattice are $s_{1}=s_{3}=-1$ and $s_{2}=1$.}
    (Insets of b,c) While the boundary of the \{$4,8$\} lattice is dominated by length-$2$ and several length-$4$ geodesics, the \{$6,4$\} lattice has a relatively smoother boundary comprising pairs of intersecting length-$2$ geodesics. 		
(d) The {extent of boundary accumulation} under increasing non-Hermiticity, $\alpha$, as shown through their boundary IPRs [see \cref{iprb}]. 
     The $\{4,4\}_{\text{A}}$ (blue) lattice shows the cleanest NHSE with rapidly increasing boundary IPR, strongly contrasting with the vanishing ${\rm IPR_{\rm boundary}}$ of $\{4,4\}_{\text{B}}$ (orange) lattice where the NHSE completely cancels. 
 While both $\{4,8\}$ and $\{6,4\}$ hyperbolic lattices exhibit some NHSE accumulation, they also display nonzero boundary IPR in their Hermitian limit, which can be attributed to their significant number of boundary sites.     The hyperbolic $\{4,8\}$ (green) lattice demonstrates stronger boundary NHSE localization than its \{$6,4$\} (red) counterpart as $\alpha$ increases. 
   }\label{fig:lattice-structure}
\end{figure}

\begin{figure*}[htbp!]
	\centering
	\includegraphics[width=0.8\linewidth]{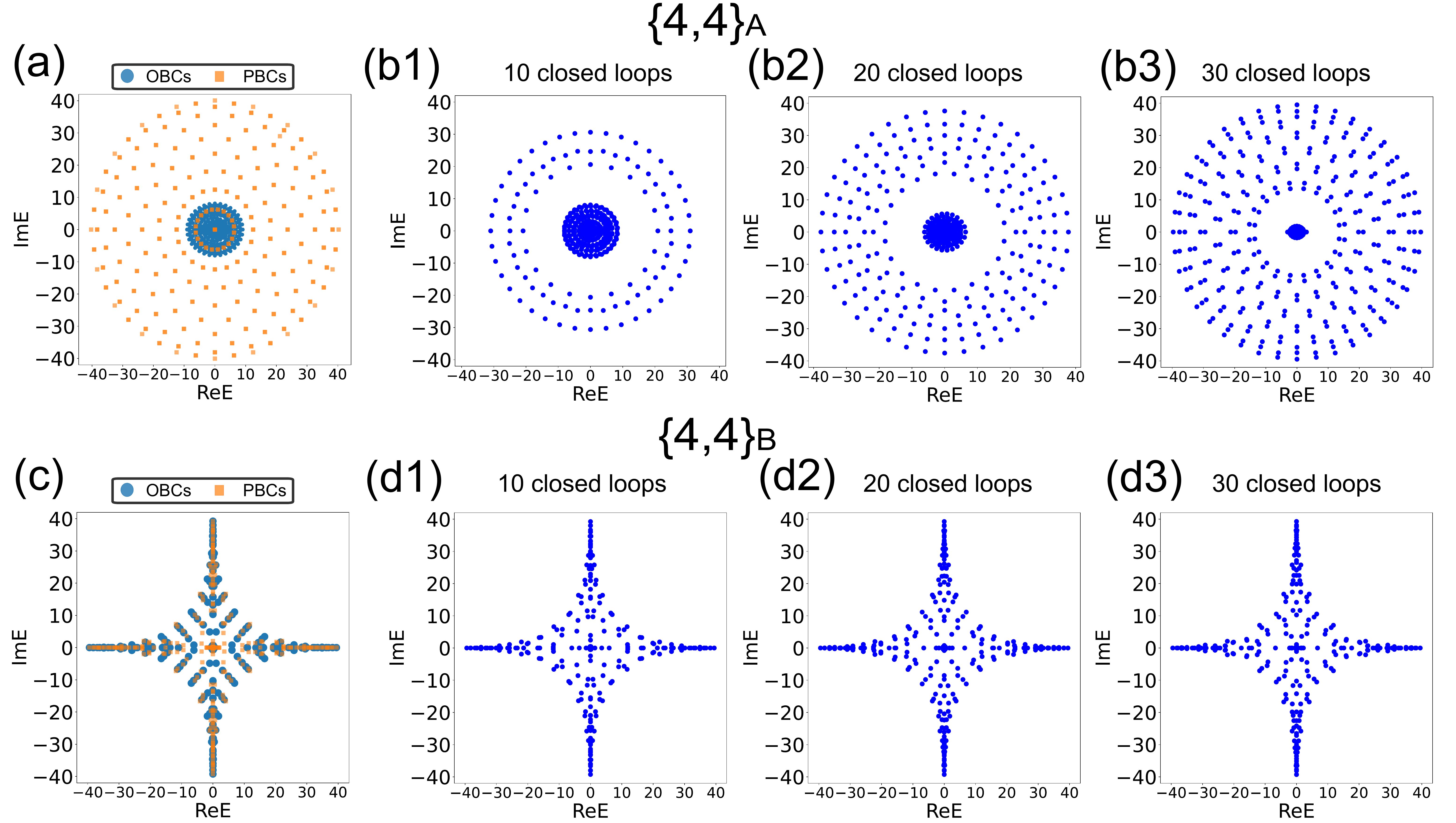}
	\caption{Comparison of energy spectra under OBC, partial PBC, and PBC in Euclidean lattices. 
    The Euclidean lattices are the $\{4,4\}_\text{A}$ [\cref{flat,fig:lattice-structure}(a)] lattice in (a) and (b1-b3), and the $\{4,4\}_\text{B}$ [\cref{flat2,fig:lattice-structure}(b)] lattice in (c) and (d1-d3). 
    (a) and (c) show the spectra under OBC (blue) and PBC (yellow). 
    The spectral transition as the number of closed loops is increased (or more specifically, partial PBCs for the Euclidean models) is also shown, with $10$ in (b1) and (d1), $20$ (b2) and (d2), and $30$ (b3) and (d3) closed loops.
    These results consistently demonstrate that the constructiveness and destructiveness in the bulk ($4$-gon) non-reciprocity lead to the distinctive spectral sensitivity to non-Hermitian parameters. 
    The non-Hermitian parameter is set as $\alpha=3$ and the system size is set to $20$ by $20$. 
		}
	\label{fig:loops}
\end{figure*}
\section{Non-Hermitian Skin Effect along Hyperbolic Geodesics}\label{sec2}
\begin{figure*}
	\centering
	\includegraphics[width=0.9\linewidth]{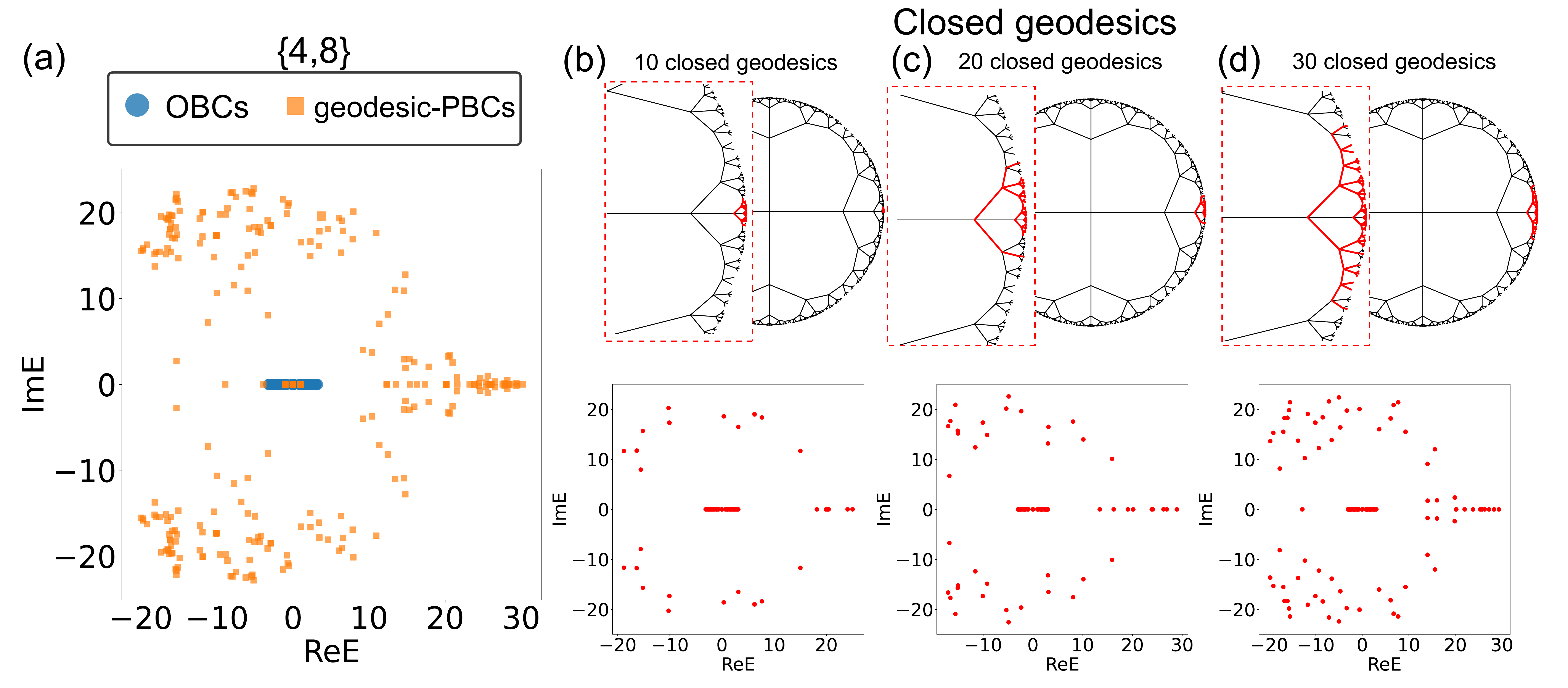}
	\caption{Spectral transition of the hyperbolic \{$4,8$\} lattice under various boundary conditions. (a) The OBC (blue) to geodesic-PBC (orange) spectral transition. (b)-(d) demonstrate the gradual spectral transition under increasing ($10$, $20$, and $30$) closed geodesic loops. The number of closed geodesics (loops) is marked by red edges in the insets of the lattices in (b-d), and it should converge to the geodesic-PBC limit as the closure of geodesics increases. Analogous to the Euclidean $\{4,4\}_\text{A}$ lattice shown in \cref{fig:lattice-structure}(a) and (b), the constructive $8$-gon non-reciprocity contributes to the heightened 
    spectral sensitivity to non-Hermiticity, as evident by the emerging complex eigenvalues shown in (b)-(d). For all results, we set  $\alpha=3$.
	}
	\label{fig:48loops}
\end{figure*}
\subsection{NHSE Sensitivity of Constructed Hyperbolic Lattices}
\begin{figure*}
	\centering
	\includegraphics[width=0.9\linewidth]{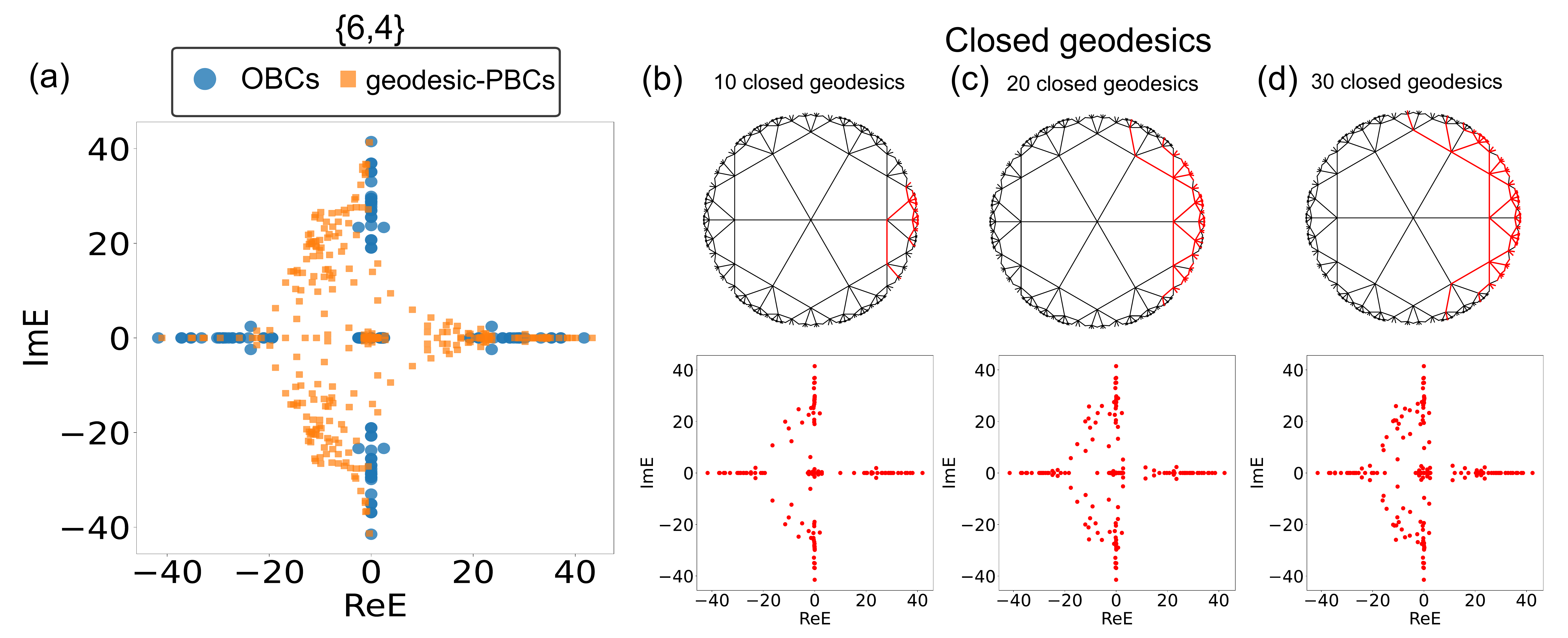}
	\caption{Spectral transition of the hyperbolic \{$6,4$\} lattice under various boundary conditions. (a) shows the OBCs (blue) to geodesic-PBCs (orange) spectral transition. (b)-(d) demonstrate the gradual spectral transition under increasing ($10$, $20$, and $30$) closed geodesic loops. The number of closed geodesics (loops) is marked by red edges in the zoomed-in views of lattices in (b-d), and it should converge to the geodesic-PBC limit as the closure of geodesics increases. Analogous to the Euclidean $\{4,4\}_\text{B}$ lattice, the destructive $8$-gon non-reciprocity contributes to the weak spectra sensitivity to non-Hermiticity, as evident in (b)-(d).  For all results, we set $\alpha=3$.
}
	\label{fig:64loops}
\end{figure*}

In a lattice experiencing the NHSE, the actual extent of skin state accumulation depends not just on the nature of the lattice geometry and boundaries, but also on how the asymmetric hoppings are juxtaposed relative to each other. Specifically, whether the NHSE pumping from nearby hoppings constructively or destructively interferes, greatly impacts the extent and scaling behavior of the NHSE \cite{Lee2019b,Zhu2023}.
As a simplest illustration, the two Euclidean lattices \{$4,4$\}$_\text{A}$ [see \cref{flat}] and \{$4,4$\}$_\text{B}$ [see \cref{flat2}] in \cref{fig:lattice-structure}(a) have identical structures, but differently oriented hoppings. Each square plaquette exhibits constructive and destructive bulk non-reciprocity respectively, leading to the cancellation of the NHSE in \{$4,4$\}$_\text{B}$.

For a lattice embedded in curved space, such as the hyperbolic lattices that we consider, the bulk non-reciprocity can be analogously encoded in the directionalities of the hoppings that make up the sides of each $q$-gon in the \{$p,q$\} hyperbolic lattice. 
However, this seemingly straightforward generalization is not without caveats, since it is insufficient to extrapolate the global non-reciprocity from a single $q$-gon in a hyperbolic lattice due to the non-commutative nature of its generators \cref{ge1}. 
Indeed, relating hyperbolic lattices to their Euclidean analogs involves two important factors: non-reciprocity within their respective $q$-gon and their geodesic-based boundaries.
Notably, some stray ends are unavoidable as a by-product of employing a standardized geodesic construction without resorting to boundary modifications for each instance.

Visually, the $q$-gon non-reciprocity of the hyperbolic \{$4,8$\} lattice \wj{(the chosen hopping orientations are $s_1=s_2=-1$)} in \cref{fig:lattice-structure}(b) can be seen to accumulate the state at two specific sites around a $q$-gon plaquette. 
Furthermore, distinct localization sites appear on its boundary, which is dominated by length-$2$ and several length-$4$ geodesics that resemble $1$D Hatano-Nelson chains under OBC \cite{Hatano1996}.
Thus, its spectrum is expected to be highly sensitive to non-Hermiticity.
This is analogous to the \{$4,4$\}$_{\text{A}}$ lattice, where its directed amplification within its bulk and boundary accumulates at the top-right corner of the square lattice.
As such, we expect some higher-order NHSE accumulation like in the \{$4,4$\}$_{\text{A}}$ lattice, albeit not as strongly due to the non-parallel nature of the geodesics due to the curvature of hyperbolic space. 

In contrast, the directed cycle \wj{(the chosen hopping orientations are $s_1=s_3=-1$ and $s_2=1$)} in each plaquette ($q=4$) of the hyperbolic \{$6,4$\} lattice in \cref{fig:lattice-structure}(c) suggests the presence of extended bulk states and not strong skin accumulation. 
Furthermore, its boundary comprises intertwined non-reciprocal length-$2$ geodesics that form a relatively smoother boundary as compared to its \{$4,8$\} counterpart.
Due to the formation of a connected boundary with conflicting non-reciprocal hopping, minor modifications to its spectra are expected and are independent of the number of closed geodesics.
This is analogous to the \{$4,4$\}$_{\text{B}}$ lattice, where there are directed cycles in both its bulk and boundary.
These cycles result in a weak NHSE sensitivity to non-Hermiticity.

Before delving into the spectral behavior of the hyperbolic lattices under the NHSE, we shall first quantify the above discussions by considering the boundary inverse participation ratio (IPR$_{\rm boundary}$), 
\begin{align}\label{iprb}
 	{\rm IPR}_{\rm boundary}=\frac{1}{\mathcal{N}}\sum_{j}\sum_{z_i\in {\rm\{ Boundary\}}}\abs{\psi_{j}(z_{i})}^{4},
\end{align}
where $\mathcal{N}$ represents the total number of eigenstates. 
This measure quantifies the localization of eigenstates near the boundary, \CH{which we use to compare the extent of boundary state accumulation} in the \{$4,4$\}$_\text{A}$, \{$4,4$\}$_\text{B}$, \{$4,8$\} and \{$6,4$\} lattices, as shown in \cref{fig:lattice-structure} (d).
As expected, the \{$4,4$\}$_\text{A}$ lattice (illustrated in blue) exhibits a significant increase in boundary IPR with non-Hermiticity, $\alpha$ due to its unfettered bulk non-reciprocity. 
Conversely, the \{$4,4$\}$_\text{B}$ lattice (orange) \CH{consistently} shows an almost vanishing boundary IPR, indicative of the cancellation of NHSE pumping. 
Notably, both hyperbolic lattices exhibit non-zero boundary IPRs in the Hermitian limit but also experience increasing boundary IPRs with increasing $\alpha$. This boundary "anomaly" is due to their significant boundary volume, and suggests that more care is needed to isolate the actual extent of NHSE accumulation from the presence of a large number of trivial (Hermitian) boundary states.


While the boundary IPR can detect the possible presence of the NHSE, it cannot unambiguously quantify its strength and extent. 
\CH{For that, we examine how the spectra of these hyperbolic lattices are sensitive to their boundary conditions as we go from PBCs to OBCs\cite{kunst2018biorthogonal,xiong2018does,xiao2020non,helbig2020generalized,zhang2022review,li2020critical,kawabata2020higher,shen2022non,Lee2019b,lin2023topological,tai2023zoology}, a key characteristic of the NHSE which also exists in Euclidean space.} However, since we do not have exact PBCs for the hyperbolic lattices, 
we shall instead use the geodesic-PBCs defined earlier. 

\begin{figure*}[htbp!]
	\centering
	\includegraphics[width=0.99\linewidth]{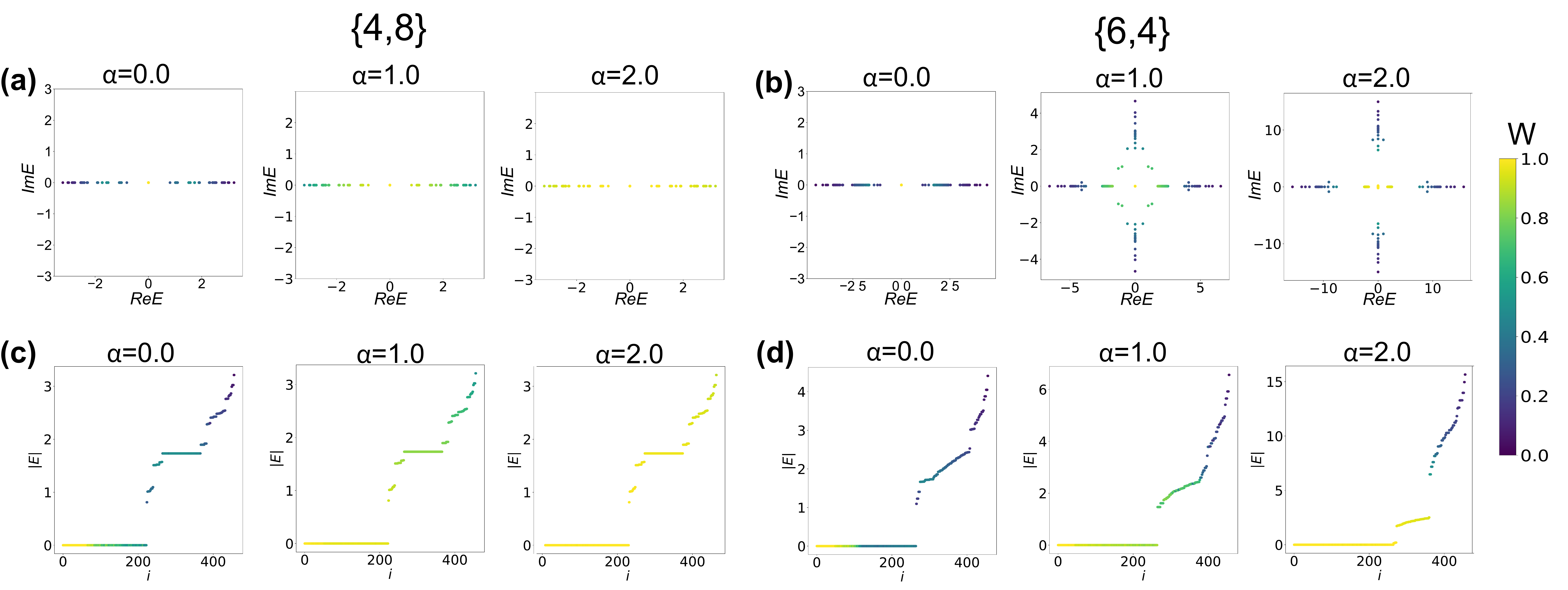}
	\caption{Boundary localization $W$ [\cref{w}] as a measure of spectral sensitivity for hyperbolic lattices.
   The OBC spectra for the $\{4,8\}$ (a) and $\{6,4\}$ (b) lattices are shown under varying non-Hermiticity levels $\alpha=0,1,2$.
   While most states exhibit enhanced $W$ with increasing $\alpha$ in the $\{4,8\}$ lattice, only states near the zero-energy states experience a change in boundary localization in the $\{6,4\}$ lattice.
   (c,d) Plots of $|E|$, which shows the distribution of $W$ among the highly degenerate zero-energy modes of the $\{4,8\}$ in (c) and $\{6,4\}$ in (d) lattices. 
   The trends observed in (a) are comprehensively demonstrated in (c): apart from a few intrinsic boundary zero modes, most eigenstates show a heightened sensitivity to non-Hermiticity. 
   Conversely, in (b), only certain zero-energy states demonstrate heightened sensitivity.
   Notably, we can distinguish the bulk and boundary [see \cref{bk}] states from their low and high $W$ values respectively in the Hermitian limit.
   As the \{$6,4$\} lattice has destructive non-reciprocity interferences within each hyperbolic square ($q=4$-gon), the bulk states cannot be successfully pumped to the boundary.
 }%
	\label{fig:boundaryweight}
\end{figure*}

One advantage of the geodesic-PBCs so defined is that we can interpolate between OBCs and full (geodesic)-PBCs by closing up the geodesics one by one. To have a sense of how that affects the spectra, we first demonstrate such "partial" PBCs for the two Euclidean lattices we defined. As expected, $\{4,4\}_\text{A}$ shows strong spectral sensitivity, while $\{4,4\}_\text{B}$ shows weak spectral sensitivity under their OBC to PBC transition in \cref{fig:loops}(a) with $\alpha=3$ (full geodesic-PBCs coincide with exact PBCs for Euclidean lattices).
Furthermore, their partial PBCs spectra converge from the OBC to PBC limit with the increasing number of closed loops ($10$ in (b1) and (d1), $20$ in (b2) and (d2), and $30$ in (b3) and (d3)).

We next move on to the spectral transition of the hyperbolic \{$4,8$\} lattice from OBC (blue circles) and geodesic-PBCs (orange squares) in \cref{fig:48loops}(a). 
The OBC spectrum of the \{$4,8$\} lattice is expected to be real, as the boundary [\cref{bk}] consists of the length-$2$ and several length-$4$ Hatano-Nelson chains, all of which individually host real skin spectra. 
Indeed, the spectrum exhibits strong sensitivity when transitioning between OBC and geodesic-PBC, analogous to the constructive bulk non-reciprocity in \{$4,4$\}$_\text{A}$ lattice, i.e. constructive $8$-gon non-reciprocity for the \{$4,8$\} lattice.
This sensitivity is further demonstrated by closing $10$ (\cref{fig:48loops}(b)), $20$ (\cref{fig:48loops}(c)), and $30$ (\cref{fig:48loops}(d)) open geodesics, as shown. 
Notably, \CH{complex eigenspectra} emerge upon the closing of geodesics, which resembles $1$D Hatano-Nelson chains under PBCs [see \cref{bk,fig:pbc}].

Qualitatively different behaviors dominate that of the \{$6,4$\} lattice with the previously assigned hopping directionalities. Even under OBCs, the combination of its destructive $4$-gon non-reciprocity and a boundary dominated by intertwined length-$2$ geodesics in the \{$6,4$\} lattice renders the spectrum complex [\cref{fig:64loops}(a)]. That also makes it exhibit generally weaker spectral sensitivity to non-Hermiticity under the geodesic-PBC transition in \cref{fig:64loops}(a), as compared to its \{$4,8$\} counterpart.
While the closure of $10$ (\cref{fig:64loops}(b)), $20$ (\cref{fig:64loops}(c)), and $30$ (\cref{fig:64loops}(d)) geodesics is noticeable compared to the $\{4,4\}_\text{B}$ lattice, its response is still weaker as compared to \CH{that in} the \{$4,8$\} lattice.

\subsection{Skin and Non-Skin Modes in Hyperbolic Lattices}
Previously, we demonstrated that geodesics-based boundary conditions and $q$-gon non-reciprocity can lead to unconventional NHSE behavior. 
However, the distinction between skin and non-skin modes remains unclear due to the extensive boundary volume of hyperbolic lattices [see \cref{fig:totalsize} in \cref{scale}].
In the following section, we will examine the distinction between these modes.

To distinguish between skin and non-skin modes, we define the following OBC boundary localization of each eigenstate,
\begin{equation}\label{w}
	W=\sum_{z_{i}\in {\rm boundary}}\abs{\psi(z_{i})}^{2},
\end{equation}
where the boundary sites [see \cref{fig:pbc,bk}] are \CH{the $L$-th generation nodes} in the hyperbolic lattices.
Notably, the strength of the NHSE \CH{can be quantified by how much $W$ changes as $\alpha$ is increased} 
from the Hermitian limit. 
Here, the boundary localization $W$ for each \CH{eigenstate of both hyperbolic lattices are shown in their} complex spectra in \cref{fig:boundaryweight} (a) and (b). 
For the $\{4,8\}$ lattice of \cref{fig:boundaryweight} (a), apart from the zero eigenstates, the remaining states exhibit a pronounced increase \CH{in} boundary localization $W$ as non-Hermiticity increases. 
This is visually shown as $W$ transitions from dark blue to yellow.
However, the situation is very different for the $\{6,4\}$ lattice in \cref{fig:boundaryweight} (b), where a weak boundary localization response is observed \CH{for most states as $\alpha$ increases}. 
This shows that the $\{6,4\}$ lattice does not experience as much NHSE as its \{$4,8$\} counterpart.

Other than the skin modes, there exist highly degenerate zero energy modes that appear as a point at $E=0$ in \cref{fig:boundaryweight} (a,b). To check for their boundary localization $W$, we also plot the $\abs{E}$ of the eigenenergies in \cref{fig:boundaryweight} (c) and (d) for $\alpha =0$, $1.0$ and $2.0$ for both the $\{4,8\}$ (c) and $\{6,4\}$ (d) lattices. Most zero modes possess inherently high boundary localization due to the stray boundary sites, such that their $W$ remain unchanged for $\alpha=0$ to $2$ for both these hyperbolic lattices, as shown in \cref{fig:boundaryweight} (c) and (d).  
This highlights that \CH{hyperbolic lattices can generally possess a large number of non-NHSE boundary states.} 
The $\abs{E}$ plots also reveal that,   unlike the $\{4,8\}$ lattice, the bulk [see \cref{bk}] states (low $W$ in the Hermitian limit) in the $\{6,4\}$ lattice do not localize at the boundary even as $\alpha$ increases.
This is directly attributed to the $4$-gon non-reciprocity, where destructive non-reciprocal interference hinders the accumulation of bulk states at the boundary, thereby significantly weakening the NHSE.

%

\subsection{Relation between Finite-size Scaling and Skin Modes}
\begin{figure}[ht!]
	\centering
	\includegraphics[width=0.99\linewidth]{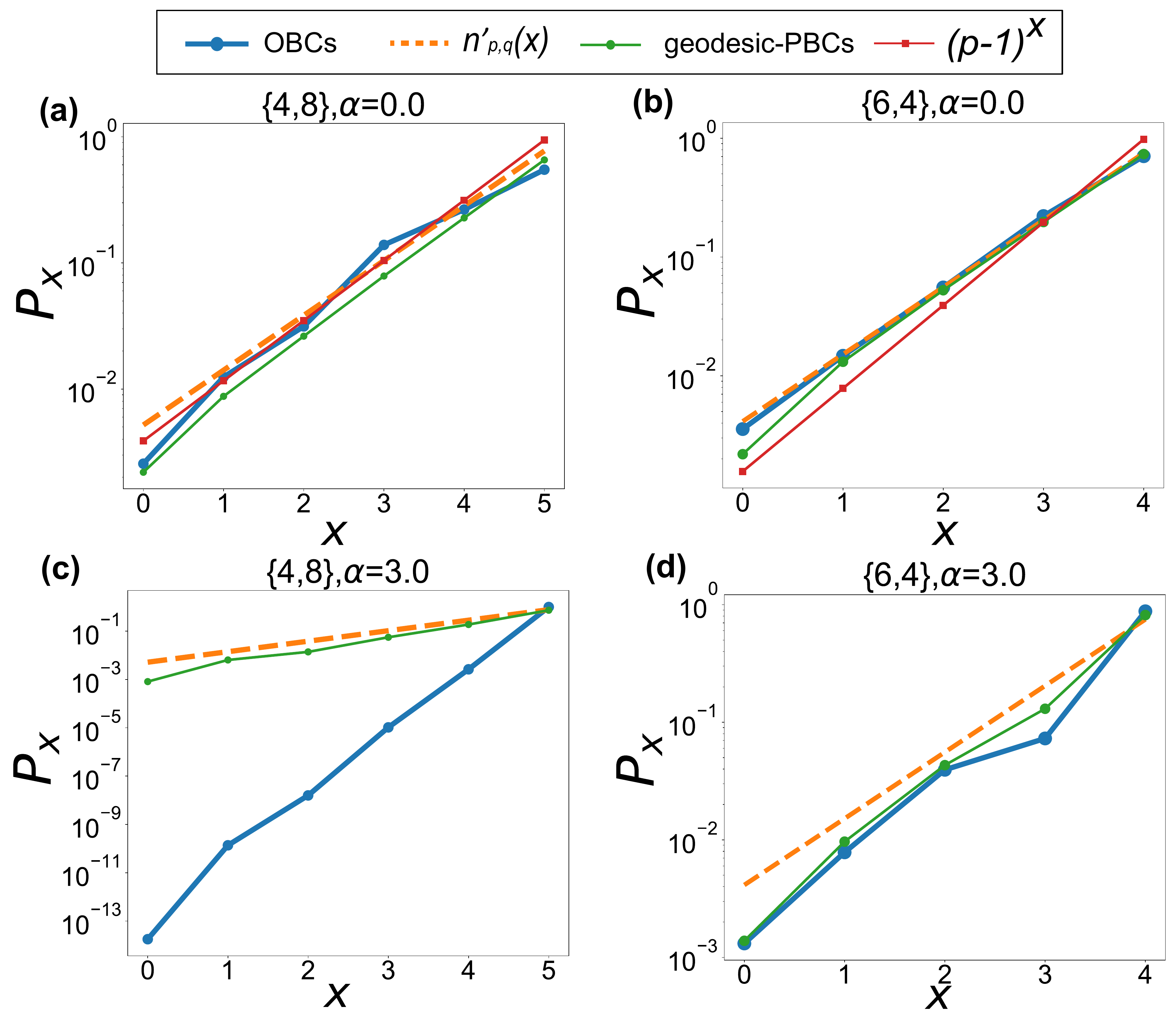}
	\caption{The spatial state distribution $P_{x}$ [\cref{p}] along $x$, which is the distance from any site to the origin of the Poincar\'e disk. We present $P_x$ under both OBC (blue) and geodesic-PBC (green) for the hyperbolic models \{$4,8$\} in (a,c) and \{$6,4$\} in (b,d). $P_x$ is proportional to both $(p-1)^x$ (red) [\cref{eqn:p-1}] and the normalized exact number of sites $n'_{p,q}(x)$ [\cref{n}] (orange) in the Hermitian limit in (a) and (b).
    A notable deviation is observed under OBC due to the strong skin effect in the \{$4,8$\} (c) lattice, while the \{$6,4$\} (d) lattice shows only minor deviations when $\alpha$ is increased from $0$ to $3$.
    Notably, the boundary states have the longest translation path with $x=L$, while the bulk states have $x<L$.
    The exponential scaling law of lattice sizes is discussed in \cref{scale}.}
    \label{fig:distribution}
\end{figure}


Alternatively, skin modes can be distinguished from the intrinsic boundary modes by examining the spatial distributions of their wavefunctions. 
Since the bulk and boundary states of our constructed hyperbolic lattices [see \cref{bk}] are related to the minimal number of translation operators, $x$, it is natural to examine this distribution along the shortest translation path from the origin of the Poincar\'e disk at $z_0 = (0,0)$.
Notably, this shortest translation path coincides with the minimal number of translation operators, $x$ in \cref{bk}, and is consistently upper-bounded by the number of total generations, $L$.
Thus, any pair of sites, $z_i$ and $z_{i'}$ will have the same shortest translation path if they are exactly $x$ translations away from $z_0$ [see \cref{gen}].
This can be captured by the set containing all shortest translation paths and is defined as
\begin{align}\label{eqn:Z_Ls}
	Z(x) = \{z_i \mid z_i = \overbrace{\gamma_{\mu} \cdots \gamma_{\mu'}}^{x} z_{0}, 0\leq x\leq L\},
\end{align}
with the distribution over $Z_x$ as
\begin{equation}\label{p}
	P_{x}=\frac{1}{\mathcal{N}}\sum_{j}\sum_{z_i \in Z(x)}\abs{\psi_{j}(z_{i})}^{2},
\end{equation}
where $P_x$ averages over all eigenstates with the total number of eigenstates as $\mathcal{N}$.

\begin{align}\label{Nmain}
    n_{p,q}(x)\approx(p-1)n_{p,q}(x-1).
\end{align}
Although \cref{Nmain} does not account for the removal of duplicated sites, this estimation is valid when compared to the exact $n_{p,q}(x)$ in \cref{N}. 
More detailed elaborations and numerical comparisons are shown in \cref{fig:totalsize} of \cref{scale}.

In the Hermitian limit where no preferred spatial accumulation is expected, the extensive boundary volume of hyperbolic lattices indicates that $P_x$ is expected to scale proportionally to $n_{p,q}(x)$, i.e.,
\begin{align}\label{eqn:p-1}
    \frac{P_{x+1}}{P_{x}}\propto \frac{n_{p,q}(x+1)}{n_{p,q}(x)}\approx p-1,
\end{align}
thus leading to $P_{x}\propto (p-1)^{x}$.
The validation of \cref{p} under OBCs (blue) and geodesic-PBCs (green), \cref{eqn:p-1} (red) and the normalized exact number of sites (orange) [see \cref{scale}], i.e., 
\begin{align}\label{n}
    n^{\prime}_{p,q}(x)=\frac{n_{p,q}(x)}{\sum_{x}n_{p,q}(x)}
\end{align}
are shown in \cref{fig:distribution} of both the \{$4,8$\} (a) and \{$6,4$\} (b) in the Hermitian limit. 
Due to the extensive boundary volume, majority of the geodesics reside on the boundary [see \cref{bk}]. 
Consequently, most eigenstates are localized on the boundary at $x=L$. 
As a result, no significant modifications in $P_x$ are observed for both hyperbolic lattices between the OBC and geodesic-PBC at $x=L$, as these states are unable to reach the bulk by construction.

The extent of the NHSE can then be read off by how the exponential scaling of $P_x$ changes as $\alpha$ is varied.
For the \{$4$,$8$\} lattice under $\alpha=3$, $P_x$ exhibits a very different gradient in \cref{fig:distribution}(c) as compared to its Hermitian counterpart in \cref{fig:distribution}(a).
Moreover, a significant deviation emerges between $n^\prime_{p,q}(x)$ and $P_x$ which is induced by the presence of numerous non-trivial skin modes.
In contrast, only a small number of eigenstates are enhanced by non-Hermiticity in the \{$6$,$4$\} lattice as shown in \cref{fig:boundaryweight} (d).
This results in a small but noticeable deviation between $n^\prime_{p,q}(x)$ and $P_x$.
Notably, the exponential scaling is not present \CH{regardless of the value of} $\alpha$ for the Euclidean cases [see \cref{px} in \cref{appx:manhattan}] due to their non-exponential finite-size scaling [see \cref{fig:totalsize} in \cref{scale}].


\section{Conclusions and Outlook\label{sec:con}}
In this work, we introduced a geodesic-based method to generate finite-sized hyperbolic lattices, enabling the tracking of non-reciprocal hopping directions and facilitating the study of the NHSE with respect to boundary and curvature effects. 
By forming closed geodesics, we establish geodesic-PBCs akin to PBCs in Euclidean lattices.
This new construction enables us to embed tight-binding chains along all generated hyperbolic geodesic paths. 
By comparing different Euclidean and hyperbolic lattice models, we unravel that $q$-gon non-reciprocity and geodesic-based boundaries are two crucial factors determining spectral sensitivity, and consequently, the NHSE.
These factors also affect the spectral transitions between OBCs, geodesic-partial-PBCs, and geodesic-PBCs.

Additionally, we demonstrated how to distinguish the non-trivial skin modes from the extensive boundary modes by utilizing the boundary localization $W$ as a measure with zero and finite non-Hermiticity $\alpha$.
This approach not only allows us to pinpoint the exact number of states exhibiting the NHSE but also enables the distinction between bulk and boundary states. 

Furthermore, the spatial density profile along the shortest path allows us to relate to the finite-size scaling law in hyperbolic lattices. 
Based on this, we demonstrate how the deviation of the spatial density profile under non-Hermiticity indicates the presence or absence of the NHSE in the hyperbolic \{$4,8$\} and \{$6,4$\} lattices.
Our findings highlight the potential to explore non-Hermitian behaviors in hyperbolic lattices, with the potential realization in platforms like electronic and quantum circuits \cite{Kollar2019,lenggenhager2022simulating,Zou2023wbo,Hofmann2020pzl,Zhang2022,Yuan2024,Pei2023,Zhang2023}, photonics \cite{zhu2020photonic,song2020two,lin2024observation,Huang2024}, and programmable quantum circuit simulators \cite{smith2019simulating,shen2023observation,xin2020quantum,mei2020digital,Chen2023zlo,Chen2024ull,zhang2024observation,Koh2023ohr,Chen2022owo,koh2022simulation}. 

\section{Acknowledgements} 
This work is supported by the QEP$2.0$ Grant from the Singapore National Research Foundation (Grant No. NRF$2021$-QEP$2$-$02$-P$09$) and the Singapore Ministry of Education Academic Research Fund  Tier-II Grant (Award No. MOE-T$2$EP$50222$-$0003$). 
We thank Ronny Thomale and Truman Ng Yu for the helpful discussions. 



\appendix
\section{Non-Hermitian Models in Euclidean Space}\label{ap1}

In this section, we provide the explicit forms of the two non-Hermitian Euclidean square lattice models used in \cref{fig:lattice-structure} of the main text. For $\{p,q\}=\{4,4\}$, the operations $\gamma_\mu$ give\CH{n} in \cref{ge1} reduce to translations in either the $x$ or $y$ directions. 
Following these, we present two representative models, as shown in \cref{fig:lattice-structure} (a) and (b): 
\begin{widetext}
\begin{align}\label{flat}
\{4,4\}_{\text{A}}:	\hat{H}_{\{4,4\}_{\text{A}}}=-\sum_{x,y}(e^{\alpha} \hat{a}_{x+1,y}^{\dagger} \hat{a}_{x,y}+e^{-\alpha} \hat{a}_{x,y}^{\dagger} \hat{a}_{x+1,y} +e^{\alpha} \hat{a}_{x,y+1}^{\dagger} \hat{a}_{x,y}+e^{-\alpha} \hat{a}_{x,y}^{\dagger} \hat{a}_{x,y+1}).
\end{align}	
\begin{align}\label{flat2}	
\{4,4\}_{\text{B}}:\hat{H}_{\{4,4\}_{\text{B}}}=-&\sum_{x,y\in even}(e^{\alpha} \hat{a}_{x+1,y}^{\dagger} \hat{a}_{x,y}+e^{-\alpha} \hat{a}_{x,y}^{\dagger} \hat{a}_{x+1,y}+e^{-\alpha} \hat{a}_{x,y+1}^{\dagger} \hat{a}_{x,y}+e^{\alpha} \hat{a}_{x,y}^{\dagger} \hat{a}_{x,y+1})\\
-&\sum_{x,y\in odd}(e^{-\alpha} \hat{a}_{x+1,y}^{\dagger} \hat{a}_{x,y}+e^{\alpha} \hat{a}_{x,y}^{\dagger} \hat{a}_{x+1,y}+e^{\alpha} \hat{a}_{x,y+1}^{\dagger} \hat{a}_{x,y}+e^{-\alpha} \hat{a}_{x,y}^{\dagger} \hat{a}_{x,y+1}).
\end{align}
\end{widetext}
or the $\{4,4\}_\text{A}$ \CH{lattice} described by \cref{flat}, constructive \CH{interference} between its non-reciprocal hopping \CH{give rise to corner skin modes} \cite{Lee2019b,Zhu2022,Li2022d,Zhu2023,kawabata2020higher,zhang2021observation,zou2021observation,zhong2024higher,Sun2023}.
In contrast, the destructive interference among pairs of bulk non-reciprocal hoppings in $\{4,4\}_\text{B}$, described by \cref{flat2}, generates no net skin effect. In the following subsection, we examine the extent of NHSE accumulation in these two lattices, based on the measure $P_x$ introduced in the main text.

\begin{figure}
	\centering
	\includegraphics[width=1\linewidth]{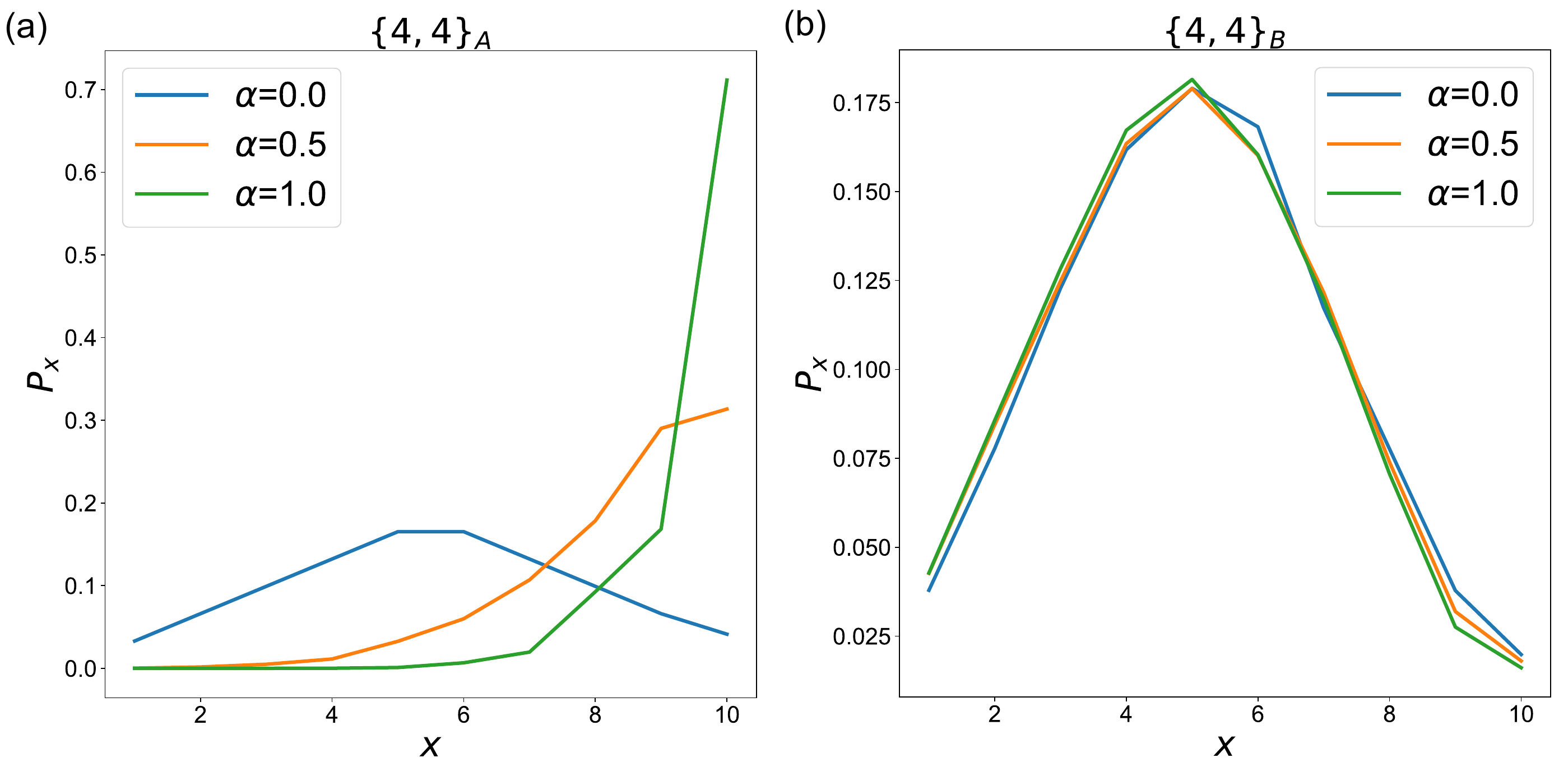}
	\caption{The state accumulation distribution $P_{x}$ (\cref{p} in the main text) based on the Manhattan distance for Euclidean $\{4,4\}_{A}$ and $\{4,4\}_{B}$ lattices. We set the NHSE strength $\alpha$ to $0.0$ (blue), $0.5$ (yellow), and $1.0$ (green). The systems sizes are $L=11$ for the $\{4,4\}_{A}$ lattice and $L=10$ for the $\{4,4\}_{B}$, with the center at $(5,5)$ for both. 
    The NHSE leads to the right-\CH{skewed} $P_x$ profile in (a), while $P_x$ in (b) show resilience to the NHSE. The sensitivity of these two lattices is consistent with the results shown in \cref{fig:lattice-structure}(d).
}
	\label{px}
\end{figure}
\subsection{Distribution of skin accumulation based on the Manhattan Distance in Euclidean lattices \label{appx:manhattan}}
\CH{The skin accumulation distribution $P_{x}$ [\cref{p}] for Euclidean lattices is examined with $x$ representing the Manhattan distance from the center, analogous to the shortest path} from the origin of the Poincar\'e disk.
For the results presented in \cref{px}, we set the size \CH{of} the $\{4,4\}_{A}$ lattice to be $L=11$,  with the center positioned at $(5,5)$.
\CH{For the $\{4,4\}_{B}$ lattice, an even} length $L$ is required due to sublattice symmetry. 
Hence, we fix the system size $L=10$ with the center at $(5,5)$ for this lattice. 
The distributions $P_x$ are presented in \cref{px}. 
Due to the \CH{NHSE} in the $\{4,4\}_{A}$ lattice, there is a noticeable \CH{shift to the right} as $\alpha$ increases, corresponding to corner skin accumulation at $(L,L)$, which is at the maximum value of $x$. 
In contrast, the $\{4,4\}_{B}$ lattice is resilient against the NHSE and hence $\alpha$. 
More details \CH{on the} boundary localization are shown in the boundary IPR plotted in \cref{fig:lattice-structure}(d).

\begin{figure*}
	\centering
	\includegraphics[width=0.7\linewidth]{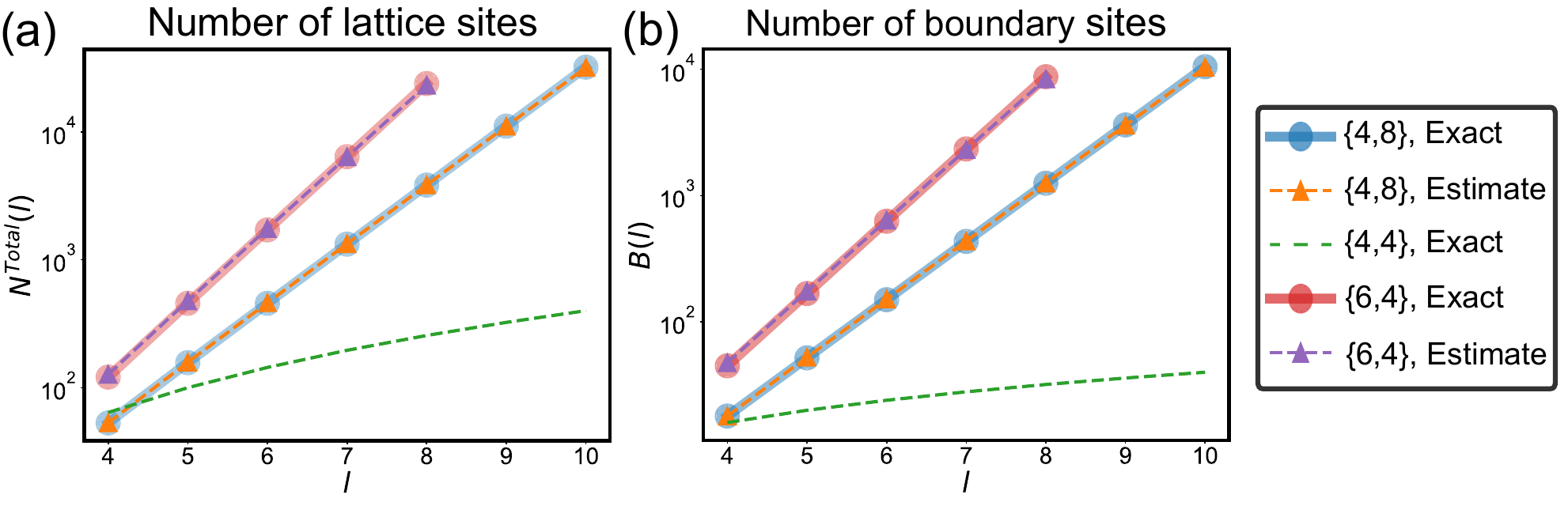}
	\caption{Verification of the exponential scaling of \{$4,8$\}, \{$6,4$\} lattices. 
    We present the number of lattice sites $N^{\mathrm{Total}}(l)$ in (a), and boundary sites $B(l)$ in (b), both in log scale. Here, $l$ denotes the lattice generation. The solid lines marked by circles and the dashed lines marked by triangles depict the exact and estimated $N^{\mathrm{Total}}(l)$ [\cref{N}] and $B(l)$ for finite-size hyperbolic lattices, respectively, showing excellent agreement. 
    Unlike the Euclidean \{$4,4$\} lattice (green dashed \CH{curves), $N^{\mathrm{Total}}(l)$ and $B(l)$ for the }hyperbolic lattices both exhibit exponential growth to excellent approximation.} 
	\label{fig:totalsize}
\end{figure*}

\section{Hyperbolic Scaling Law \label{scale}}
The spatial characteristics, geometries, and boundary configurations of a lattice profoundly influence the nature of the NHSE in a lattice \cite{shen2022non,schindler2021dislocation,zhang2022review,shang2022experimental,Jiang2023,manna2022inner,liu2021exact,guo2021exact}. 
As hyperbolic lattices have fundamentally differing connectivity properties, they are expected to have a profoundly differing NHSE. 
Below, we present finite-size and boundary scaling properties of hyperbolic lattices [\cref{fig:lattices}] and compare them with their Euclidean counterparts.

With the addition of a layer of sites $x$ to an existing hyperbolic \{$p,q$\} lattice with $x-1$ layers (or generations), the lattice grows by a certain number of sites which we call $n_{p,q}(x)$, for generic $p$ and $q$. 
While the exact solution of $n_{p,q}(x)$ is complicated, we can estimate it by considering the Bethe lattice \{$p,\infty$\} with apeirogons (polygon\CH{s with infinite number of sides}) that contain vertices of coordination numbers $p$ \cite{bradley1985directed,saberi2013growth}.
For such a Bethe lattice, each additional layer iteration generates $p$ number of sites. Hence the number of uniquely generated sites in a given $l$-th iteration is given by 
\begin{align}\label{eqn:n_exact}
	n_{p,\infty}(x)=(p-1)n_{p,\infty}(x-1),
\end{align}
where the factor $p-1$ \CH{takes into consideration duplicate} sites generated before the $x$-th iteration.
However, \cref{eqn:n_exact} does not simply apply to any \{$p,q$\} lattice with finite $q$ due to the formation of closed $q$-gons at some iteration $x>2$.
In particular, the number of duplicated sites that must be removed increases for each $q$-gon that emerges after $q/2$ generations.
Nevertheless, the exponential growth of hyperbolic lattices \cite{boettcher2022crystallography} allows us to make a reasonable approximation even if some duplicated sites are disregarded. Hence an estimate of $n_{p,q}(x)$ can be obtained by through a similar recursion relation:
\begin{align}\label{N}
    \wj{n_{p,q}(x)\approx(p-1)n_{p,q}(x-1),\quad x>2,}
\end{align}
with the initial conditions as $n_{p,q}(1)=1$ and $n_{p,q}(2)=p$. 
\wj{Additionally, this exponential growth allow us to further approximate the $n_{p,q}(x)$ by the number of newly generated non-duplicated sites:
\begin{align}\label{suppn}
    n_{p,q}(x)\approx (p-1)^{x}.
\end{align}
This scaling law is clearly demonstrated in \cref{fig:distribution}.}


With that, the lattice volume is given by
\begin{align}\label{NT}
    N^{\rm Total}_{p,q}(L)=\sum^{L}_{x=1}n_{p,q}(x).
\end{align}
The number of boundary sites is given by the final generation $B_{p,q}(L)=n_{p,q}(L)$.

In \cref{fig:totalsize}(a), we present comparisons of the estimated [\cref{suppn}] and exact number of nodes for our constructed hyperbolic \{$4$,$8$\} and \{$6$,$4$\} lattices. 
The strong agreement observed between the dashed (estimated) and solid (exact) curves in \cref{fig:totalsize} validates \cref{N}. 
Similar agreements hold for the boundary volume, $B_{p,q}(L)$, as illustrated in \cref{fig:totalsize}(b). 
Unlike their Euclidean counterparts (green), which do not grow exponentially due to substantial numbers of duplicate sites, hyperbolic \{$4$,$8$\} and \{$6$,$4$\} lattices \CH{grow exponentially to a very good approximation}. 

\bibliographystyle{apsrev4-1}

\bibliography{Paper}

\begin{thebibliography}{119}%
\makeatletter
\providecommand \@ifxundefined [1]{%
 \@ifx{#1\undefined}
}%
\providecommand \@ifnum [1]{%
 \ifnum #1\expandafter \@firstoftwo
 \else \expandafter \@secondoftwo
 \fi
}%
\providecommand \@ifx [1]{%
 \ifx #1\expandafter \@firstoftwo
 \else \expandafter \@secondoftwo
 \fi
}%
\providecommand \natexlab [1]{#1}%
\providecommand \enquote  [1]{``#1''}%
\providecommand \bibnamefont  [1]{#1}%
\providecommand \bibfnamefont [1]{#1}%
\providecommand \citenamefont [1]{#1}%
\providecommand \href@noop [0]{\@secondoftwo}%
\providecommand \href [0]{\begingroup \@sanitize@url \@href}%
\providecommand \@href[1]{\@@startlink{#1}\@@href}%
\providecommand \@@href[1]{\endgroup#1\@@endlink}%
\providecommand \@sanitize@url [0]{\catcode `\\12\catcode `\$12\catcode `\&12\catcode `\#12\catcode `\^12\catcode `\_12\catcode `\%12\relax}%
\providecommand \@@startlink[1]{}%
\providecommand \@@endlink[0]{}%
\providecommand \url  [0]{\begingroup\@sanitize@url \@url }%
\providecommand \@url [1]{\endgroup\@href {#1}{\urlprefix }}%
\providecommand \urlprefix  [0]{URL }%
\providecommand \Eprint [0]{\href }%
\providecommand \doibase [0]{http://dx.doi.org/}%
\providecommand \selectlanguage [0]{\@gobble}%
\providecommand \bibinfo  [0]{\@secondoftwo}%
\providecommand \bibfield  [0]{\@secondoftwo}%
\providecommand \translation [1]{[#1]}%
\providecommand \BibitemOpen [0]{}%
\providecommand \bibitemStop [0]{}%
\providecommand \bibitemNoStop [0]{.\EOS\space}%
\providecommand \EOS [0]{\spacefactor3000\relax}%
\providecommand \BibitemShut  [1]{\csname bibitem#1\endcsname}%
\let\auto@bib@innerbib\@empty
\bibitem [{\citenamefont {Kunst}\ \emph {et~al.}(2018)\citenamefont {Kunst}, \citenamefont {Edvardsson}, \citenamefont {Budich},\ and\ \citenamefont {Bergholtz}}]{kunst2018biorthogonal}%
  \BibitemOpen
  \bibfield  {author} {\bibinfo {author} {\bibfnamefont {F.~K.}\ \bibnamefont {Kunst}}, \bibinfo {author} {\bibfnamefont {E.}~\bibnamefont {Edvardsson}}, \bibinfo {author} {\bibfnamefont {J.~C.}\ \bibnamefont {Budich}}, \ and\ \bibinfo {author} {\bibfnamefont {E.~J.}\ \bibnamefont {Bergholtz}},\ }\href@noop {} {\bibfield  {journal} {\bibinfo  {journal} {Physical review letters}\ }\textbf {\bibinfo {volume} {121}},\ \bibinfo {pages} {026808} (\bibinfo {year} {2018})}\BibitemShut {NoStop}%
\bibitem [{\citenamefont {Xiong}(2018)}]{xiong2018does}%
  \BibitemOpen
  \bibfield  {author} {\bibinfo {author} {\bibfnamefont {Y.}~\bibnamefont {Xiong}},\ }\href@noop {} {\bibfield  {journal} {\bibinfo  {journal} {Journal of Physics Communications}\ }\textbf {\bibinfo {volume} {2}},\ \bibinfo {pages} {035043} (\bibinfo {year} {2018})}\BibitemShut {NoStop}%
\bibitem [{\citenamefont {Song}\ \emph {et~al.}(2019)\citenamefont {Song}, \citenamefont {Yao},\ and\ \citenamefont {Wang}}]{song2019non}%
  \BibitemOpen
  \bibfield  {author} {\bibinfo {author} {\bibfnamefont {F.}~\bibnamefont {Song}}, \bibinfo {author} {\bibfnamefont {S.}~\bibnamefont {Yao}}, \ and\ \bibinfo {author} {\bibfnamefont {Z.}~\bibnamefont {Wang}},\ }\href@noop {} {\bibfield  {journal} {\bibinfo  {journal} {Physical review letters}\ }\textbf {\bibinfo {volume} {123}},\ \bibinfo {pages} {170401} (\bibinfo {year} {2019})}\BibitemShut {NoStop}%
\bibitem [{\citenamefont {Longhi}(2019)}]{longhi2019probing}%
  \BibitemOpen
  \bibfield  {author} {\bibinfo {author} {\bibfnamefont {S.}~\bibnamefont {Longhi}},\ }\href@noop {} {\bibfield  {journal} {\bibinfo  {journal} {Physical Review Research}\ }\textbf {\bibinfo {volume} {1}},\ \bibinfo {pages} {023013} (\bibinfo {year} {2019})}\BibitemShut {NoStop}%
\bibitem [{\citenamefont {Xiao}\ \emph {et~al.}(2020)\citenamefont {Xiao}, \citenamefont {Deng}, \citenamefont {Wang}, \citenamefont {Zhu}, \citenamefont {Wang}, \citenamefont {Yi},\ and\ \citenamefont {Xue}}]{xiao2020non}%
  \BibitemOpen
  \bibfield  {author} {\bibinfo {author} {\bibfnamefont {L.}~\bibnamefont {Xiao}}, \bibinfo {author} {\bibfnamefont {T.}~\bibnamefont {Deng}}, \bibinfo {author} {\bibfnamefont {K.}~\bibnamefont {Wang}}, \bibinfo {author} {\bibfnamefont {G.}~\bibnamefont {Zhu}}, \bibinfo {author} {\bibfnamefont {Z.}~\bibnamefont {Wang}}, \bibinfo {author} {\bibfnamefont {W.}~\bibnamefont {Yi}}, \ and\ \bibinfo {author} {\bibfnamefont {P.}~\bibnamefont {Xue}},\ }\href@noop {} {\bibfield  {journal} {\bibinfo  {journal} {Nature Physics}\ }\textbf {\bibinfo {volume} {16}},\ \bibinfo {pages} {761} (\bibinfo {year} {2020})}\BibitemShut {NoStop}%
\bibitem [{\citenamefont {Helbig}\ \emph {et~al.}(2020)\citenamefont {Helbig}, \citenamefont {Hofmann}, \citenamefont {Imhof}, \citenamefont {Abdelghany}, \citenamefont {Kiessling}, \citenamefont {Molenkamp}, \citenamefont {Lee}, \citenamefont {Szameit}, \citenamefont {Greiter},\ and\ \citenamefont {Thomale}}]{helbig2020generalized}%
  \BibitemOpen
  \bibfield  {author} {\bibinfo {author} {\bibfnamefont {T.}~\bibnamefont {Helbig}}, \bibinfo {author} {\bibfnamefont {T.}~\bibnamefont {Hofmann}}, \bibinfo {author} {\bibfnamefont {S.}~\bibnamefont {Imhof}}, \bibinfo {author} {\bibfnamefont {M.}~\bibnamefont {Abdelghany}}, \bibinfo {author} {\bibfnamefont {T.}~\bibnamefont {Kiessling}}, \bibinfo {author} {\bibfnamefont {L.}~\bibnamefont {Molenkamp}}, \bibinfo {author} {\bibfnamefont {C.}~\bibnamefont {Lee}}, \bibinfo {author} {\bibfnamefont {A.}~\bibnamefont {Szameit}}, \bibinfo {author} {\bibfnamefont {M.}~\bibnamefont {Greiter}}, \ and\ \bibinfo {author} {\bibfnamefont {R.}~\bibnamefont {Thomale}},\ }\href@noop {} {\bibfield  {journal} {\bibinfo  {journal} {Nature Physics}\ }\textbf {\bibinfo {volume} {16}},\ \bibinfo {pages} {747} (\bibinfo {year} {2020})}\BibitemShut {NoStop}%
\bibitem [{\citenamefont {Okuma}\ \emph {et~al.}(2020)\citenamefont {Okuma}, \citenamefont {Kawabata}, \citenamefont {Shiozaki},\ and\ \citenamefont {Sato}}]{okuma2020topological}%
  \BibitemOpen
  \bibfield  {author} {\bibinfo {author} {\bibfnamefont {N.}~\bibnamefont {Okuma}}, \bibinfo {author} {\bibfnamefont {K.}~\bibnamefont {Kawabata}}, \bibinfo {author} {\bibfnamefont {K.}~\bibnamefont {Shiozaki}}, \ and\ \bibinfo {author} {\bibfnamefont {M.}~\bibnamefont {Sato}},\ }\href@noop {} {\bibfield  {journal} {\bibinfo  {journal} {Physical review letters}\ }\textbf {\bibinfo {volume} {124}},\ \bibinfo {pages} {086801} (\bibinfo {year} {2020})}\BibitemShut {NoStop}%
\bibitem [{\citenamefont {Li}\ \emph {et~al.}(2020)\citenamefont {Li}, \citenamefont {Lee}, \citenamefont {Mu},\ and\ \citenamefont {Gong}}]{li2020critical}%
  \BibitemOpen
  \bibfield  {author} {\bibinfo {author} {\bibfnamefont {L.}~\bibnamefont {Li}}, \bibinfo {author} {\bibfnamefont {C.~H.}\ \bibnamefont {Lee}}, \bibinfo {author} {\bibfnamefont {S.}~\bibnamefont {Mu}}, \ and\ \bibinfo {author} {\bibfnamefont {J.}~\bibnamefont {Gong}},\ }\href@noop {} {\bibfield  {journal} {\bibinfo  {journal} {Nature communications}\ }\textbf {\bibinfo {volume} {11}},\ \bibinfo {pages} {5491} (\bibinfo {year} {2020})}\BibitemShut {NoStop}%
\bibitem [{\citenamefont {Arouca}\ \emph {et~al.}(2020)\citenamefont {Arouca}, \citenamefont {Lee},\ and\ \citenamefont {Morais~Smith}}]{arouca2020unconventional}%
  \BibitemOpen
  \bibfield  {author} {\bibinfo {author} {\bibfnamefont {R.}~\bibnamefont {Arouca}}, \bibinfo {author} {\bibfnamefont {C.}~\bibnamefont {Lee}}, \ and\ \bibinfo {author} {\bibfnamefont {C.}~\bibnamefont {Morais~Smith}},\ }\href@noop {} {\bibfield  {journal} {\bibinfo  {journal} {Physical Review B}\ }\textbf {\bibinfo {volume} {102}},\ \bibinfo {pages} {245145} (\bibinfo {year} {2020})}\BibitemShut {NoStop}%
\bibitem [{\citenamefont {Pan}\ \emph {et~al.}(2020)\citenamefont {Pan}, \citenamefont {Chen}, \citenamefont {Chen},\ and\ \citenamefont {Zhai}}]{pan2020non}%
  \BibitemOpen
  \bibfield  {author} {\bibinfo {author} {\bibfnamefont {L.}~\bibnamefont {Pan}}, \bibinfo {author} {\bibfnamefont {X.}~\bibnamefont {Chen}}, \bibinfo {author} {\bibfnamefont {Y.}~\bibnamefont {Chen}}, \ and\ \bibinfo {author} {\bibfnamefont {H.}~\bibnamefont {Zhai}},\ }\href@noop {} {\bibfield  {journal} {\bibinfo  {journal} {Nature Physics}\ }\textbf {\bibinfo {volume} {16}},\ \bibinfo {pages} {767} (\bibinfo {year} {2020})}\BibitemShut {NoStop}%
\bibitem [{\citenamefont {Yang}\ \emph {et~al.}(2024{\natexlab{a}})\citenamefont {Yang}, \citenamefont {Yang}, \citenamefont {Ma}, \citenamefont {Li}, \citenamefont {Zhang},\ and\ \citenamefont {Chan}}]{yang2024non}%
  \BibitemOpen
  \bibfield  {author} {\bibinfo {author} {\bibfnamefont {Y.}~\bibnamefont {Yang}}, \bibinfo {author} {\bibfnamefont {B.}~\bibnamefont {Yang}}, \bibinfo {author} {\bibfnamefont {G.}~\bibnamefont {Ma}}, \bibinfo {author} {\bibfnamefont {J.}~\bibnamefont {Li}}, \bibinfo {author} {\bibfnamefont {S.}~\bibnamefont {Zhang}}, \ and\ \bibinfo {author} {\bibfnamefont {C.}~\bibnamefont {Chan}},\ }\href@noop {} {\bibfield  {journal} {\bibinfo  {journal} {Science}\ }\textbf {\bibinfo {volume} {383}},\ \bibinfo {pages} {eadf9621} (\bibinfo {year} {2024}{\natexlab{a}})}\BibitemShut {NoStop}%
\bibitem [{\citenamefont {Shen}\ and\ \citenamefont {Lee}(2022)}]{shen2022non}%
  \BibitemOpen
  \bibfield  {author} {\bibinfo {author} {\bibfnamefont {R.}~\bibnamefont {Shen}}\ and\ \bibinfo {author} {\bibfnamefont {C.~H.}\ \bibnamefont {Lee}},\ }\href@noop {} {\bibfield  {journal} {\bibinfo  {journal} {Communications Physics}\ }\textbf {\bibinfo {volume} {5}},\ \bibinfo {pages} {238} (\bibinfo {year} {2022})}\BibitemShut {NoStop}%
\bibitem [{\citenamefont {Yang}\ \emph {et~al.}(2022)\citenamefont {Yang}, \citenamefont {Tan}, \citenamefont {Tai}, \citenamefont {Koh}, \citenamefont {Li}, \citenamefont {Longhi},\ and\ \citenamefont {Lee}}]{yang2022designing}%
  \BibitemOpen
  \bibfield  {author} {\bibinfo {author} {\bibfnamefont {R.}~\bibnamefont {Yang}}, \bibinfo {author} {\bibfnamefont {J.~W.}\ \bibnamefont {Tan}}, \bibinfo {author} {\bibfnamefont {T.}~\bibnamefont {Tai}}, \bibinfo {author} {\bibfnamefont {J.~M.}\ \bibnamefont {Koh}}, \bibinfo {author} {\bibfnamefont {L.}~\bibnamefont {Li}}, \bibinfo {author} {\bibfnamefont {S.}~\bibnamefont {Longhi}}, \ and\ \bibinfo {author} {\bibfnamefont {C.~H.}\ \bibnamefont {Lee}},\ }\href@noop {} {\bibfield  {journal} {\bibinfo  {journal} {Science Bulletin}\ }\textbf {\bibinfo {volume} {67}},\ \bibinfo {pages} {1865} (\bibinfo {year} {2022})}\BibitemShut {NoStop}%
\bibitem [{\citenamefont {Li}\ and\ \citenamefont {Lee}(2022)}]{li2022non}%
  \BibitemOpen
  \bibfield  {author} {\bibinfo {author} {\bibfnamefont {L.}~\bibnamefont {Li}}\ and\ \bibinfo {author} {\bibfnamefont {C.~H.}\ \bibnamefont {Lee}},\ }\href@noop {} {\bibfield  {journal} {\bibinfo  {journal} {Science Bulletin}\ }\textbf {\bibinfo {volume} {67}},\ \bibinfo {pages} {685} (\bibinfo {year} {2022})}\BibitemShut {NoStop}%
\bibitem [{\citenamefont {Zou}\ \emph {et~al.}(2021)\citenamefont {Zou}, \citenamefont {Chen}, \citenamefont {He}, \citenamefont {Bao}, \citenamefont {Lee}, \citenamefont {Sun},\ and\ \citenamefont {Zhang}}]{zou2021observation}%
  \BibitemOpen
  \bibfield  {author} {\bibinfo {author} {\bibfnamefont {D.}~\bibnamefont {Zou}}, \bibinfo {author} {\bibfnamefont {T.}~\bibnamefont {Chen}}, \bibinfo {author} {\bibfnamefont {W.}~\bibnamefont {He}}, \bibinfo {author} {\bibfnamefont {J.}~\bibnamefont {Bao}}, \bibinfo {author} {\bibfnamefont {C.~H.}\ \bibnamefont {Lee}}, \bibinfo {author} {\bibfnamefont {H.}~\bibnamefont {Sun}}, \ and\ \bibinfo {author} {\bibfnamefont {X.}~\bibnamefont {Zhang}},\ }\href@noop {} {\bibfield  {journal} {\bibinfo  {journal} {Nature Communications}\ }\textbf {\bibinfo {volume} {12}},\ \bibinfo {pages} {1} (\bibinfo {year} {2021})}\BibitemShut {NoStop}%
\bibitem [{\citenamefont {Zhang}\ \emph {et~al.}(2021)\citenamefont {Zhang}, \citenamefont {Tian}, \citenamefont {Jiang}, \citenamefont {Lu},\ and\ \citenamefont {Chen}}]{zhang2021observation}%
  \BibitemOpen
  \bibfield  {author} {\bibinfo {author} {\bibfnamefont {X.}~\bibnamefont {Zhang}}, \bibinfo {author} {\bibfnamefont {Y.}~\bibnamefont {Tian}}, \bibinfo {author} {\bibfnamefont {J.-H.}\ \bibnamefont {Jiang}}, \bibinfo {author} {\bibfnamefont {M.-H.}\ \bibnamefont {Lu}}, \ and\ \bibinfo {author} {\bibfnamefont {Y.-F.}\ \bibnamefont {Chen}},\ }\href@noop {} {\bibfield  {journal} {\bibinfo  {journal} {Nature communications}\ }\textbf {\bibinfo {volume} {12}},\ \bibinfo {pages} {1} (\bibinfo {year} {2021})}\BibitemShut {NoStop}%
\bibitem [{\citenamefont {Qin}\ \emph {et~al.}(2022)\citenamefont {Qin}, \citenamefont {Shen},\ and\ \citenamefont {Lee}}]{qin2022non}%
  \BibitemOpen
  \bibfield  {author} {\bibinfo {author} {\bibfnamefont {F.}~\bibnamefont {Qin}}, \bibinfo {author} {\bibfnamefont {R.}~\bibnamefont {Shen}}, \ and\ \bibinfo {author} {\bibfnamefont {C.~H.}\ \bibnamefont {Lee}},\ }\href@noop {} {\bibfield  {journal} {\bibinfo  {journal} {arXiv preprint arXiv:2202.10481}\ } (\bibinfo {year} {2022})}\BibitemShut {NoStop}%
\bibitem [{\citenamefont {Rafi-Ul-Islam}\ \emph {et~al.}(2022)\citenamefont {Rafi-Ul-Islam}, \citenamefont {Siu}, \citenamefont {Sahin}, \citenamefont {Lee},\ and\ \citenamefont {Jalil}}]{Rafi-Ul-Islam2022}%
  \BibitemOpen
  \bibfield  {author} {\bibinfo {author} {\bibfnamefont {S.~M.}\ \bibnamefont {Rafi-Ul-Islam}}, \bibinfo {author} {\bibfnamefont {Z.~B.}\ \bibnamefont {Siu}}, \bibinfo {author} {\bibfnamefont {H.}~\bibnamefont {Sahin}}, \bibinfo {author} {\bibfnamefont {C.~H.}\ \bibnamefont {Lee}}, \ and\ \bibinfo {author} {\bibfnamefont {M.~B.}\ \bibnamefont {Jalil}},\ }\href {\doibase 10.1103/PhysRevResearch.4.013243} {\bibfield  {journal} {\bibinfo  {journal} {Phys. Rev. Res.}\ }\textbf {\bibinfo {volume} {4}},\ \bibinfo {pages} {1} (\bibinfo {year} {2022})}\BibitemShut {NoStop}%
\bibitem [{\citenamefont {Jiang}\ and\ \citenamefont {Lee}(2023)}]{Jiang2023}%
  \BibitemOpen
  \bibfield  {author} {\bibinfo {author} {\bibfnamefont {H.}~\bibnamefont {Jiang}}\ and\ \bibinfo {author} {\bibfnamefont {C.~H.}\ \bibnamefont {Lee}},\ }\href {\doibase 10.1103/PhysRevLett.131.076401} {\bibfield  {journal} {\bibinfo  {journal} {Phys. Rev. Lett.}\ }\textbf {\bibinfo {volume} {131}},\ \bibinfo {pages} {076401} (\bibinfo {year} {2023})}\BibitemShut {NoStop}%
\bibitem [{\citenamefont {Zhang}\ \emph {et~al.}(2022{\natexlab{a}})\citenamefont {Zhang}, \citenamefont {Li}, \citenamefont {Zhang},\ and\ \citenamefont {Lee}}]{zhang2022real}%
  \BibitemOpen
  \bibfield  {author} {\bibinfo {author} {\bibfnamefont {B.}~\bibnamefont {Zhang}}, \bibinfo {author} {\bibfnamefont {Q.}~\bibnamefont {Li}}, \bibinfo {author} {\bibfnamefont {X.}~\bibnamefont {Zhang}}, \ and\ \bibinfo {author} {\bibfnamefont {C.~H.}\ \bibnamefont {Lee}},\ }\href@noop {} {\bibfield  {journal} {\bibinfo  {journal} {Chinese Physics B}\ } (\bibinfo {year} {2022}{\natexlab{a}})}\BibitemShut {NoStop}%
\bibitem [{\citenamefont {Qin}\ \emph {et~al.}(2024{\natexlab{a}})\citenamefont {Qin}, \citenamefont {Shen}, \citenamefont {Li},\ and\ \citenamefont {Lee}}]{qin2024kinked}%
  \BibitemOpen
  \bibfield  {author} {\bibinfo {author} {\bibfnamefont {F.}~\bibnamefont {Qin}}, \bibinfo {author} {\bibfnamefont {R.}~\bibnamefont {Shen}}, \bibinfo {author} {\bibfnamefont {L.}~\bibnamefont {Li}}, \ and\ \bibinfo {author} {\bibfnamefont {C.~H.}\ \bibnamefont {Lee}},\ }\href@noop {} {\bibfield  {journal} {\bibinfo  {journal} {Physical Review A}\ }\textbf {\bibinfo {volume} {109}},\ \bibinfo {pages} {053311} (\bibinfo {year} {2024}{\natexlab{a}})}\BibitemShut {NoStop}%
\bibitem [{\citenamefont {Qin}\ \emph {et~al.}(2023)\citenamefont {Qin}, \citenamefont {Ma}, \citenamefont {Shen},\ and\ \citenamefont {Lee}}]{qin2023universal}%
  \BibitemOpen
  \bibfield  {author} {\bibinfo {author} {\bibfnamefont {F.}~\bibnamefont {Qin}}, \bibinfo {author} {\bibfnamefont {Y.}~\bibnamefont {Ma}}, \bibinfo {author} {\bibfnamefont {R.}~\bibnamefont {Shen}}, \ and\ \bibinfo {author} {\bibfnamefont {C.~H.}\ \bibnamefont {Lee}},\ }\href@noop {} {\bibfield  {journal} {\bibinfo  {journal} {Physical Review B}\ }\textbf {\bibinfo {volume} {107}},\ \bibinfo {pages} {155430} (\bibinfo {year} {2023})}\BibitemShut {NoStop}%
\bibitem [{\citenamefont {Shen}\ \emph {et~al.}(2023)\citenamefont {Shen}, \citenamefont {Chen}, \citenamefont {Yang},\ and\ \citenamefont {Lee}}]{shen2023observation}%
  \BibitemOpen
  \bibfield  {author} {\bibinfo {author} {\bibfnamefont {R.}~\bibnamefont {Shen}}, \bibinfo {author} {\bibfnamefont {T.}~\bibnamefont {Chen}}, \bibinfo {author} {\bibfnamefont {B.}~\bibnamefont {Yang}}, \ and\ \bibinfo {author} {\bibfnamefont {C.~H.}\ \bibnamefont {Lee}},\ }\href@noop {} {\bibfield  {journal} {\bibinfo  {journal} {arXiv preprint arXiv:2311.10143}\ } (\bibinfo {year} {2023})}\BibitemShut {NoStop}%
\bibitem [{\citenamefont {Shen}\ \emph {et~al.}(2024)\citenamefont {Shen}, \citenamefont {Qin}, \citenamefont {Desaules}, \citenamefont {Papi{\'c}},\ and\ \citenamefont {Lee}}]{shen2024enhanced}%
  \BibitemOpen
  \bibfield  {author} {\bibinfo {author} {\bibfnamefont {R.}~\bibnamefont {Shen}}, \bibinfo {author} {\bibfnamefont {F.}~\bibnamefont {Qin}}, \bibinfo {author} {\bibfnamefont {J.-Y.}\ \bibnamefont {Desaules}}, \bibinfo {author} {\bibfnamefont {Z.}~\bibnamefont {Papi{\'c}}}, \ and\ \bibinfo {author} {\bibfnamefont {C.~H.}\ \bibnamefont {Lee}},\ }\href@noop {} {\bibfield  {journal} {\bibinfo  {journal} {arXiv preprint arXiv:2403.02395}\ } (\bibinfo {year} {2024})}\BibitemShut {NoStop}%
\bibitem [{\citenamefont {Liu}\ \emph {et~al.}(2024{\natexlab{a}})\citenamefont {Liu}, \citenamefont {Jiang}, \citenamefont {Xue}, \citenamefont {Li}, \citenamefont {Gong}, \citenamefont {Liu},\ and\ \citenamefont {Lee}}]{liu2024non}%
  \BibitemOpen
  \bibfield  {author} {\bibinfo {author} {\bibfnamefont {S.}~\bibnamefont {Liu}}, \bibinfo {author} {\bibfnamefont {H.}~\bibnamefont {Jiang}}, \bibinfo {author} {\bibfnamefont {W.-T.}\ \bibnamefont {Xue}}, \bibinfo {author} {\bibfnamefont {Q.}~\bibnamefont {Li}}, \bibinfo {author} {\bibfnamefont {J.}~\bibnamefont {Gong}}, \bibinfo {author} {\bibfnamefont {X.}~\bibnamefont {Liu}}, \ and\ \bibinfo {author} {\bibfnamefont {C.~H.}\ \bibnamefont {Lee}},\ }\href@noop {} {\bibfield  {journal} {\bibinfo  {journal} {arXiv preprint arXiv:2408.02736}\ } (\bibinfo {year} {2024}{\natexlab{a}})}\BibitemShut {NoStop}%
\bibitem [{\citenamefont {Yang}\ \emph {et~al.}(2024{\natexlab{b}})\citenamefont {Yang}, \citenamefont {Yuan},\ and\ \citenamefont {Lee}}]{Yang2024eud}%
  \BibitemOpen
  \bibfield  {author} {\bibinfo {author} {\bibfnamefont {M.}~\bibnamefont {Yang}}, \bibinfo {author} {\bibfnamefont {L.}~\bibnamefont {Yuan}}, \ and\ \bibinfo {author} {\bibfnamefont {C.~H.}\ \bibnamefont {Lee}},\ }\href@noop {} {\  (\bibinfo {year} {2024}{\natexlab{b}})},\ \Eprint {http://arxiv.org/abs/2410.01258} {arXiv:2410.01258 [cond-mat.other]} \BibitemShut {NoStop}%
\bibitem [{\citenamefont {Li}\ \emph {et~al.}(2024)\citenamefont {Li}, \citenamefont {Wang}, \citenamefont {Wang}, \citenamefont {Lin}, \citenamefont {Ma},\ and\ \citenamefont {Jiang}}]{li2024observation}%
  \BibitemOpen
  \bibfield  {author} {\bibinfo {author} {\bibfnamefont {Z.}~\bibnamefont {Li}}, \bibinfo {author} {\bibfnamefont {L.-W.}\ \bibnamefont {Wang}}, \bibinfo {author} {\bibfnamefont {X.}~\bibnamefont {Wang}}, \bibinfo {author} {\bibfnamefont {Z.-K.}\ \bibnamefont {Lin}}, \bibinfo {author} {\bibfnamefont {G.}~\bibnamefont {Ma}}, \ and\ \bibinfo {author} {\bibfnamefont {J.-H.}\ \bibnamefont {Jiang}},\ }\href@noop {} {\bibfield  {journal} {\bibinfo  {journal} {Nature Communications}\ }\textbf {\bibinfo {volume} {15}},\ \bibinfo {pages} {6544} (\bibinfo {year} {2024})}\BibitemShut {NoStop}%
\bibitem [{\citenamefont {Gliozzi}\ \emph {et~al.}(2024{\natexlab{a}})\citenamefont {Gliozzi}, \citenamefont {De~Tomasi},\ and\ \citenamefont {Hughes}}]{gliozzi2024many}%
  \BibitemOpen
  \bibfield  {author} {\bibinfo {author} {\bibfnamefont {J.}~\bibnamefont {Gliozzi}}, \bibinfo {author} {\bibfnamefont {G.}~\bibnamefont {De~Tomasi}}, \ and\ \bibinfo {author} {\bibfnamefont {T.~L.}\ \bibnamefont {Hughes}},\ }\href@noop {} {\bibfield  {journal} {\bibinfo  {journal} {Physical Review Letters}\ }\textbf {\bibinfo {volume} {133}},\ \bibinfo {pages} {136503} (\bibinfo {year} {2024}{\natexlab{a}})}\BibitemShut {NoStop}%
\bibitem [{\citenamefont {Liu}\ \emph {et~al.}(2024{\natexlab{b}})\citenamefont {Liu}, \citenamefont {Mandal}, \citenamefont {Zhou}, \citenamefont {Xi}, \citenamefont {Banerjee}, \citenamefont {Hu}, \citenamefont {Wei}, \citenamefont {Wang}, \citenamefont {Wang}, \citenamefont {Gao} \emph {et~al.}}]{liu2024localization}%
  \BibitemOpen
  \bibfield  {author} {\bibinfo {author} {\bibfnamefont {G.-G.}\ \bibnamefont {Liu}}, \bibinfo {author} {\bibfnamefont {S.}~\bibnamefont {Mandal}}, \bibinfo {author} {\bibfnamefont {P.}~\bibnamefont {Zhou}}, \bibinfo {author} {\bibfnamefont {X.}~\bibnamefont {Xi}}, \bibinfo {author} {\bibfnamefont {R.}~\bibnamefont {Banerjee}}, \bibinfo {author} {\bibfnamefont {Y.-H.}\ \bibnamefont {Hu}}, \bibinfo {author} {\bibfnamefont {M.}~\bibnamefont {Wei}}, \bibinfo {author} {\bibfnamefont {M.}~\bibnamefont {Wang}}, \bibinfo {author} {\bibfnamefont {Q.}~\bibnamefont {Wang}}, \bibinfo {author} {\bibfnamefont {Z.}~\bibnamefont {Gao}},  \emph {et~al.},\ }\href@noop {} {\bibfield  {journal} {\bibinfo  {journal} {Physical Review Letters}\ }\textbf {\bibinfo {volume} {132}},\ \bibinfo {pages} {113802} (\bibinfo {year} {2024}{\natexlab{b}})}\BibitemShut {NoStop}%
\bibitem [{\citenamefont {Yoshida}\ \emph {et~al.}(2024)\citenamefont {Yoshida}, \citenamefont {Zhang}, \citenamefont {Neupert},\ and\ \citenamefont {Kawakami}}]{yoshida2024non}%
  \BibitemOpen
  \bibfield  {author} {\bibinfo {author} {\bibfnamefont {T.}~\bibnamefont {Yoshida}}, \bibinfo {author} {\bibfnamefont {S.-B.}\ \bibnamefont {Zhang}}, \bibinfo {author} {\bibfnamefont {T.}~\bibnamefont {Neupert}}, \ and\ \bibinfo {author} {\bibfnamefont {N.}~\bibnamefont {Kawakami}},\ }\href@noop {} {\bibfield  {journal} {\bibinfo  {journal} {Physical Review Letters}\ }\textbf {\bibinfo {volume} {133}},\ \bibinfo {pages} {076502} (\bibinfo {year} {2024})}\BibitemShut {NoStop}%
\bibitem [{\citenamefont {Gliozzi}\ \emph {et~al.}(2024{\natexlab{b}})\citenamefont {Gliozzi}, \citenamefont {{De Tomasi}},\ and\ \citenamefont {Hughes}}]{Gliozzi2024a}%
  \BibitemOpen
  \bibfield  {author} {\bibinfo {author} {\bibfnamefont {J.}~\bibnamefont {Gliozzi}}, \bibinfo {author} {\bibfnamefont {G.}~\bibnamefont {{De Tomasi}}}, \ and\ \bibinfo {author} {\bibfnamefont {T.~L.}\ \bibnamefont {Hughes}},\ }\href {\doibase 10.1103/PhysRevLett.133.136503} {\bibfield  {journal} {\bibinfo  {journal} {Phys. Rev. Lett.}\ }\textbf {\bibinfo {volume} {133}},\ \bibinfo {pages} {136503} (\bibinfo {year} {2024}{\natexlab{b}})},\ \Eprint {http://arxiv.org/abs/2401.04162} {2401.04162} \BibitemShut {NoStop}%
\bibitem [{\citenamefont {Li}\ \emph {et~al.}(2022{\natexlab{a}})\citenamefont {Li}, \citenamefont {Trauzettel}, \citenamefont {Neupert},\ and\ \citenamefont {Zhang}}]{Li2022c}%
  \BibitemOpen
  \bibfield  {author} {\bibinfo {author} {\bibfnamefont {C.-A.}\ \bibnamefont {Li}}, \bibinfo {author} {\bibfnamefont {B.}~\bibnamefont {Trauzettel}}, \bibinfo {author} {\bibfnamefont {T.}~\bibnamefont {Neupert}}, \ and\ \bibinfo {author} {\bibfnamefont {S.-B.}\ \bibnamefont {Zhang}},\ }\href {\doibase 10.1103/PhysRevLett.131.116601} {\bibfield  {journal} {\bibinfo  {journal} {Phys. Rev. Lett.}\ }\textbf {\bibinfo {volume} {131}},\ \bibinfo {pages} {116601} (\bibinfo {year} {2022}{\natexlab{a}})}\BibitemShut {NoStop}%
\bibitem [{\citenamefont {Sun}\ \emph {et~al.}(2024)\citenamefont {Sun}, \citenamefont {Li}, \citenamefont {Feng},\ and\ \citenamefont {Guo}}]{Sun2024a}%
  \BibitemOpen
  \bibfield  {author} {\bibinfo {author} {\bibfnamefont {J.}~\bibnamefont {Sun}}, \bibinfo {author} {\bibfnamefont {C.-A.}\ \bibnamefont {Li}}, \bibinfo {author} {\bibfnamefont {S.}~\bibnamefont {Feng}}, \ and\ \bibinfo {author} {\bibfnamefont {H.}~\bibnamefont {Guo}},\ }\href {http://arxiv.org/abs/2409.07117} {\ ,\ \bibinfo {pages} {1} (\bibinfo {year} {2024})},\ \Eprint {http://arxiv.org/abs/2409.07117} {2409.07117} \BibitemShut {NoStop}%
\bibitem [{\citenamefont {Shen}\ and\ \citenamefont {Lee}(2021)}]{shen2021non}%
  \BibitemOpen
  \bibfield  {author} {\bibinfo {author} {\bibfnamefont {R.}~\bibnamefont {Shen}}\ and\ \bibinfo {author} {\bibfnamefont {C.~H.}\ \bibnamefont {Lee}},\ }\href@noop {} {\bibfield  {journal} {\bibinfo  {journal} {arXiv preprint arXiv:2107.03414}\ } (\bibinfo {year} {2021})}\BibitemShut {NoStop}%
\bibitem [{\citenamefont {Schindler}\ and\ \citenamefont {Prem}(2021)}]{schindler2021dislocation}%
  \BibitemOpen
  \bibfield  {author} {\bibinfo {author} {\bibfnamefont {F.}~\bibnamefont {Schindler}}\ and\ \bibinfo {author} {\bibfnamefont {A.}~\bibnamefont {Prem}},\ }\href@noop {} {\bibfield  {journal} {\bibinfo  {journal} {Physical Review B}\ }\textbf {\bibinfo {volume} {104}},\ \bibinfo {pages} {L161106} (\bibinfo {year} {2021})}\BibitemShut {NoStop}%
\bibitem [{\citenamefont {Zhang}\ \emph {et~al.}(2022{\natexlab{b}})\citenamefont {Zhang}, \citenamefont {Zhang}, \citenamefont {Lu},\ and\ \citenamefont {Chen}}]{zhang2022review}%
  \BibitemOpen
  \bibfield  {author} {\bibinfo {author} {\bibfnamefont {X.}~\bibnamefont {Zhang}}, \bibinfo {author} {\bibfnamefont {T.}~\bibnamefont {Zhang}}, \bibinfo {author} {\bibfnamefont {M.-H.}\ \bibnamefont {Lu}}, \ and\ \bibinfo {author} {\bibfnamefont {Y.-F.}\ \bibnamefont {Chen}},\ }\href@noop {} {\bibfield  {journal} {\bibinfo  {journal} {Advances in Physics: X}\ }\textbf {\bibinfo {volume} {7}},\ \bibinfo {pages} {2109431} (\bibinfo {year} {2022}{\natexlab{b}})}\BibitemShut {NoStop}%
\bibitem [{\citenamefont {Shang}\ \emph {et~al.}(2022)\citenamefont {Shang}, \citenamefont {Liu}, \citenamefont {Shao}, \citenamefont {Han}, \citenamefont {Zang}, \citenamefont {Zhang}, \citenamefont {Salama}, \citenamefont {Gao}, \citenamefont {Lee}, \citenamefont {Thomale} \emph {et~al.}}]{shang2022experimental}%
  \BibitemOpen
  \bibfield  {author} {\bibinfo {author} {\bibfnamefont {C.}~\bibnamefont {Shang}}, \bibinfo {author} {\bibfnamefont {S.}~\bibnamefont {Liu}}, \bibinfo {author} {\bibfnamefont {R.}~\bibnamefont {Shao}}, \bibinfo {author} {\bibfnamefont {P.}~\bibnamefont {Han}}, \bibinfo {author} {\bibfnamefont {X.}~\bibnamefont {Zang}}, \bibinfo {author} {\bibfnamefont {X.}~\bibnamefont {Zhang}}, \bibinfo {author} {\bibfnamefont {K.~N.}\ \bibnamefont {Salama}}, \bibinfo {author} {\bibfnamefont {W.}~\bibnamefont {Gao}}, \bibinfo {author} {\bibfnamefont {C.~H.}\ \bibnamefont {Lee}}, \bibinfo {author} {\bibfnamefont {R.}~\bibnamefont {Thomale}},  \emph {et~al.},\ }\href@noop {} {\bibfield  {journal} {\bibinfo  {journal} {arXiv preprint arXiv:2203.00484}\ } (\bibinfo {year} {2022})}\BibitemShut {NoStop}%
\bibitem [{\citenamefont {Manna}\ and\ \citenamefont {Roy}(2022)}]{manna2022inner}%
  \BibitemOpen
  \bibfield  {author} {\bibinfo {author} {\bibfnamefont {S.}~\bibnamefont {Manna}}\ and\ \bibinfo {author} {\bibfnamefont {B.}~\bibnamefont {Roy}},\ }\href@noop {} {\bibfield  {journal} {\bibinfo  {journal} {arXiv preprint arXiv:2202.07658}\ } (\bibinfo {year} {2022})}\BibitemShut {NoStop}%
\bibitem [{\citenamefont {Gu}\ \emph {et~al.}(2016)\citenamefont {Gu}, \citenamefont {Lee}, \citenamefont {Wen}, \citenamefont {Cho}, \citenamefont {Ryu},\ and\ \citenamefont {Qi}}]{gu2016holographic}%
  \BibitemOpen
  \bibfield  {author} {\bibinfo {author} {\bibfnamefont {Y.}~\bibnamefont {Gu}}, \bibinfo {author} {\bibfnamefont {C.~H.}\ \bibnamefont {Lee}}, \bibinfo {author} {\bibfnamefont {X.}~\bibnamefont {Wen}}, \bibinfo {author} {\bibfnamefont {G.~Y.}\ \bibnamefont {Cho}}, \bibinfo {author} {\bibfnamefont {S.}~\bibnamefont {Ryu}}, \ and\ \bibinfo {author} {\bibfnamefont {X.-L.}\ \bibnamefont {Qi}},\ }\href@noop {} {\bibfield  {journal} {\bibinfo  {journal} {Physical Review B}\ }\textbf {\bibinfo {volume} {94}},\ \bibinfo {pages} {125107} (\bibinfo {year} {2016})}\BibitemShut {NoStop}%
\bibitem [{\citenamefont {Zhong}\ \emph {et~al.}(2024)\citenamefont {Zhong}, \citenamefont {de~Castro}, \citenamefont {Lu}, \citenamefont {Kim}, \citenamefont {Oudich}, \citenamefont {Ji}, \citenamefont {Shi}, \citenamefont {Chen}, \citenamefont {Lu}, \citenamefont {Jing} \emph {et~al.}}]{zhong2024higher}%
  \BibitemOpen
  \bibfield  {author} {\bibinfo {author} {\bibfnamefont {J.-X.}\ \bibnamefont {Zhong}}, \bibinfo {author} {\bibfnamefont {P.~F.}\ \bibnamefont {de~Castro}}, \bibinfo {author} {\bibfnamefont {T.}~\bibnamefont {Lu}}, \bibinfo {author} {\bibfnamefont {J.}~\bibnamefont {Kim}}, \bibinfo {author} {\bibfnamefont {M.}~\bibnamefont {Oudich}}, \bibinfo {author} {\bibfnamefont {J.}~\bibnamefont {Ji}}, \bibinfo {author} {\bibfnamefont {L.}~\bibnamefont {Shi}}, \bibinfo {author} {\bibfnamefont {K.}~\bibnamefont {Chen}}, \bibinfo {author} {\bibfnamefont {J.}~\bibnamefont {Lu}}, \bibinfo {author} {\bibfnamefont {Y.}~\bibnamefont {Jing}},  \emph {et~al.},\ }\href@noop {} {\bibfield  {journal} {\bibinfo  {journal} {arXiv preprint arXiv:2409.01516}\ } (\bibinfo {year} {2024})}\BibitemShut {NoStop}%
\bibitem [{\citenamefont {Qin}\ \emph {et~al.}(2024{\natexlab{b}})\citenamefont {Qin}, \citenamefont {Zhang},\ and\ \citenamefont {Li}}]{qin2024geometry}%
  \BibitemOpen
  \bibfield  {author} {\bibinfo {author} {\bibfnamefont {Y.}~\bibnamefont {Qin}}, \bibinfo {author} {\bibfnamefont {K.}~\bibnamefont {Zhang}}, \ and\ \bibinfo {author} {\bibfnamefont {L.}~\bibnamefont {Li}},\ }\href@noop {} {\bibfield  {journal} {\bibinfo  {journal} {Physical Review A}\ }\textbf {\bibinfo {volume} {109}},\ \bibinfo {pages} {023317} (\bibinfo {year} {2024}{\natexlab{b}})}\BibitemShut {NoStop}%
\bibitem [{\citenamefont {Qin}\ \emph {et~al.}(2024{\natexlab{c}})\citenamefont {Qin}, \citenamefont {Lee},\ and\ \citenamefont {Li}}]{qin2024dynamical}%
  \BibitemOpen
  \bibfield  {author} {\bibinfo {author} {\bibfnamefont {Y.}~\bibnamefont {Qin}}, \bibinfo {author} {\bibfnamefont {C.~H.}\ \bibnamefont {Lee}}, \ and\ \bibinfo {author} {\bibfnamefont {L.}~\bibnamefont {Li}},\ }\href@noop {} {\bibfield  {journal} {\bibinfo  {journal} {arXiv preprint arXiv:2405.12288}\ } (\bibinfo {year} {2024}{\natexlab{c}})}\BibitemShut {NoStop}%
\bibitem [{\citenamefont {Xiong}\ \emph {et~al.}(2024)\citenamefont {Xiong}, \citenamefont {Xing},\ and\ \citenamefont {Hu}}]{xiong2024non}%
  \BibitemOpen
  \bibfield  {author} {\bibinfo {author} {\bibfnamefont {Y.}~\bibnamefont {Xiong}}, \bibinfo {author} {\bibfnamefont {Z.-Y.}\ \bibnamefont {Xing}}, \ and\ \bibinfo {author} {\bibfnamefont {H.}~\bibnamefont {Hu}},\ }\href@noop {} {\bibfield  {journal} {\bibinfo  {journal} {arXiv preprint arXiv:2407.01296}\ } (\bibinfo {year} {2024})}\BibitemShut {NoStop}%
\bibitem [{\citenamefont {Shang}\ \emph {et~al.}(2024)\citenamefont {Shang}, \citenamefont {Liu}, \citenamefont {Jiang}, \citenamefont {Shao}, \citenamefont {Zang}, \citenamefont {Lee}, \citenamefont {Thomale}, \citenamefont {Manchon}, \citenamefont {Cui},\ and\ \citenamefont {Schwingenschl{\"o}gl}}]{shang2024observation}%
  \BibitemOpen
  \bibfield  {author} {\bibinfo {author} {\bibfnamefont {C.}~\bibnamefont {Shang}}, \bibinfo {author} {\bibfnamefont {S.}~\bibnamefont {Liu}}, \bibinfo {author} {\bibfnamefont {C.}~\bibnamefont {Jiang}}, \bibinfo {author} {\bibfnamefont {R.}~\bibnamefont {Shao}}, \bibinfo {author} {\bibfnamefont {X.}~\bibnamefont {Zang}}, \bibinfo {author} {\bibfnamefont {C.~H.}\ \bibnamefont {Lee}}, \bibinfo {author} {\bibfnamefont {R.}~\bibnamefont {Thomale}}, \bibinfo {author} {\bibfnamefont {A.}~\bibnamefont {Manchon}}, \bibinfo {author} {\bibfnamefont {T.~J.}\ \bibnamefont {Cui}}, \ and\ \bibinfo {author} {\bibfnamefont {U.}~\bibnamefont {Schwingenschl{\"o}gl}},\ }\href@noop {} {\bibfield  {journal} {\bibinfo  {journal} {Advanced Science}\ }\textbf {\bibinfo {volume} {11}},\ \bibinfo {pages} {2303222} (\bibinfo {year} {2024})}\BibitemShut {NoStop}%
\bibitem [{\citenamefont {Hamanaka}\ and\ \citenamefont {Kawabata}(2024)}]{Hamanaka2024}%
  \BibitemOpen
  \bibfield  {author} {\bibinfo {author} {\bibfnamefont {S.}~\bibnamefont {Hamanaka}}\ and\ \bibinfo {author} {\bibfnamefont {K.}~\bibnamefont {Kawabata}},\ }\href {http://arxiv.org/abs/2401.08304} {\ ,\ \bibinfo {pages} {1} (\bibinfo {year} {2024})},\ \Eprint {http://arxiv.org/abs/2401.08304} {arXiv:2401.08304} \BibitemShut {NoStop}%
\bibitem [{\citenamefont {Hamanaka}\ \emph {et~al.}(2024)\citenamefont {Hamanaka}, \citenamefont {Iliasov}, \citenamefont {Neupert},\ and\ \citenamefont {Yoshida}}]{Hamanaka2024a}%
  \BibitemOpen
  \bibfield  {author} {\bibinfo {author} {\bibfnamefont {S.}~\bibnamefont {Hamanaka}}, \bibinfo {author} {\bibfnamefont {A.~A.}\ \bibnamefont {Iliasov}}, \bibinfo {author} {\bibfnamefont {T.}~\bibnamefont {Neupert}}, \ and\ \bibinfo {author} {\bibfnamefont {T.}~\bibnamefont {Yoshida}},\ }\href@noop {} {\ ,\ \bibinfo {pages} {1} (\bibinfo {year} {2024})},\ \Eprint {http://arxiv.org/abs/arXiv:2408.11024v1} {arXiv:arXiv:2408.11024v1} \BibitemShut {NoStop}%
\bibitem [{\citenamefont {Okugawa}\ \emph {et~al.}(2020)\citenamefont {Okugawa}, \citenamefont {Takahashi},\ and\ \citenamefont {Yokomizo}}]{okugawa2020second}%
  \BibitemOpen
  \bibfield  {author} {\bibinfo {author} {\bibfnamefont {R.}~\bibnamefont {Okugawa}}, \bibinfo {author} {\bibfnamefont {R.}~\bibnamefont {Takahashi}}, \ and\ \bibinfo {author} {\bibfnamefont {K.}~\bibnamefont {Yokomizo}},\ }\href@noop {} {\bibfield  {journal} {\bibinfo  {journal} {Physical Review B}\ }\textbf {\bibinfo {volume} {102}},\ \bibinfo {pages} {241202} (\bibinfo {year} {2020})}\BibitemShut {NoStop}%
\bibitem [{\citenamefont {Kawabata}\ \emph {et~al.}(2020)\citenamefont {Kawabata}, \citenamefont {Sato},\ and\ \citenamefont {Shiozaki}}]{kawabata2020higher}%
  \BibitemOpen
  \bibfield  {author} {\bibinfo {author} {\bibfnamefont {K.}~\bibnamefont {Kawabata}}, \bibinfo {author} {\bibfnamefont {M.}~\bibnamefont {Sato}}, \ and\ \bibinfo {author} {\bibfnamefont {K.}~\bibnamefont {Shiozaki}},\ }\href@noop {} {\bibfield  {journal} {\bibinfo  {journal} {Physical Review B}\ }\textbf {\bibinfo {volume} {102}},\ \bibinfo {pages} {205118} (\bibinfo {year} {2020})}\BibitemShut {NoStop}%
\bibitem [{\citenamefont {Zhang}\ \emph {et~al.}(2022{\natexlab{c}})\citenamefont {Zhang}, \citenamefont {Yang},\ and\ \citenamefont {Fang}}]{zhang2022universal}%
  \BibitemOpen
  \bibfield  {author} {\bibinfo {author} {\bibfnamefont {K.}~\bibnamefont {Zhang}}, \bibinfo {author} {\bibfnamefont {Z.}~\bibnamefont {Yang}}, \ and\ \bibinfo {author} {\bibfnamefont {C.}~\bibnamefont {Fang}},\ }\href@noop {} {\bibfield  {journal} {\bibinfo  {journal} {Nature communications}\ }\textbf {\bibinfo {volume} {13}},\ \bibinfo {pages} {2496} (\bibinfo {year} {2022}{\natexlab{c}})}\BibitemShut {NoStop}%
\bibitem [{\citenamefont {Zhang}\ and\ \citenamefont {Guo}(2024)}]{Zhang2024}%
  \BibitemOpen
  \bibfield  {author} {\bibinfo {author} {\bibfnamefont {X.}~\bibnamefont {Zhang}}\ and\ \bibinfo {author} {\bibfnamefont {H.}~\bibnamefont {Guo}},\ }\href@noop {} {\ \textbf {\bibinfo {volume} {1}} (\bibinfo {year} {2024})},\ \Eprint {http://arxiv.org/abs/arXiv:2408.07355v1} {arXiv:arXiv:2408.07355v1} \BibitemShut {NoStop}%
\bibitem [{\citenamefont {Fang}\ \emph {et~al.}(2022)\citenamefont {Fang}, \citenamefont {Hu}, \citenamefont {Zhou},\ and\ \citenamefont {Ding}}]{fang2022geometry}%
  \BibitemOpen
  \bibfield  {author} {\bibinfo {author} {\bibfnamefont {Z.}~\bibnamefont {Fang}}, \bibinfo {author} {\bibfnamefont {M.}~\bibnamefont {Hu}}, \bibinfo {author} {\bibfnamefont {L.}~\bibnamefont {Zhou}}, \ and\ \bibinfo {author} {\bibfnamefont {K.}~\bibnamefont {Ding}},\ }\href@noop {} {\bibfield  {journal} {\bibinfo  {journal} {Nanophotonics}\ }\textbf {\bibinfo {volume} {11}},\ \bibinfo {pages} {3447} (\bibinfo {year} {2022})}\BibitemShut {NoStop}%
\bibitem [{\citenamefont {Zhou}\ \emph {et~al.}(2023)\citenamefont {Zhou}, \citenamefont {Wu}, \citenamefont {Pu}, \citenamefont {Lu}, \citenamefont {Huang}, \citenamefont {Deng}, \citenamefont {Ke},\ and\ \citenamefont {Liu}}]{zhou2023observation}%
  \BibitemOpen
  \bibfield  {author} {\bibinfo {author} {\bibfnamefont {Q.}~\bibnamefont {Zhou}}, \bibinfo {author} {\bibfnamefont {J.}~\bibnamefont {Wu}}, \bibinfo {author} {\bibfnamefont {Z.}~\bibnamefont {Pu}}, \bibinfo {author} {\bibfnamefont {J.}~\bibnamefont {Lu}}, \bibinfo {author} {\bibfnamefont {X.}~\bibnamefont {Huang}}, \bibinfo {author} {\bibfnamefont {W.}~\bibnamefont {Deng}}, \bibinfo {author} {\bibfnamefont {M.}~\bibnamefont {Ke}}, \ and\ \bibinfo {author} {\bibfnamefont {Z.}~\bibnamefont {Liu}},\ }\href@noop {} {\bibfield  {journal} {\bibinfo  {journal} {Nature Communications}\ }\textbf {\bibinfo {volume} {14}},\ \bibinfo {pages} {4569} (\bibinfo {year} {2023})}\BibitemShut {NoStop}%
\bibitem [{\citenamefont {Wang}\ \emph {et~al.}(2023)\citenamefont {Wang}, \citenamefont {Hu}, \citenamefont {Wang}, \citenamefont {Ma},\ and\ \citenamefont {Ding}}]{wang2023experimental}%
  \BibitemOpen
  \bibfield  {author} {\bibinfo {author} {\bibfnamefont {W.}~\bibnamefont {Wang}}, \bibinfo {author} {\bibfnamefont {M.}~\bibnamefont {Hu}}, \bibinfo {author} {\bibfnamefont {X.}~\bibnamefont {Wang}}, \bibinfo {author} {\bibfnamefont {G.}~\bibnamefont {Ma}}, \ and\ \bibinfo {author} {\bibfnamefont {K.}~\bibnamefont {Ding}},\ }\href@noop {} {\bibfield  {journal} {\bibinfo  {journal} {arXiv preprint arXiv:2302.06314}\ } (\bibinfo {year} {2023})}\BibitemShut {NoStop}%
\bibitem [{\citenamefont {Song}\ \emph {et~al.}(2021)\citenamefont {Song}, \citenamefont {Monceaux}, \citenamefont {Bittner}, \citenamefont {Chao}, \citenamefont {de~la Cruz}, \citenamefont {Lafargue}, \citenamefont {Decanini}, \citenamefont {Dietz}, \citenamefont {Zyss}, \citenamefont {Grigis} \emph {et~al.}}]{song2021mobius}%
  \BibitemOpen
  \bibfield  {author} {\bibinfo {author} {\bibfnamefont {Y.}~\bibnamefont {Song}}, \bibinfo {author} {\bibfnamefont {Y.}~\bibnamefont {Monceaux}}, \bibinfo {author} {\bibfnamefont {S.}~\bibnamefont {Bittner}}, \bibinfo {author} {\bibfnamefont {K.}~\bibnamefont {Chao}}, \bibinfo {author} {\bibfnamefont {H.~M.~R.}\ \bibnamefont {de~la Cruz}}, \bibinfo {author} {\bibfnamefont {C.}~\bibnamefont {Lafargue}}, \bibinfo {author} {\bibfnamefont {D.}~\bibnamefont {Decanini}}, \bibinfo {author} {\bibfnamefont {B.}~\bibnamefont {Dietz}}, \bibinfo {author} {\bibfnamefont {J.}~\bibnamefont {Zyss}}, \bibinfo {author} {\bibfnamefont {A.}~\bibnamefont {Grigis}},  \emph {et~al.},\ }\href@noop {} {\bibfield  {journal} {\bibinfo  {journal} {Physical Review Letters}\ }\textbf {\bibinfo {volume} {127}},\ \bibinfo {pages} {203901} (\bibinfo {year} {2021})}\BibitemShut {NoStop}%
\bibitem [{\citenamefont {Maciejko}\ and\ \citenamefont {Rayan}(2021)}]{maciejko2021hyperbolic}%
  \BibitemOpen
  \bibfield  {author} {\bibinfo {author} {\bibfnamefont {J.}~\bibnamefont {Maciejko}}\ and\ \bibinfo {author} {\bibfnamefont {S.}~\bibnamefont {Rayan}},\ }\href@noop {} {\bibfield  {journal} {\bibinfo  {journal} {Science advances}\ }\textbf {\bibinfo {volume} {7}},\ \bibinfo {pages} {eabe9170} (\bibinfo {year} {2021})}\BibitemShut {NoStop}%
\bibitem [{\citenamefont {Stegmaier}\ \emph {et~al.}(2022)\citenamefont {Stegmaier}, \citenamefont {Upreti}, \citenamefont {Thomale},\ and\ \citenamefont {Boettcher}}]{stegmaier2022universality}%
  \BibitemOpen
  \bibfield  {author} {\bibinfo {author} {\bibfnamefont {A.}~\bibnamefont {Stegmaier}}, \bibinfo {author} {\bibfnamefont {L.~K.}\ \bibnamefont {Upreti}}, \bibinfo {author} {\bibfnamefont {R.}~\bibnamefont {Thomale}}, \ and\ \bibinfo {author} {\bibfnamefont {I.}~\bibnamefont {Boettcher}},\ }\href@noop {} {\bibfield  {journal} {\bibinfo  {journal} {Physical Review Letters}\ }\textbf {\bibinfo {volume} {128}},\ \bibinfo {pages} {166402} (\bibinfo {year} {2022})}\BibitemShut {NoStop}%
\bibitem [{\citenamefont {Liu}\ \emph {et~al.}(2022)\citenamefont {Liu}, \citenamefont {Hua}, \citenamefont {Peng},\ and\ \citenamefont {Zhou}}]{liu2022chern}%
  \BibitemOpen
  \bibfield  {author} {\bibinfo {author} {\bibfnamefont {Z.-R.}\ \bibnamefont {Liu}}, \bibinfo {author} {\bibfnamefont {C.-B.}\ \bibnamefont {Hua}}, \bibinfo {author} {\bibfnamefont {T.}~\bibnamefont {Peng}}, \ and\ \bibinfo {author} {\bibfnamefont {B.}~\bibnamefont {Zhou}},\ }\href@noop {} {\bibfield  {journal} {\bibinfo  {journal} {Physical Review B}\ }\textbf {\bibinfo {volume} {105}},\ \bibinfo {pages} {245301} (\bibinfo {year} {2022})}\BibitemShut {NoStop}%
\bibitem [{\citenamefont {Boettcher}\ \emph {et~al.}(2022)\citenamefont {Boettcher}, \citenamefont {Gorshkov}, \citenamefont {Koll{\'a}r}, \citenamefont {Maciejko}, \citenamefont {Rayan},\ and\ \citenamefont {Thomale}}]{boettcher2022crystallography}%
  \BibitemOpen
  \bibfield  {author} {\bibinfo {author} {\bibfnamefont {I.}~\bibnamefont {Boettcher}}, \bibinfo {author} {\bibfnamefont {A.~V.}\ \bibnamefont {Gorshkov}}, \bibinfo {author} {\bibfnamefont {A.~J.}\ \bibnamefont {Koll{\'a}r}}, \bibinfo {author} {\bibfnamefont {J.}~\bibnamefont {Maciejko}}, \bibinfo {author} {\bibfnamefont {S.}~\bibnamefont {Rayan}}, \ and\ \bibinfo {author} {\bibfnamefont {R.}~\bibnamefont {Thomale}},\ }\href@noop {} {\bibfield  {journal} {\bibinfo  {journal} {Physical Review B}\ }\textbf {\bibinfo {volume} {105}},\ \bibinfo {pages} {125118} (\bibinfo {year} {2022})}\BibitemShut {NoStop}%
\bibitem [{\citenamefont {Maciejko}\ and\ \citenamefont {Rayan}(2022)}]{maciejko2022automorphic}%
  \BibitemOpen
  \bibfield  {author} {\bibinfo {author} {\bibfnamefont {J.}~\bibnamefont {Maciejko}}\ and\ \bibinfo {author} {\bibfnamefont {S.}~\bibnamefont {Rayan}},\ }\href@noop {} {\bibfield  {journal} {\bibinfo  {journal} {Proceedings of the National Academy of Sciences}\ }\textbf {\bibinfo {volume} {119}},\ \bibinfo {pages} {e2116869119} (\bibinfo {year} {2022})}\BibitemShut {NoStop}%
\bibitem [{\citenamefont {Cheng}\ \emph {et~al.}(2022)\citenamefont {Cheng}, \citenamefont {Serafin}, \citenamefont {McInerney}, \citenamefont {Rocklin}, \citenamefont {Sun},\ and\ \citenamefont {Mao}}]{cheng2022band}%
  \BibitemOpen
  \bibfield  {author} {\bibinfo {author} {\bibfnamefont {N.}~\bibnamefont {Cheng}}, \bibinfo {author} {\bibfnamefont {F.}~\bibnamefont {Serafin}}, \bibinfo {author} {\bibfnamefont {J.}~\bibnamefont {McInerney}}, \bibinfo {author} {\bibfnamefont {Z.}~\bibnamefont {Rocklin}}, \bibinfo {author} {\bibfnamefont {K.}~\bibnamefont {Sun}}, \ and\ \bibinfo {author} {\bibfnamefont {X.}~\bibnamefont {Mao}},\ }\href@noop {} {\bibfield  {journal} {\bibinfo  {journal} {arXiv preprint arXiv:2203.15208}\ } (\bibinfo {year} {2022})}\BibitemShut {NoStop}%
\bibitem [{\citenamefont {Bienias}\ \emph {et~al.}(2022)\citenamefont {Bienias}, \citenamefont {Boettcher}, \citenamefont {Belyansky}, \citenamefont {Koll{\'a}r},\ and\ \citenamefont {Gorshkov}}]{bienias2022circuit}%
  \BibitemOpen
  \bibfield  {author} {\bibinfo {author} {\bibfnamefont {P.}~\bibnamefont {Bienias}}, \bibinfo {author} {\bibfnamefont {I.}~\bibnamefont {Boettcher}}, \bibinfo {author} {\bibfnamefont {R.}~\bibnamefont {Belyansky}}, \bibinfo {author} {\bibfnamefont {A.~J.}\ \bibnamefont {Koll{\'a}r}}, \ and\ \bibinfo {author} {\bibfnamefont {A.~V.}\ \bibnamefont {Gorshkov}},\ }\href@noop {} {\bibfield  {journal} {\bibinfo  {journal} {Physical Review Letters}\ }\textbf {\bibinfo {volume} {128}},\ \bibinfo {pages} {013601} (\bibinfo {year} {2022})}\BibitemShut {NoStop}%
\bibitem [{\citenamefont {Zhang}\ \emph {et~al.}(2022{\natexlab{d}})\citenamefont {Zhang}, \citenamefont {Yuan}, \citenamefont {Sun}, \citenamefont {Sun},\ and\ \citenamefont {Zhang}}]{zhang2022observation}%
  \BibitemOpen
  \bibfield  {author} {\bibinfo {author} {\bibfnamefont {W.}~\bibnamefont {Zhang}}, \bibinfo {author} {\bibfnamefont {H.}~\bibnamefont {Yuan}}, \bibinfo {author} {\bibfnamefont {N.}~\bibnamefont {Sun}}, \bibinfo {author} {\bibfnamefont {H.}~\bibnamefont {Sun}}, \ and\ \bibinfo {author} {\bibfnamefont {X.}~\bibnamefont {Zhang}},\ }\href@noop {} {\bibfield  {journal} {\bibinfo  {journal} {Nature communications}\ }\textbf {\bibinfo {volume} {13}},\ \bibinfo {pages} {1} (\bibinfo {year} {2022}{\natexlab{d}})}\BibitemShut {NoStop}%
\bibitem [{\citenamefont {Lenggenhager}\ \emph {et~al.}(2022)\citenamefont {Lenggenhager}, \citenamefont {Stegmaier}, \citenamefont {Upreti}, \citenamefont {Hofmann}, \citenamefont {Helbig}, \citenamefont {Vollhardt}, \citenamefont {Greiter}, \citenamefont {Lee}, \citenamefont {Imhof}, \citenamefont {Brand} \emph {et~al.}}]{lenggenhager2022simulating}%
  \BibitemOpen
  \bibfield  {author} {\bibinfo {author} {\bibfnamefont {P.~M.}\ \bibnamefont {Lenggenhager}}, \bibinfo {author} {\bibfnamefont {A.}~\bibnamefont {Stegmaier}}, \bibinfo {author} {\bibfnamefont {L.~K.}\ \bibnamefont {Upreti}}, \bibinfo {author} {\bibfnamefont {T.}~\bibnamefont {Hofmann}}, \bibinfo {author} {\bibfnamefont {T.}~\bibnamefont {Helbig}}, \bibinfo {author} {\bibfnamefont {A.}~\bibnamefont {Vollhardt}}, \bibinfo {author} {\bibfnamefont {M.}~\bibnamefont {Greiter}}, \bibinfo {author} {\bibfnamefont {C.~H.}\ \bibnamefont {Lee}}, \bibinfo {author} {\bibfnamefont {S.}~\bibnamefont {Imhof}}, \bibinfo {author} {\bibfnamefont {H.}~\bibnamefont {Brand}},  \emph {et~al.},\ }\href@noop {} {\bibfield  {journal} {\bibinfo  {journal} {Nature communications}\ }\textbf {\bibinfo {volume} {13}},\ \bibinfo {pages} {1} (\bibinfo {year} {2022})}\BibitemShut {NoStop}%
\bibitem [{\citenamefont {Chen}\ \emph {et~al.}(2022)\citenamefont {Chen}, \citenamefont {Brand}, \citenamefont {Helbig}, \citenamefont {Hofmann}, \citenamefont {Imhof}, \citenamefont {Fritzsche}, \citenamefont {Kie{\ss}ling}, \citenamefont {Stegmaier}, \citenamefont {Upreti}, \citenamefont {Neupert} \emph {et~al.}}]{chen2022hyperbolic}%
  \BibitemOpen
  \bibfield  {author} {\bibinfo {author} {\bibfnamefont {A.}~\bibnamefont {Chen}}, \bibinfo {author} {\bibfnamefont {H.}~\bibnamefont {Brand}}, \bibinfo {author} {\bibfnamefont {T.}~\bibnamefont {Helbig}}, \bibinfo {author} {\bibfnamefont {T.}~\bibnamefont {Hofmann}}, \bibinfo {author} {\bibfnamefont {S.}~\bibnamefont {Imhof}}, \bibinfo {author} {\bibfnamefont {A.}~\bibnamefont {Fritzsche}}, \bibinfo {author} {\bibfnamefont {T.}~\bibnamefont {Kie{\ss}ling}}, \bibinfo {author} {\bibfnamefont {A.}~\bibnamefont {Stegmaier}}, \bibinfo {author} {\bibfnamefont {L.~K.}\ \bibnamefont {Upreti}}, \bibinfo {author} {\bibfnamefont {T.}~\bibnamefont {Neupert}},  \emph {et~al.},\ }\href@noop {} {\bibfield  {journal} {\bibinfo  {journal} {arXiv preprint arXiv:2205.05106}\ } (\bibinfo {year} {2022})}\BibitemShut {NoStop}%
\bibitem [{\citenamefont {Sun}\ \emph {et~al.}(2023{\natexlab{a}})\citenamefont {Sun}, \citenamefont {Li}, \citenamefont {Feng},\ and\ \citenamefont {Guo}}]{Sun2023}%
  \BibitemOpen
  \bibfield  {author} {\bibinfo {author} {\bibfnamefont {J.}~\bibnamefont {Sun}}, \bibinfo {author} {\bibfnamefont {C.-A.}\ \bibnamefont {Li}}, \bibinfo {author} {\bibfnamefont {S.}~\bibnamefont {Feng}}, \ and\ \bibinfo {author} {\bibfnamefont {H.}~\bibnamefont {Guo}},\ }\href {\doibase doi/10.1103/PhysRevB.108.075122} {\bibfield  {journal} {\bibinfo  {journal} {Physical Review B}\ }\textbf {\bibinfo {volume} {108}},\ \bibinfo {pages} {075122} (\bibinfo {year} {2023}{\natexlab{a}})}\BibitemShut {NoStop}%
\bibitem [{\citenamefont {Yu}\ \emph {et~al.}(2020)\citenamefont {Yu}, \citenamefont {Piao},\ and\ \citenamefont {Park}}]{yu2020topological}%
  \BibitemOpen
  \bibfield  {author} {\bibinfo {author} {\bibfnamefont {S.}~\bibnamefont {Yu}}, \bibinfo {author} {\bibfnamefont {X.}~\bibnamefont {Piao}}, \ and\ \bibinfo {author} {\bibfnamefont {N.}~\bibnamefont {Park}},\ }\href@noop {} {\bibfield  {journal} {\bibinfo  {journal} {Physical Review Letters}\ }\textbf {\bibinfo {volume} {125}},\ \bibinfo {pages} {053901} (\bibinfo {year} {2020})}\BibitemShut {NoStop}%
\bibitem [{\citenamefont {Lux}\ and\ \citenamefont {Prodan}(2023{\natexlab{a}})}]{Lux2023}%
  \BibitemOpen
  \bibfield  {author} {\bibinfo {author} {\bibfnamefont {F.~R.}\ \bibnamefont {Lux}}\ and\ \bibinfo {author} {\bibfnamefont {E.}~\bibnamefont {Prodan}},\ }\href {\doibase 10.1103/PhysRevLett.131.176603} {\bibfield  {journal} {\bibinfo  {journal} {Physical Review Letters}\ }\textbf {\bibinfo {volume} {131}},\ \bibinfo {pages} {176603} (\bibinfo {year} {2023}{\natexlab{a}})}\BibitemShut {NoStop}%
\bibitem [{\citenamefont {Urwyler}\ \emph {et~al.}(2022)\citenamefont {Urwyler}, \citenamefont {Lenggenhager}, \citenamefont {Boettcher}, \citenamefont {Thomale}, \citenamefont {Neupert},\ and\ \citenamefont {Bzdu{\v{s}}ek}}]{Urwyler2022}%
  \BibitemOpen
  \bibfield  {author} {\bibinfo {author} {\bibfnamefont {D.~M.}\ \bibnamefont {Urwyler}}, \bibinfo {author} {\bibfnamefont {P.~M.}\ \bibnamefont {Lenggenhager}}, \bibinfo {author} {\bibfnamefont {I.}~\bibnamefont {Boettcher}}, \bibinfo {author} {\bibfnamefont {R.}~\bibnamefont {Thomale}}, \bibinfo {author} {\bibfnamefont {T.}~\bibnamefont {Neupert}}, \ and\ \bibinfo {author} {\bibfnamefont {T.}~\bibnamefont {Bzdu{\v{s}}ek}},\ }\href {\doibase 10.1103/PhysRevLett.129.246402} {\bibfield  {journal} {\bibinfo  {journal} {Phys. Rev. Lett.}\ }\textbf {\bibinfo {volume} {129}},\ \bibinfo {pages} {1} (\bibinfo {year} {2022})}\BibitemShut {NoStop}%
\bibitem [{\citenamefont {Lenggenhager}\ \emph {et~al.}(2024)\citenamefont {Lenggenhager}, \citenamefont {Dey}, \citenamefont {Bzdu{\v{s}}ek},\ and\ \citenamefont {Maciejko}}]{Lenggenhager2024}%
  \BibitemOpen
  \bibfield  {author} {\bibinfo {author} {\bibfnamefont {P.~M.}\ \bibnamefont {Lenggenhager}}, \bibinfo {author} {\bibfnamefont {S.}~\bibnamefont {Dey}}, \bibinfo {author} {\bibfnamefont {T.}~\bibnamefont {Bzdu{\v{s}}ek}}, \ and\ \bibinfo {author} {\bibfnamefont {J.}~\bibnamefont {Maciejko}},\ }\href {http://arxiv.org/abs/2407.09601} {\bibfield  {journal} {\bibinfo  {journal} {arXiv preprint arXiv:2407.09601}\ } (\bibinfo {year} {2024})}\BibitemShut {NoStop}%
\bibitem [{\citenamefont {Tummuru}\ \emph {et~al.}(2024)\citenamefont {Tummuru}, \citenamefont {Chen}, \citenamefont {Lenggenhager}, \citenamefont {Neupert}, \citenamefont {Maciejko},\ and\ \citenamefont {Bzdu{\v{s}}ek}}]{Tummuru2023}%
  \BibitemOpen
  \bibfield  {author} {\bibinfo {author} {\bibfnamefont {T.}~\bibnamefont {Tummuru}}, \bibinfo {author} {\bibfnamefont {A.}~\bibnamefont {Chen}}, \bibinfo {author} {\bibfnamefont {P.~M.}\ \bibnamefont {Lenggenhager}}, \bibinfo {author} {\bibfnamefont {T.}~\bibnamefont {Neupert}}, \bibinfo {author} {\bibfnamefont {J.}~\bibnamefont {Maciejko}}, \ and\ \bibinfo {author} {\bibfnamefont {T.}~\bibnamefont {Bzdu{\v{s}}ek}},\ }\href {\doibase 10.1103/PhysRevLett.132.206601} {\bibfield  {journal} {\bibinfo  {journal} {Phys. Rev. Lett.}\ }\textbf {\bibinfo {volume} {132}},\ \bibinfo {pages} {206601} (\bibinfo {year} {2024})}\BibitemShut {NoStop}%
\bibitem [{\citenamefont {Koll{\'{a}}r}\ \emph {et~al.}(2019)\citenamefont {Koll{\'{a}}r}, \citenamefont {Fitzpatrick},\ and\ \citenamefont {Houck}}]{Kollar2019}%
  \BibitemOpen
  \bibfield  {author} {\bibinfo {author} {\bibfnamefont {A.~J.}\ \bibnamefont {Koll{\'{a}}r}}, \bibinfo {author} {\bibfnamefont {M.}~\bibnamefont {Fitzpatrick}}, \ and\ \bibinfo {author} {\bibfnamefont {A.~A.}\ \bibnamefont {Houck}},\ }\href {\doibase 10.1038/s41586-019-1348-3} {\bibfield  {journal} {\bibinfo  {journal} {Nature}\ }\textbf {\bibinfo {volume} {571}},\ \bibinfo {pages} {45} (\bibinfo {year} {2019})}\BibitemShut {NoStop}%
\bibitem [{\citenamefont {Zhang}\ \emph {et~al.}(2022{\natexlab{e}})\citenamefont {Zhang}, \citenamefont {Yuan}, \citenamefont {Sun}, \citenamefont {Sun},\ and\ \citenamefont {Zhang}}]{Zhang2022}%
  \BibitemOpen
  \bibfield  {author} {\bibinfo {author} {\bibfnamefont {W.}~\bibnamefont {Zhang}}, \bibinfo {author} {\bibfnamefont {H.}~\bibnamefont {Yuan}}, \bibinfo {author} {\bibfnamefont {N.}~\bibnamefont {Sun}}, \bibinfo {author} {\bibfnamefont {H.}~\bibnamefont {Sun}}, \ and\ \bibinfo {author} {\bibfnamefont {X.}~\bibnamefont {Zhang}},\ }\href {\doibase 10.1038/s41467-022-30631-x} {\bibfield  {journal} {\bibinfo  {journal} {Nature Communications}\ }\textbf {\bibinfo {volume} {13}},\ \bibinfo {pages} {2937} (\bibinfo {year} {2022}{\natexlab{e}})}\BibitemShut {NoStop}%
\bibitem [{\citenamefont {Yuan}\ \emph {et~al.}(2024)\citenamefont {Yuan}, \citenamefont {Zhang}, \citenamefont {Pei},\ and\ \citenamefont {Zhang}}]{Yuan2024}%
  \BibitemOpen
  \bibfield  {author} {\bibinfo {author} {\bibfnamefont {H.}~\bibnamefont {Yuan}}, \bibinfo {author} {\bibfnamefont {W.}~\bibnamefont {Zhang}}, \bibinfo {author} {\bibfnamefont {Q.}~\bibnamefont {Pei}}, \ and\ \bibinfo {author} {\bibfnamefont {X.}~\bibnamefont {Zhang}},\ }\href {\doibase 10.1103/PhysRevB.109.L041109} {\bibfield  {journal} {\bibinfo  {journal} {Physical Review B}\ }\textbf {\bibinfo {volume} {109}},\ \bibinfo {pages} {L041109} (\bibinfo {year} {2024})}\BibitemShut {NoStop}%
\bibitem [{\citenamefont {Pei}\ \emph {et~al.}(2023)\citenamefont {Pei}, \citenamefont {Yuan}, \citenamefont {Zhang},\ and\ \citenamefont {Zhang}}]{Pei2023}%
  \BibitemOpen
  \bibfield  {author} {\bibinfo {author} {\bibfnamefont {Q.}~\bibnamefont {Pei}}, \bibinfo {author} {\bibfnamefont {H.}~\bibnamefont {Yuan}}, \bibinfo {author} {\bibfnamefont {W.}~\bibnamefont {Zhang}}, \ and\ \bibinfo {author} {\bibfnamefont {X.}~\bibnamefont {Zhang}},\ }\href {\doibase 10.1103/PhysRevB.107.165145} {\bibfield  {journal} {\bibinfo  {journal} {Physical Review B}\ }\textbf {\bibinfo {volume} {107}},\ \bibinfo {pages} {165145} (\bibinfo {year} {2023})}\BibitemShut {NoStop}%
\bibitem [{\citenamefont {Zhang}\ \emph {et~al.}(2023)\citenamefont {Zhang}, \citenamefont {Di}, \citenamefont {Zheng}, \citenamefont {Sun},\ and\ \citenamefont {Zhang}}]{Zhang2023}%
  \BibitemOpen
  \bibfield  {author} {\bibinfo {author} {\bibfnamefont {W.}~\bibnamefont {Zhang}}, \bibinfo {author} {\bibfnamefont {F.}~\bibnamefont {Di}}, \bibinfo {author} {\bibfnamefont {X.}~\bibnamefont {Zheng}}, \bibinfo {author} {\bibfnamefont {H.}~\bibnamefont {Sun}}, \ and\ \bibinfo {author} {\bibfnamefont {X.}~\bibnamefont {Zhang}},\ }\href {\doibase 10.1038/s41467-023-36767-8} {\bibfield  {journal} {\bibinfo  {journal} {Nature Communications}\ }\textbf {\bibinfo {volume} {14}},\ \bibinfo {pages} {1083} (\bibinfo {year} {2023})}\BibitemShut {NoStop}%
\bibitem [{\citenamefont {Huang}\ \emph {et~al.}(2024)\citenamefont {Huang}, \citenamefont {He}, \citenamefont {Zhang}, \citenamefont {Zhang}, \citenamefont {Liu}, \citenamefont {Feng}, \citenamefont {Liu}, \citenamefont {Cui}, \citenamefont {Huang}, \citenamefont {Zhang},\ and\ \citenamefont {Zhang}}]{Huang2024}%
  \BibitemOpen
  \bibfield  {author} {\bibinfo {author} {\bibfnamefont {L.}~\bibnamefont {Huang}}, \bibinfo {author} {\bibfnamefont {L.}~\bibnamefont {He}}, \bibinfo {author} {\bibfnamefont {W.}~\bibnamefont {Zhang}}, \bibinfo {author} {\bibfnamefont {H.}~\bibnamefont {Zhang}}, \bibinfo {author} {\bibfnamefont {D.}~\bibnamefont {Liu}}, \bibinfo {author} {\bibfnamefont {X.}~\bibnamefont {Feng}}, \bibinfo {author} {\bibfnamefont {F.}~\bibnamefont {Liu}}, \bibinfo {author} {\bibfnamefont {K.}~\bibnamefont {Cui}}, \bibinfo {author} {\bibfnamefont {Y.}~\bibnamefont {Huang}}, \bibinfo {author} {\bibfnamefont {W.}~\bibnamefont {Zhang}}, \ and\ \bibinfo {author} {\bibfnamefont {X.}~\bibnamefont {Zhang}},\ }\href {\doibase 10.1038/s41467-024-46035-y} {\bibfield  {journal} {\bibinfo  {journal} {Nature Communications}\ }\textbf {\bibinfo {volume} {15}},\ \bibinfo {pages} {1647} (\bibinfo {year} {2024})}\BibitemShut {NoStop}%
\bibitem [{\citenamefont {Lv}\ \emph {et~al.}(2022)\citenamefont {Lv}, \citenamefont {Zhang}, \citenamefont {Zhai},\ and\ \citenamefont {Zhou}}]{Lv2022}%
  \BibitemOpen
  \bibfield  {author} {\bibinfo {author} {\bibfnamefont {C.}~\bibnamefont {Lv}}, \bibinfo {author} {\bibfnamefont {R.}~\bibnamefont {Zhang}}, \bibinfo {author} {\bibfnamefont {Z.}~\bibnamefont {Zhai}}, \ and\ \bibinfo {author} {\bibfnamefont {Q.}~\bibnamefont {Zhou}},\ }\href {\doibase 10.1038/s41467-022-29774-8} {\bibfield  {journal} {\bibinfo  {journal} {Nat. Commun.}\ }\textbf {\bibinfo {volume} {13}},\ \bibinfo {pages} {2184} (\bibinfo {year} {2022})}\BibitemShut {NoStop}%
\bibitem [{\citenamefont {Chadha}\ and\ \citenamefont {Narayan}(2024)}]{Chadha2024}%
  \BibitemOpen
  \bibfield  {author} {\bibinfo {author} {\bibfnamefont {N.}~\bibnamefont {Chadha}}\ and\ \bibinfo {author} {\bibfnamefont {A.}~\bibnamefont {Narayan}},\ }\href {\doibase 10.1088/1751-8121/ad2cb1} {\bibfield  {journal} {\bibinfo  {journal} {J. Phys. A Math. Theor.}\ }\textbf {\bibinfo {volume} {57}},\ \bibinfo {pages} {115203} (\bibinfo {year} {2024})}\BibitemShut {NoStop}%
\bibitem [{\citenamefont {Lux}\ and\ \citenamefont {Prodan}(2023{\natexlab{b}})}]{lux2023converging}%
  \BibitemOpen
  \bibfield  {author} {\bibinfo {author} {\bibfnamefont {F.~R.}\ \bibnamefont {Lux}}\ and\ \bibinfo {author} {\bibfnamefont {E.}~\bibnamefont {Prodan}},\ }\href@noop {} {\bibfield  {journal} {\bibinfo  {journal} {Physical Review Letters}\ }\textbf {\bibinfo {volume} {131}},\ \bibinfo {pages} {176603} (\bibinfo {year} {2023}{\natexlab{b}})}\BibitemShut {NoStop}%
\bibitem [{\citenamefont {Sun}\ \emph {et~al.}(2023{\natexlab{b}})\citenamefont {Sun}, \citenamefont {Li}, \citenamefont {Feng},\ and\ \citenamefont {Guo}}]{sun2023hybrid}%
  \BibitemOpen
  \bibfield  {author} {\bibinfo {author} {\bibfnamefont {J.}~\bibnamefont {Sun}}, \bibinfo {author} {\bibfnamefont {C.-A.}\ \bibnamefont {Li}}, \bibinfo {author} {\bibfnamefont {S.}~\bibnamefont {Feng}}, \ and\ \bibinfo {author} {\bibfnamefont {H.}~\bibnamefont {Guo}},\ }\href@noop {} {\bibfield  {journal} {\bibinfo  {journal} {Physical Review B}\ }\textbf {\bibinfo {volume} {108}},\ \bibinfo {pages} {075122} (\bibinfo {year} {2023}{\natexlab{b}})}\BibitemShut {NoStop}%
\bibitem [{\citenamefont {Lv}\ and\ \citenamefont {Zhou}(2024)}]{lv2024hidden}%
  \BibitemOpen
  \bibfield  {author} {\bibinfo {author} {\bibfnamefont {C.}~\bibnamefont {Lv}}\ and\ \bibinfo {author} {\bibfnamefont {Q.}~\bibnamefont {Zhou}},\ }\href@noop {} {\bibfield  {journal} {\bibinfo  {journal} {arXiv preprint arXiv:2408.05132}\ } (\bibinfo {year} {2024})}\BibitemShut {NoStop}%
\bibitem [{\citenamefont {Cannon}\ \emph {et~al.}(1997)\citenamefont {Cannon}, \citenamefont {Floyd}, \citenamefont {Kenyon}, \citenamefont {Parry} \emph {et~al.}}]{cannon1997hyperbolic}%
  \BibitemOpen
  \bibfield  {author} {\bibinfo {author} {\bibfnamefont {J.~W.}\ \bibnamefont {Cannon}}, \bibinfo {author} {\bibfnamefont {W.~J.}\ \bibnamefont {Floyd}}, \bibinfo {author} {\bibfnamefont {R.}~\bibnamefont {Kenyon}}, \bibinfo {author} {\bibfnamefont {W.~R.}\ \bibnamefont {Parry}},  \emph {et~al.},\ }\href@noop {} {\bibfield  {journal} {\bibinfo  {journal} {Flavors of geometry}\ }\textbf {\bibinfo {volume} {31}},\ \bibinfo {pages} {2} (\bibinfo {year} {1997})}\BibitemShut {NoStop}%
\bibitem [{\citenamefont {Milnor}(1982)}]{milnor1982hyperbolic}%
  \BibitemOpen
  \bibfield  {author} {\bibinfo {author} {\bibfnamefont {J.~W.}\ \bibnamefont {Milnor}},\ }\href@noop {} {\bibfield  {journal} {\bibinfo  {journal} {Bulletin of the American Mathematical Society}\ }\textbf {\bibinfo {volume} {6}},\ \bibinfo {pages} {9} (\bibinfo {year} {1982})}\BibitemShut {NoStop}%
\bibitem [{\citenamefont {Benedetti}\ and\ \citenamefont {Petronio}(1992)}]{benedetti1992lectures}%
  \BibitemOpen
  \bibfield  {author} {\bibinfo {author} {\bibfnamefont {R.}~\bibnamefont {Benedetti}}\ and\ \bibinfo {author} {\bibfnamefont {C.}~\bibnamefont {Petronio}},\ }\href@noop {} {\emph {\bibinfo {title} {Lectures on hyperbolic geometry}}}\ (\bibinfo  {publisher} {Springer Science \& Business Media},\ \bibinfo {year} {1992})\BibitemShut {NoStop}%
\bibitem [{\citenamefont {Ramsay}\ and\ \citenamefont {Richtmyer}(1995)}]{ramsay1995introduction}%
  \BibitemOpen
  \bibfield  {author} {\bibinfo {author} {\bibfnamefont {A.}~\bibnamefont {Ramsay}}\ and\ \bibinfo {author} {\bibfnamefont {R.~D.}\ \bibnamefont {Richtmyer}},\ }\href@noop {} {\emph {\bibinfo {title} {Introduction to hyperbolic geometry}}}\ (\bibinfo  {publisher} {Springer Science \& Business Media},\ \bibinfo {year} {1995})\BibitemShut {NoStop}%
\bibitem [{\citenamefont {Brower}\ \emph {et~al.}(2022)\citenamefont {Brower}, \citenamefont {Cogburn},\ and\ \citenamefont {Owen}}]{brower2022hyperbolic}%
  \BibitemOpen
  \bibfield  {author} {\bibinfo {author} {\bibfnamefont {R.~C.}\ \bibnamefont {Brower}}, \bibinfo {author} {\bibfnamefont {C.~V.}\ \bibnamefont {Cogburn}}, \ and\ \bibinfo {author} {\bibfnamefont {E.}~\bibnamefont {Owen}},\ }\href@noop {} {\bibfield  {journal} {\bibinfo  {journal} {Physical Review D}\ }\textbf {\bibinfo {volume} {105}},\ \bibinfo {pages} {114503} (\bibinfo {year} {2022})}\BibitemShut {NoStop}%
\bibitem [{\citenamefont {Zhu}\ \emph {et~al.}(2021)\citenamefont {Zhu}, \citenamefont {Guo}, \citenamefont {Breuckmann}, \citenamefont {Guo},\ and\ \citenamefont {Feng}}]{zhu2021quantum}%
  \BibitemOpen
  \bibfield  {author} {\bibinfo {author} {\bibfnamefont {X.}~\bibnamefont {Zhu}}, \bibinfo {author} {\bibfnamefont {J.}~\bibnamefont {Guo}}, \bibinfo {author} {\bibfnamefont {N.~P.}\ \bibnamefont {Breuckmann}}, \bibinfo {author} {\bibfnamefont {H.}~\bibnamefont {Guo}}, \ and\ \bibinfo {author} {\bibfnamefont {S.}~\bibnamefont {Feng}},\ }\href@noop {} {\bibfield  {journal} {\bibinfo  {journal} {Journal of Physics: Condensed Matter}\ }\textbf {\bibinfo {volume} {33}},\ \bibinfo {pages} {335602} (\bibinfo {year} {2021})}\BibitemShut {NoStop}%
\bibitem [{\citenamefont {Cassels}\ \emph {et~al.}(2001)\citenamefont {Cassels}, \citenamefont {Hitchin} \emph {et~al.}}]{cassels2001topics}%
  \BibitemOpen
  \bibfield  {author} {\bibinfo {author} {\bibfnamefont {J.}~\bibnamefont {Cassels}}, \bibinfo {author} {\bibfnamefont {N.}~\bibnamefont {Hitchin}},  \emph {et~al.},\ }\href@noop {} {\emph {\bibinfo {title} {Topics on Riemann surfaces and Fuchsian groups}}},\ Vol.\ \bibinfo {volume} {287}\ (\bibinfo  {publisher} {Cambridge University Press},\ \bibinfo {year} {2001})\BibitemShut {NoStop}%
\bibitem [{\citenamefont {Coxeter}\ and\ \citenamefont {Moser}(2013)}]{coxeter2013generators}%
  \BibitemOpen
  \bibfield  {author} {\bibinfo {author} {\bibfnamefont {H.~S.}\ \bibnamefont {Coxeter}}\ and\ \bibinfo {author} {\bibfnamefont {W.~O.}\ \bibnamefont {Moser}},\ }\href@noop {} {\emph {\bibinfo {title} {Generators and relations for discrete groups}}},\ Vol.~\bibinfo {volume} {14}\ (\bibinfo  {publisher} {Springer Science \& Business Media},\ \bibinfo {year} {2013})\BibitemShut {NoStop}%
\bibitem [{\citenamefont {Magnus}(1974)}]{magnus1974noneuclidean}%
  \BibitemOpen
  \bibfield  {author} {\bibinfo {author} {\bibfnamefont {W.}~\bibnamefont {Magnus}},\ }\href@noop {} {\emph {\bibinfo {title} {Noneuclidean tesselations and their groups}}}\ (\bibinfo  {publisher} {Academic Press},\ \bibinfo {year} {1974})\BibitemShut {NoStop}%
\bibitem [{\citenamefont {Balazs}\ and\ \citenamefont {Voros}(1986)}]{balazs1986chaos}%
  \BibitemOpen
  \bibfield  {author} {\bibinfo {author} {\bibfnamefont {N.}~\bibnamefont {Balazs}}\ and\ \bibinfo {author} {\bibfnamefont {A.}~\bibnamefont {Voros}},\ }\href@noop {} {\bibfield  {journal} {\bibinfo  {journal} {Physics reports}\ }\textbf {\bibinfo {volume} {143}},\ \bibinfo {pages} {109} (\bibinfo {year} {1986})}\BibitemShut {NoStop}%
\bibitem [{\citenamefont {Lenggenhager}\ \emph {et~al.}(2023)\citenamefont {Lenggenhager}, \citenamefont {Maciejko},\ and\ \citenamefont {Bzdu{\v{s}}ek}}]{Lenggenhager2023}%
  \BibitemOpen
  \bibfield  {author} {\bibinfo {author} {\bibfnamefont {P.~M.}\ \bibnamefont {Lenggenhager}}, \bibinfo {author} {\bibfnamefont {J.}~\bibnamefont {Maciejko}}, \ and\ \bibinfo {author} {\bibfnamefont {T.}~\bibnamefont {Bzdu{\v{s}}ek}},\ }\href {\doibase 10.1103/PhysRevLett.131.226401} {\bibfield  {journal} {\bibinfo  {journal} {Physical Review Letters}\ }\textbf {\bibinfo {volume} {131}},\ \bibinfo {pages} {226401} (\bibinfo {year} {2023})}\BibitemShut {NoStop}%
\bibitem [{\citenamefont {Mosseri}\ and\ \citenamefont {Vidal}(2023)}]{Mosseri2023}%
  \BibitemOpen
  \bibfield  {author} {\bibinfo {author} {\bibfnamefont {R.}~\bibnamefont {Mosseri}}\ and\ \bibinfo {author} {\bibfnamefont {J.}~\bibnamefont {Vidal}},\ }\href {\doibase 10.1103/PhysRevB.108.035154} {\bibfield  {journal} {\bibinfo  {journal} {Physical Review B}\ }\textbf {\bibinfo {volume} {108}},\ \bibinfo {pages} {035154} (\bibinfo {year} {2023})}\BibitemShut {NoStop}%
\bibitem [{\citenamefont {Lux}\ and\ \citenamefont {Prodan}(2023{\natexlab{c}})}]{Lux2023a}%
  \BibitemOpen
  \bibfield  {author} {\bibinfo {author} {\bibfnamefont {F.~R.}\ \bibnamefont {Lux}}\ and\ \bibinfo {author} {\bibfnamefont {E.}~\bibnamefont {Prodan}},\ }\href {\doibase 10.1007/s00023-023-01373-3} {\bibfield  {journal} {\bibinfo  {journal} {Annales Henri Poincar{\'{e}}}\ } (\bibinfo {year} {2023}{\natexlab{c}}),\ 10.1007/s00023-023-01373-3}\BibitemShut {NoStop}%
\bibitem [{\citenamefont {Lee}\ \emph {et~al.}(2019)\citenamefont {Lee}, \citenamefont {Li},\ and\ \citenamefont {Gong}}]{Lee2019b}%
  \BibitemOpen
  \bibfield  {author} {\bibinfo {author} {\bibfnamefont {C.~H.}\ \bibnamefont {Lee}}, \bibinfo {author} {\bibfnamefont {L.}~\bibnamefont {Li}}, \ and\ \bibinfo {author} {\bibfnamefont {J.}~\bibnamefont {Gong}},\ }\href {\doibase 10.1103/PhysRevLett.123.016805} {\bibfield  {journal} {\bibinfo  {journal} {Phys. Rev. Lett.}\ }\textbf {\bibinfo {volume} {123}},\ \bibinfo {pages} {016805} (\bibinfo {year} {2019})}\BibitemShut {NoStop}%
\bibitem [{\citenamefont {Zhu}\ and\ \citenamefont {Li}(2024)}]{Zhu2023}%
  \BibitemOpen
  \bibfield  {author} {\bibinfo {author} {\bibfnamefont {W.}~\bibnamefont {Zhu}}\ and\ \bibinfo {author} {\bibfnamefont {L.}~\bibnamefont {Li}},\ }\href {\doibase 10.1088/1361-648X/ad3593} {\bibfield  {journal} {\bibinfo  {journal} {J. Phys. Condens. Matter}\ }\textbf {\bibinfo {volume} {36}},\ \bibinfo {pages} {253003} (\bibinfo {year} {2024})}\BibitemShut {NoStop}%
\bibitem [{\citenamefont {Hatano}\ and\ \citenamefont {Nelson}(1996)}]{Hatano1996}%
  \BibitemOpen
  \bibfield  {author} {\bibinfo {author} {\bibfnamefont {N.}~\bibnamefont {Hatano}}\ and\ \bibinfo {author} {\bibfnamefont {D.~R.}\ \bibnamefont {Nelson}},\ }\href {\doibase 10.1103/PhysRevLett.77.570} {\bibfield  {journal} {\bibinfo  {journal} {Phys. Rev. Lett.}\ }\textbf {\bibinfo {volume} {77}},\ \bibinfo {pages} {570} (\bibinfo {year} {1996})}\BibitemShut {NoStop}%
\bibitem [{\citenamefont {Lin}\ \emph {et~al.}(2023)\citenamefont {Lin}, \citenamefont {Tai}, \citenamefont {Li},\ and\ \citenamefont {Lee}}]{lin2023topological}%
  \BibitemOpen
  \bibfield  {author} {\bibinfo {author} {\bibfnamefont {R.}~\bibnamefont {Lin}}, \bibinfo {author} {\bibfnamefont {T.}~\bibnamefont {Tai}}, \bibinfo {author} {\bibfnamefont {L.}~\bibnamefont {Li}}, \ and\ \bibinfo {author} {\bibfnamefont {C.~H.}\ \bibnamefont {Lee}},\ }\href@noop {} {\bibfield  {journal} {\bibinfo  {journal} {Frontiers of Physics}\ }\textbf {\bibinfo {volume} {18}},\ \bibinfo {pages} {53605} (\bibinfo {year} {2023})}\BibitemShut {NoStop}%
\bibitem [{\citenamefont {Tai}\ and\ \citenamefont {Lee}(2023)}]{tai2023zoology}%
  \BibitemOpen
  \bibfield  {author} {\bibinfo {author} {\bibfnamefont {T.}~\bibnamefont {Tai}}\ and\ \bibinfo {author} {\bibfnamefont {C.~H.}\ \bibnamefont {Lee}},\ }\href@noop {} {\bibfield  {journal} {\bibinfo  {journal} {Physical Review B}\ }\textbf {\bibinfo {volume} {107}},\ \bibinfo {pages} {L220301} (\bibinfo {year} {2023})}\BibitemShut {NoStop}%
\bibitem [{\citenamefont {Zou}\ \emph {et~al.}(2024)\citenamefont {Zou}, \citenamefont {Chen}, \citenamefont {Meng}, \citenamefont {Ang}, \citenamefont {Zhang},\ and\ \citenamefont {Lee}}]{Zou2023wbo}%
  \BibitemOpen
  \bibfield  {author} {\bibinfo {author} {\bibfnamefont {D.}~\bibnamefont {Zou}}, \bibinfo {author} {\bibfnamefont {T.}~\bibnamefont {Chen}}, \bibinfo {author} {\bibfnamefont {H.}~\bibnamefont {Meng}}, \bibinfo {author} {\bibfnamefont {Y.~S.}\ \bibnamefont {Ang}}, \bibinfo {author} {\bibfnamefont {X.}~\bibnamefont {Zhang}}, \ and\ \bibinfo {author} {\bibfnamefont {C.~H.}\ \bibnamefont {Lee}},\ }\href {\doibase 10.1016/j.scib.2024.05.036} {\bibfield  {journal} {\bibinfo  {journal} {Sci. Bull.}\ }\textbf {\bibinfo {volume} {69}},\ \bibinfo {pages} {2194} (\bibinfo {year} {2024})},\ \Eprint {http://arxiv.org/abs/2308.01970} {arXiv:2308.01970 [quant-ph]} \BibitemShut {NoStop}%
\bibitem [{\citenamefont {Hofmann}\ \emph {et~al.}(2020)\citenamefont {Hofmann} \emph {et~al.}}]{Hofmann2020pzl}%
  \BibitemOpen
  \bibfield  {author} {\bibinfo {author} {\bibfnamefont {T.}~\bibnamefont {Hofmann}} \emph {et~al.},\ }\href {\doibase 10.1103/PhysRevResearch.2.023265} {\bibfield  {journal} {\bibinfo  {journal} {Phys. Rev. Res.}\ }\textbf {\bibinfo {volume} {2}},\ \bibinfo {pages} {023265} (\bibinfo {year} {2020})}\BibitemShut {NoStop}%
\bibitem [{\citenamefont {Zhu}\ \emph {et~al.}(2020)\citenamefont {Zhu}, \citenamefont {Wang}, \citenamefont {Gupta}, \citenamefont {Zhang}, \citenamefont {Xie}, \citenamefont {Lu},\ and\ \citenamefont {Chen}}]{zhu2020photonic}%
  \BibitemOpen
  \bibfield  {author} {\bibinfo {author} {\bibfnamefont {X.}~\bibnamefont {Zhu}}, \bibinfo {author} {\bibfnamefont {H.}~\bibnamefont {Wang}}, \bibinfo {author} {\bibfnamefont {S.~K.}\ \bibnamefont {Gupta}}, \bibinfo {author} {\bibfnamefont {H.}~\bibnamefont {Zhang}}, \bibinfo {author} {\bibfnamefont {B.}~\bibnamefont {Xie}}, \bibinfo {author} {\bibfnamefont {M.}~\bibnamefont {Lu}}, \ and\ \bibinfo {author} {\bibfnamefont {Y.}~\bibnamefont {Chen}},\ }\href@noop {} {\bibfield  {journal} {\bibinfo  {journal} {Physical Review Research}\ }\textbf {\bibinfo {volume} {2}},\ \bibinfo {pages} {013280} (\bibinfo {year} {2020})}\BibitemShut {NoStop}%
\bibitem [{\citenamefont {Song}\ \emph {et~al.}(2020)\citenamefont {Song}, \citenamefont {Liu}, \citenamefont {Zheng}, \citenamefont {Zhang}, \citenamefont {Wang},\ and\ \citenamefont {Lu}}]{song2020two}%
  \BibitemOpen
  \bibfield  {author} {\bibinfo {author} {\bibfnamefont {Y.}~\bibnamefont {Song}}, \bibinfo {author} {\bibfnamefont {W.}~\bibnamefont {Liu}}, \bibinfo {author} {\bibfnamefont {L.}~\bibnamefont {Zheng}}, \bibinfo {author} {\bibfnamefont {Y.}~\bibnamefont {Zhang}}, \bibinfo {author} {\bibfnamefont {B.}~\bibnamefont {Wang}}, \ and\ \bibinfo {author} {\bibfnamefont {P.}~\bibnamefont {Lu}},\ }\href@noop {} {\bibfield  {journal} {\bibinfo  {journal} {Physical Review Applied}\ }\textbf {\bibinfo {volume} {14}},\ \bibinfo {pages} {064076} (\bibinfo {year} {2020})}\BibitemShut {NoStop}%
\bibitem [{\citenamefont {Lin}\ \emph {et~al.}(2024)\citenamefont {Lin}, \citenamefont {Song}, \citenamefont {Wang}, \citenamefont {Xin}, \citenamefont {Sun}, \citenamefont {Wu}, \citenamefont {Huang}, \citenamefont {Zhu}, \citenamefont {Jiang},\ and\ \citenamefont {Li}}]{lin2024observation}%
  \BibitemOpen
  \bibfield  {author} {\bibinfo {author} {\bibfnamefont {Z.}~\bibnamefont {Lin}}, \bibinfo {author} {\bibfnamefont {W.}~\bibnamefont {Song}}, \bibinfo {author} {\bibfnamefont {L.-W.}\ \bibnamefont {Wang}}, \bibinfo {author} {\bibfnamefont {H.}~\bibnamefont {Xin}}, \bibinfo {author} {\bibfnamefont {J.}~\bibnamefont {Sun}}, \bibinfo {author} {\bibfnamefont {S.}~\bibnamefont {Wu}}, \bibinfo {author} {\bibfnamefont {C.}~\bibnamefont {Huang}}, \bibinfo {author} {\bibfnamefont {S.}~\bibnamefont {Zhu}}, \bibinfo {author} {\bibfnamefont {J.-H.}\ \bibnamefont {Jiang}}, \ and\ \bibinfo {author} {\bibfnamefont {T.}~\bibnamefont {Li}},\ }\href@noop {} {\bibfield  {journal} {\bibinfo  {journal} {Physical Review Letters}\ }\textbf {\bibinfo {volume} {133}},\ \bibinfo {pages} {073803} (\bibinfo {year} {2024})}\BibitemShut {NoStop}%
\bibitem [{\citenamefont {Smith}\ \emph {et~al.}(2019)\citenamefont {Smith}, \citenamefont {Kim}, \citenamefont {Pollmann},\ and\ \citenamefont {Knolle}}]{smith2019simulating}%
  \BibitemOpen
  \bibfield  {author} {\bibinfo {author} {\bibfnamefont {A.}~\bibnamefont {Smith}}, \bibinfo {author} {\bibfnamefont {M.}~\bibnamefont {Kim}}, \bibinfo {author} {\bibfnamefont {F.}~\bibnamefont {Pollmann}}, \ and\ \bibinfo {author} {\bibfnamefont {J.}~\bibnamefont {Knolle}},\ }\href@noop {} {\bibfield  {journal} {\bibinfo  {journal} {npj Quantum Information}\ }\textbf {\bibinfo {volume} {5}},\ \bibinfo {pages} {106} (\bibinfo {year} {2019})}\BibitemShut {NoStop}%
\bibitem [{\citenamefont {Xin}\ \emph {et~al.}(2020)\citenamefont {Xin}, \citenamefont {Li}, \citenamefont {Fan}, \citenamefont {Zhu}, \citenamefont {Zhang}, \citenamefont {Nie}, \citenamefont {Li}, \citenamefont {Liu},\ and\ \citenamefont {Lu}}]{xin2020quantum}%
  \BibitemOpen
  \bibfield  {author} {\bibinfo {author} {\bibfnamefont {T.}~\bibnamefont {Xin}}, \bibinfo {author} {\bibfnamefont {Y.}~\bibnamefont {Li}}, \bibinfo {author} {\bibfnamefont {Y.-a.}\ \bibnamefont {Fan}}, \bibinfo {author} {\bibfnamefont {X.}~\bibnamefont {Zhu}}, \bibinfo {author} {\bibfnamefont {Y.}~\bibnamefont {Zhang}}, \bibinfo {author} {\bibfnamefont {X.}~\bibnamefont {Nie}}, \bibinfo {author} {\bibfnamefont {J.}~\bibnamefont {Li}}, \bibinfo {author} {\bibfnamefont {Q.}~\bibnamefont {Liu}}, \ and\ \bibinfo {author} {\bibfnamefont {D.}~\bibnamefont {Lu}},\ }\href@noop {} {\bibfield  {journal} {\bibinfo  {journal} {Physical Review Letters}\ }\textbf {\bibinfo {volume} {125}},\ \bibinfo {pages} {090502} (\bibinfo {year} {2020})}\BibitemShut {NoStop}%
\bibitem [{\citenamefont {Mei}\ \emph {et~al.}(2020)\citenamefont {Mei}, \citenamefont {Guo}, \citenamefont {Yu}, \citenamefont {Xiao}, \citenamefont {Zhu},\ and\ \citenamefont {Jia}}]{mei2020digital}%
  \BibitemOpen
  \bibfield  {author} {\bibinfo {author} {\bibfnamefont {F.}~\bibnamefont {Mei}}, \bibinfo {author} {\bibfnamefont {Q.}~\bibnamefont {Guo}}, \bibinfo {author} {\bibfnamefont {Y.-F.}\ \bibnamefont {Yu}}, \bibinfo {author} {\bibfnamefont {L.}~\bibnamefont {Xiao}}, \bibinfo {author} {\bibfnamefont {S.-L.}\ \bibnamefont {Zhu}}, \ and\ \bibinfo {author} {\bibfnamefont {S.}~\bibnamefont {Jia}},\ }\href@noop {} {\bibfield  {journal} {\bibinfo  {journal} {Physical Review Letters}\ }\textbf {\bibinfo {volume} {125}},\ \bibinfo {pages} {160503} (\bibinfo {year} {2020})}\BibitemShut {NoStop}%
\bibitem [{\citenamefont {Chen}\ \emph {et~al.}(2023{\natexlab{a}})\citenamefont {Chen}, \citenamefont {Shen}, \citenamefont {Lee}, \citenamefont {Yang},\ and\ \citenamefont {Bomantara}}]{Chen2023zlo}%
  \BibitemOpen
  \bibfield  {author} {\bibinfo {author} {\bibfnamefont {T.}~\bibnamefont {Chen}}, \bibinfo {author} {\bibfnamefont {R.}~\bibnamefont {Shen}}, \bibinfo {author} {\bibfnamefont {C.~H.}\ \bibnamefont {Lee}}, \bibinfo {author} {\bibfnamefont {B.}~\bibnamefont {Yang}}, \ and\ \bibinfo {author} {\bibfnamefont {R.~W.}\ \bibnamefont {Bomantara}},\ }\href@noop {} {\  (\bibinfo {year} {2023}{\natexlab{a}})},\ \Eprint {http://arxiv.org/abs/2309.11560} {arXiv:2309.11560 [quant-ph]} \BibitemShut {NoStop}%
\bibitem [{\citenamefont {Chen}\ \emph {et~al.}(2024)\citenamefont {Chen}, \citenamefont {Ding}, \citenamefont {Shen}, \citenamefont {Zhu},\ and\ \citenamefont {Gong}}]{Chen2024ull}%
  \BibitemOpen
  \bibfield  {author} {\bibinfo {author} {\bibfnamefont {T.}~\bibnamefont {Chen}}, \bibinfo {author} {\bibfnamefont {H.-T.}\ \bibnamefont {Ding}}, \bibinfo {author} {\bibfnamefont {R.}~\bibnamefont {Shen}}, \bibinfo {author} {\bibfnamefont {S.-L.}\ \bibnamefont {Zhu}}, \ and\ \bibinfo {author} {\bibfnamefont {J.}~\bibnamefont {Gong}},\ }\href@noop {} {\  (\bibinfo {year} {2024})},\ \Eprint {http://arxiv.org/abs/2403.14249} {arXiv:2403.14249 [quant-ph]} \BibitemShut {NoStop}%
\bibitem [{\citenamefont {Zhang}\ \emph {et~al.}(2024)\citenamefont {Zhang}, \citenamefont {Carrasquilla},\ and\ \citenamefont {Kim}}]{zhang2024observation}%
  \BibitemOpen
  \bibfield  {author} {\bibinfo {author} {\bibfnamefont {Y.}~\bibnamefont {Zhang}}, \bibinfo {author} {\bibfnamefont {J.}~\bibnamefont {Carrasquilla}}, \ and\ \bibinfo {author} {\bibfnamefont {Y.~B.}\ \bibnamefont {Kim}},\ }\href@noop {} {\bibfield  {journal} {\bibinfo  {journal} {arXiv preprint arXiv:2406.15557}\ } (\bibinfo {year} {2024})}\BibitemShut {NoStop}%
\bibitem [{\citenamefont {Koh}\ \emph {et~al.}(2024)\citenamefont {Koh}, \citenamefont {Tai},\ and\ \citenamefont {Lee}}]{Koh2023ohr}%
  \BibitemOpen
  \bibfield  {author} {\bibinfo {author} {\bibfnamefont {J.~M.}\ \bibnamefont {Koh}}, \bibinfo {author} {\bibfnamefont {T.}~\bibnamefont {Tai}}, \ and\ \bibinfo {author} {\bibfnamefont {C.~H.}\ \bibnamefont {Lee}},\ }\href {\doibase 10.1038/s41467-024-49648-5} {\bibfield  {journal} {\bibinfo  {journal} {Nature Commun.}\ }\textbf {\bibinfo {volume} {15}},\ \bibinfo {pages} {5807} (\bibinfo {year} {2024})},\ \Eprint {http://arxiv.org/abs/2303.02179} {arXiv:2303.02179 [cond-mat.str-el]} \BibitemShut {NoStop}%
\bibitem [{\citenamefont {Chen}\ \emph {et~al.}(2023{\natexlab{b}})\citenamefont {Chen}, \citenamefont {Shen}, \citenamefont {Lee},\ and\ \citenamefont {Yang}}]{Chen2022owo}%
  \BibitemOpen
  \bibfield  {author} {\bibinfo {author} {\bibfnamefont {T.}~\bibnamefont {Chen}}, \bibinfo {author} {\bibfnamefont {R.}~\bibnamefont {Shen}}, \bibinfo {author} {\bibfnamefont {C.~H.}\ \bibnamefont {Lee}}, \ and\ \bibinfo {author} {\bibfnamefont {B.}~\bibnamefont {Yang}},\ }\href {\doibase 10.21468/SciPostPhys.15.4.170} {\bibfield  {journal} {\bibinfo  {journal} {SciPost Phys.}\ }\textbf {\bibinfo {volume} {15}},\ \bibinfo {pages} {170} (\bibinfo {year} {2023}{\natexlab{b}})},\ \Eprint {http://arxiv.org/abs/2210.13840} {arXiv:2210.13840 [quant-ph]} \BibitemShut {NoStop}%
\bibitem [{\citenamefont {Koh}\ \emph {et~al.}(2022)\citenamefont {Koh}, \citenamefont {Tai},\ and\ \citenamefont {Lee}}]{koh2022simulation}%
  \BibitemOpen
  \bibfield  {author} {\bibinfo {author} {\bibfnamefont {J.~M.}\ \bibnamefont {Koh}}, \bibinfo {author} {\bibfnamefont {T.}~\bibnamefont {Tai}}, \ and\ \bibinfo {author} {\bibfnamefont {C.~H.}\ \bibnamefont {Lee}},\ }\href@noop {} {\bibfield  {journal} {\bibinfo  {journal} {Physical Review Letters}\ }\textbf {\bibinfo {volume} {129}},\ \bibinfo {pages} {140502} (\bibinfo {year} {2022})}\BibitemShut {NoStop}%
\bibitem [{\citenamefont {Zhu}\ and\ \citenamefont {Gong}(2022)}]{Zhu2022}%
  \BibitemOpen
  \bibfield  {author} {\bibinfo {author} {\bibfnamefont {W.}~\bibnamefont {Zhu}}\ and\ \bibinfo {author} {\bibfnamefont {J.}~\bibnamefont {Gong}},\ }\href {\doibase 10.1103/PhysRevB.106.035425} {\bibfield  {journal} {\bibinfo  {journal} {Phys. Rev. B}\ }\textbf {\bibinfo {volume} {106}},\ \bibinfo {pages} {1} (\bibinfo {year} {2022})}\BibitemShut {NoStop}%
\bibitem [{\citenamefont {Li}\ \emph {et~al.}(2022{\natexlab{b}})\citenamefont {Li}, \citenamefont {Liang}, \citenamefont {Wang}, \citenamefont {Lu},\ and\ \citenamefont {Liu}}]{Li2022d}%
  \BibitemOpen
  \bibfield  {author} {\bibinfo {author} {\bibfnamefont {Y.}~\bibnamefont {Li}}, \bibinfo {author} {\bibfnamefont {C.}~\bibnamefont {Liang}}, \bibinfo {author} {\bibfnamefont {C.}~\bibnamefont {Wang}}, \bibinfo {author} {\bibfnamefont {C.}~\bibnamefont {Lu}}, \ and\ \bibinfo {author} {\bibfnamefont {Y.~C.}\ \bibnamefont {Liu}},\ }\href {\doibase 10.1103/PhysRevLett.128.223903} {\bibfield  {journal} {\bibinfo  {journal} {Phys. Rev. Lett.}\ }\textbf {\bibinfo {volume} {128}},\ \bibinfo {pages} {223903} (\bibinfo {year} {2022}{\natexlab{b}})}\BibitemShut {NoStop}%
\bibitem [{\citenamefont {Liu}\ \emph {et~al.}(2021)\citenamefont {Liu}, \citenamefont {Zeng}, \citenamefont {Li},\ and\ \citenamefont {Chen}}]{liu2021exact}%
  \BibitemOpen
  \bibfield  {author} {\bibinfo {author} {\bibfnamefont {Y.}~\bibnamefont {Liu}}, \bibinfo {author} {\bibfnamefont {Y.}~\bibnamefont {Zeng}}, \bibinfo {author} {\bibfnamefont {L.}~\bibnamefont {Li}}, \ and\ \bibinfo {author} {\bibfnamefont {S.}~\bibnamefont {Chen}},\ }\href@noop {} {\bibfield  {journal} {\bibinfo  {journal} {Physical Review B}\ }\textbf {\bibinfo {volume} {104}},\ \bibinfo {pages} {085401} (\bibinfo {year} {2021})}\BibitemShut {NoStop}%
\bibitem [{\citenamefont {Guo}\ \emph {et~al.}(2021)\citenamefont {Guo}, \citenamefont {Liu}, \citenamefont {Zhao}, \citenamefont {Liu},\ and\ \citenamefont {Chen}}]{guo2021exact}%
  \BibitemOpen
  \bibfield  {author} {\bibinfo {author} {\bibfnamefont {C.-X.}\ \bibnamefont {Guo}}, \bibinfo {author} {\bibfnamefont {C.-H.}\ \bibnamefont {Liu}}, \bibinfo {author} {\bibfnamefont {X.-M.}\ \bibnamefont {Zhao}}, \bibinfo {author} {\bibfnamefont {Y.}~\bibnamefont {Liu}}, \ and\ \bibinfo {author} {\bibfnamefont {S.}~\bibnamefont {Chen}},\ }\href@noop {} {\bibfield  {journal} {\bibinfo  {journal} {Physical Review Letters}\ }\textbf {\bibinfo {volume} {127}},\ \bibinfo {pages} {116801} (\bibinfo {year} {2021})}\BibitemShut {NoStop}%
\bibitem [{\citenamefont {Bradley}\ and\ \citenamefont {Strenski}(1985)}]{bradley1985directed}%
  \BibitemOpen
  \bibfield  {author} {\bibinfo {author} {\bibfnamefont {R.}~\bibnamefont {Bradley}}\ and\ \bibinfo {author} {\bibfnamefont {P.}~\bibnamefont {Strenski}},\ }\href@noop {} {\bibfield  {journal} {\bibinfo  {journal} {Physical Review B}\ }\textbf {\bibinfo {volume} {31}},\ \bibinfo {pages} {4319} (\bibinfo {year} {1985})}\BibitemShut {NoStop}%
\bibitem [{\citenamefont {Saberi}(2013)}]{saberi2013growth}%
  \BibitemOpen
  \bibfield  {author} {\bibinfo {author} {\bibfnamefont {A.~A.}\ \bibnamefont {Saberi}},\ }\href@noop {} {\bibfield  {journal} {\bibinfo  {journal} {EPL (Europhysics Letters)}\ }\textbf {\bibinfo {volume} {103}},\ \bibinfo {pages} {10005} (\bibinfo {year} {2013})}\BibitemShut {NoStop}%
\end{thebibliography}%
\end{document}